\documentclass[aps,rmp,reprint,nolongbibliography,amsmath,amssymb]{revtex4-2}
\usepackage[T1]{fontenc}
\usepackage[table]{xcolor}
\usepackage{braket}
\usepackage{enumerate}
\usepackage{csquotes}
\usepackage{mathtools}
\definecolor{customdarkblue}{rgb}{0.0, 0.0, 0.75}  
\usepackage{graphicx}
\usepackage{hyperref}
\usepackage{multirow}
\usepackage{comment}
\hypersetup{
    colorlinks=true,    
    urlcolor=blue,      
    linkcolor=blue,    
    citecolor=customdarkblue     
}
\usepackage[colorinlistoftodos,prependcaption]{todonotes} 

\usepackage{array}        
\usepackage{makecell}     
\usepackage{enumitem}     
\usepackage{ragged2e}     

\usepackage[percent]{overpic}

\usepackage{tcolorbox}

\newcolumntype{C}[1]{>{\centering\arraybackslash}m{#1}}  
\newcolumntype{L}[1]{>{\raggedright\arraybackslash}m{#1}}  

\setlist[itemize]{leftmargin=*,nosep}

\usepackage{tabularray}          
\UseTblrLibrary{booktabs,varwidth}
\usepackage{hhline}

\usepackage{microtype}
\definecolor{darkgreen}{RGB}{0, 146, 69}

\definecolor{oilightorange}{HTML}{FCE8D2}
\definecolor{oilightblue}{HTML}{DCEFF7}

\newcommand{\tr}{\mathrm{tr}}

\newcommand{\implementations}{\noindent \textbf{Experimental progress}.\qquad}

\usepackage{etoolbox}
\makeatletter
\newcommand*{\sortentry}[1]{%
  \if@filesw
    \immediate\write\@auxout{\string\scNAT@aux@sortentry{#1}}%
  \fi}
\newcommand*{\scNAT@aux@sortentry}{\listgadd{\scNAT@bibsortlist}}
\newcommand*{\scNAT@bibsortlist}{}
\newcommand*{\scNAT@citekeys}{}
\newcommand*{\scNAT@foundkeys}{}

\newcommand*{\scNAT@writetocitelistsort}[1]{%
  \ifinlist{#1}{\scNAT@citekeys}
    {\ifdefvoid{\NAT@cite@list}
       {\def\NAT@cite@list{#1}}
       {\expandafter\def\expandafter\NAT@cite@list\expandafter{\NAT@cite@list,#1}}%
     \listgadd{\scNAT@foundkeys}{#1}}{}}

\newcommand*{\scNAT@writetocitelistforgotten}[1]{%
  \ifinlist{#1}{\scNAT@foundkeys}{}{%
    \ifdefvoid{\NAT@cite@list}
      {\def\NAT@cite@list{#1}}
      {\expandafter\def\expandafter\NAT@cite@list\expandafter{\NAT@cite@list,#1}}}}

\newcommand*{\scNAT@sortcites}[1]{%
  \let\NAT@cite@list\@empty
  \let\scNAT@citekeys\@empty
  \let\scNAT@foundkeys\@empty
  \forcsvlist{\listadd{\scNAT@citekeys}}{#1}%
  \forlistloop{\scNAT@writetocitelistsort}{\scNAT@bibsortlist}%
  \forlistloop{\scNAT@writetocitelistforgotten}{\scNAT@citekeys}%
}

\patchcmd{\NAT@citex}{\NAT@reset@citea}{\scNAT@sortcites{#3}\NAT@reset@citea}{}{}
\makeatother

\begin{document}

\title[Distributed Quantum Information Processing --- Review Outline]{Distributed Quantum Information Processing: \\ A Review of Recent Progress}

\author{J.~Knörzer$^{1,2,3}$, X.~Liu$^{4,5}$, B.~F.~Schiffer$^6$, J.~Tura$^{4,5}$}

\address{\parbox{\textwidth}{%
$^1$Department of Physics, ETH Zürich, CH-8093 Zürich, Switzerland; \href{mailto:jknoerzer@ethz.ch}{jknoerzer@ethz.ch}\\
$^2$Quantum Center, ETH Zürich, CH-8093 Zürich, Switzerland\\
$^3$ETH Zürich---PSI Quantum Computing Hub, Paul Scherrer Institut, CH-5232 Villigen, Switzerland\\
$^4$Instituut-Lorentz, Universiteit Leiden, P.O. Box 9506, 2300 RA Leiden, The Netherlands\\
$^5$$\langle aQa ^L\rangle $ Applied Quantum Algorithms, Universiteit Leiden\\
$^6$Max-Planck-Institut für Quantenoptik, Hans-Kopfermann-Str.~1, D-85748 Garching, Germany}
}


\begin{abstract}
Distributed quantum information processing seeks to overcome the scalability limitations of monolithic quantum devices by interconnecting multiple quantum processing nodes via classical and quantum communication.
This approach extends the capabilities of individual devices, enabling access to larger problem instances and novel algorithmic techniques.
Beyond increasing qubit counts, it also enables qualitatively new capabilities, such as joint measurements on multiple copies of high-dimensional quantum states.
The distinction between single-copy and multi-copy access reveals important differences in task complexity and helps identify which computational problems stand to benefit from distributed quantum resources.
At the same time, it highlights trade-offs between classical and quantum communication models and the practical challenges involved in realizing them experimentally.
In this review, we contextualize recent developments by surveying the theoretical foundations of distributed quantum protocols and examining the experimental platforms and algorithmic applications that realize them in practice.
\end{abstract}

\date{\today}

\maketitle

\tableofcontents

\section{Introduction}

\subsection{Background}

Distributed computing encompasses methods that enable multiple computers to solve complex problems together.
This cooperation relies on communication between individual computing nodes over a shared network.
Early distributed systems, such as Ethernet, were local-area networks designed to interconnect computers within confined areas such as hospitals, schools or office buildings.
By the late 1970s and early 1980s, the study of distributed computing had established itself as a veritable branch of computer science~\cite{steen2017distributed}.
Today it underpins numerous applications in telecommunications, scientific computing and banking.

Distributed architectures may benefit not only classical computing but also quantum information processing (QIP) and, in particular, quantum computing.
The term \emph{distributed QIP} refers to scenarios in which quantum devices can exchange either classical or quantum information.
This distinction is crucial:
while classical communication is much easier to realize, many applications in quantum science and technology benefit significantly from the ability to transfer quantum information.
Networks that enable transmission of quantum information are known as \emph{quantum networks}~\cite{kimble2008quantum}.

Quantum networks are built on three essential components:
(i) \emph{nodes}, which are quantum processors that generate, manipulate, and measure quantum states;
(ii) \emph{communication channels}, which are used to transfer quantum and classical data between nodes;
and (iii) \emph{quantum repeaters}, which extend the range of quantum communication~\cite{azuma2023quantum}.
Unlike classical signals, quantum states cannot be directly amplified due to the no-cloning theorem~\cite{wootters1982single,dieks1982communication}.
Instead, quantum repeaters mitigate photon loss and decoherence through specialized techniques such as entanglement swapping~\cite{zukowski1993eventreadydetectors} and entanglement distillation~\cite{bennett1996purification,bennett1996concentrating}.

The development of quantum networks represents a crucial step toward scaling quantum technologies beyond the limits of individual devices.
Through quantum communication between processors, quantum networks allow one to perform joint computations and measurements, coordinate quantum error correction across devices, and share entanglement as a resource for cryptography and sensing.
The idea of distributed quantum computing dates back to the late 1990s, when theoretical proposals showed how entanglement could be harnessed to perform quantum computations across spatially separated nodes~\cite{cirac1999distributed,eisert2000optimal}.
At the time, these proposals were far ahead of what experimental hardware could support.
However, in recent years, significant progress has brought them closer to reality.
Key capabilities have been demonstrated in laboratory-scale systems~\cite{magnard2020microwave,alshowkan2021reconfigurable,yam2025cryogenic}, metropolitan-scale networks~\cite{yu2020entanglement,stolk2024metropolitanscale,liu2024creation}, and even through satellite-based communication channels~\cite{deforgesdeparny2023satellitebased}.
Moreover, theoretical proposals have been further improved and distributed quantum protocols have been studied on more formal grounds.
With these advances, it is timely to explore the new possibilities that quantum communication enables for information science, and for distributed quantum computing, in particular~\cite{vanmeter2016path,cuomo2020distributed}.
In this review, we survey the theoretical foundations and applications of distributed QIP, and examine the experimental progress that is beginning to translate these concepts into working systems.

\subsection{Scope of this review}

Although integrating multiple quantum systems, from local modules to geographically separated nodes, is a key objective of several quantum technologies, this review focuses mainly on distributed quantum algorithms and the computational tasks they enable.
We do not provide a comprehensive overview of the broader field of quantum communication and networking, which has been extensively covered in reviews dedicated to quantum memories, entanglement distribution, quantum repeaters, and fault-tolerant networking protocols~\cite{simon2010quantum,sangouard2011quantum,terhal2015quantum,azuma2023quantum}.
For overviews of broader architectural visions, including the quantum internet, we refer the reader to~\cite{reiserer2022colloquium,wehner2018quantum,chung2025interqnet} and references therein.
Also beyond the scope of this work is a detailed overview of progress in quantum cryptography; we refer to the comprehensive reviews~\cite{gisin2002quantum, pirandola2020advances}.

We begin by reviewing key protocols and the most advanced physical realizations of quantum networks in Sec.~\ref{sec:implementations}, where our focus is on high-level principles and design considerations rather than technical specifics.
Across different platforms, light-matter interactions underpin the fundamental processes that enable quantum state transfer and remote entanglement generation between spatially separated nodes.
We conclude the section by surveying recent advances in implementing these concepts on various physical platforms.

Sec.~\ref{sec:concepts} outlines the theoretical framework for modeling and understanding the generation, transfer, and manipulation of quantum information in distributed systems.
We begin with an introduction to classical and quantum communication channels, which can be used to model experiments such as the transmission of single photons through optical fibers.
The discussion then turns to quantum teleportation, quantum circuits, and quantum state learning, together with key metrics for assessing quantum experiments, such as state and process fidelities.
For more comprehensive treatments of the fundamental theory of quantum information, we refer the reader to standard textbooks~\cite{nielsen2012quantum} and recent reviews~\cite{horodecki2021quantum}.

Sec.~\ref{sec:algorithms-applications} presents algorithms and applications tailored to distributed quantum settings.
It begins with Sec.~\ref{ssec:distributed-adaptations}, where we summarize adaptations of known algorithms to modular architectures.
Whereas Sec.~\ref{sec:concepts} focuses on established principles, this section highlights more recent developments that illustrate the breadth and growing relevance of the field.  
A central insight driving much of this research is the advantage of coherent access to multiple copies of a quantum state $\rho$~\cite{aharonov2022quantum}, which may be achieved in monolithic devices but can benefit from modular setups.  
For instance, two copies of $\rho$ enable resource-efficient estimation of its purity, $\tr(\rho^2)$, and the evaluation of arbitrary $n$-qubit Pauli observables via Bell sampling~\cite{huang2022quantum,king2024triply,chen2024optimal}.  
Sec.~\ref{ssec:two-copies} surveys such problems and two-copy algorithms, discussing their significance and resource requirements.

Beyond the advantages of two-copy measurements, there are strong motivations to explore applications that exploit access to multiple copies of a quantum state $\rho$, despite the added experimental complexity.  
Such access enables, for example, the estimation of traces of higher powers, $\tr(\rho^K)$ for $K\geq 2$, which are essential for characterizing nonlinear state properties and quantifying correlations within the state~\cite{shin2025resourceefficient,quek2024multivariate,liu2024estimating}.  
It also allows the evaluation of entanglement measures such as Rényi entropies~\cite{linke2018measuring} and concentratable entanglement~\cite{beckey2021computable,liu2025generalized}.
Multi-copy strategies further underpin techniques like virtual distillation, a key approach to quantum error mitigation~\cite{cai2023quantum}, and virtual cooling~\cite{cotler2019quantum}, where collective measurements on finite-temperature systems simulate observations at lower effective temperatures.  
These and related algorithms, along with their significance and resource requirements, are discussed in Secs.~\ref{ssec:multi-copy} and~\ref{ssec:memory-queries}.

In Sec.~\ref{ssec:complexity}, we will provide a comprehensive summary of the known complexity results for the tasks presented in this review, comparing the scenarios where one has access to multiple copies of a quantum state versus being restricted to single-copy access.
Additionally, we address situations involving limited quantum communication, where, for instance, one might have access to more than a single copy but less than two exact copies.
These intermediate cases reveal unique computational trade-offs and can be critical for practical applications where full two-copy access is challenging to implement.
Furthermore, we will outline what is known about tasks for which no distributed quantum advantage exists, offering insights into the boundaries of quantum advantage within these settings.
By exploring these distinctions, we aim to clarify the resource requirements and limitations for a variety of quantum information processing tasks.

As discussed in Sec.~\ref{ssec:verification}, distributed settings enable cross-platform state and process verification, blind quantum computing, and interactive-proof protocols, allowing correctness to be established even when computations are delegated to untrusted devices.

Sec.~\ref{ssec:qem-qec} reviews error mitigation and correction strategies, from virtual distillation to full quantum error-correcting codes coordinated across network links, which are essential for overcoming noise limitations.

Distributed quantum machine learning, covered in Sec.~\ref{ssec:qml}, can leverage parallel access to quantum data and modular execution to accelerate training and inference, and to integrate hybrid classical–quantum resources across nodes.

In state preparation, described in Sec.~\ref{ssec:state-preparation}, distributed protocols can accelerate convergence to eigenstates of a target Hamiltonian, for example via adiabatic methods enhanced by multi-device hypothesis testing or by distributed filtering schemes that iteratively reduce energy variance using entanglement between auxiliary qubits across nodes.

Sec.~\ref{ssec:q-metrology} outlines how quantum metrology can benefit from a quantum network of entangled sensors.
Leveraging entanglement across multiple probes allows for precision measurements beyond the standard quantum limit, or shot-noise limit, and to probe nonlocal observables encoded across spatially separated nodes.

Finally, in Sec.~\ref{sec:outlook} we discuss current challenges and future directions for distributed quantum computing.
We identify key open questions and bottlenecks in the field, both of theoretical and experimental nature.
Answering these will set the stage for future advancements with scalable modular architectures for quantum information processing.

\begin{table*}[t]
\centering
\begin{tblr}{
  colspec = {Q[l,2.7cm]|Q[l,6cm]|Q[l,6cm]},
  row{1} = {font=\bfseries, valign=t},
  row{2-Z} = {valign=t},          
  stretch = 1.2,
}
 & Classical Data & Quantum Data \\\hline

\textbf{Classical\\Communication} &
\textbf{Classical distributed computing}, examples of tasks \& protocols:
\begin{itemize}
  \item MapReduce
  \item Byzantine agreement
  \item Leader election, consensus
\end{itemize} &
\SetCell{bg=oilightblue}
\textbf{With entanglement:}
\begin{itemize}
  \item Quantum teleportation (Sec.~\ref{ssec:teleportation})
  \item CHSH game (Sec.~\ref{ssec:bell-nonlocality})
  \item Entanglement distillation (Sec.~\ref{ssec:ent-dist})
\end{itemize}
\textbf{No entanglement:}
\begin{itemize}
  \item Circuit knitting (Sec.~\ref{ssec:quantum-circuits})
  \item Shadow tomography (Sec.~\ref{ssec:randomized-measurements})
\end{itemize}
\\\hline

\textbf{Quantum\\Communication} &
\textbf{Communication protocols:}
\begin{itemize}
  \item Superdense coding
  \item Quantum Key Distribution
\end{itemize}
&
\SetCell{bg=oilightorange}
\textbf{Quantum algorithms:}
\begin{itemize}
  \item Quantum state transfer (Sec.~\ref{ssec:interconnects-protocols})
  \item Bell sampling (Sec.~\ref{ssec:two-copies})
  \item Multi-copy measurements (Sec.~\ref{ssec:multi-copy})
  \item Memory-usage queries (Sec.~\ref{ssec:memory-queries})
  \item Verification (Sec.~\ref{ssec:verification})
\end{itemize}
\\
\end{tblr}
\caption{Overview of communication settings and data types, each classified as classical or quantum. 
A non-exhaustive selection of tasks and frameworks are listed at the intersection of these categories and discussed in detail later in this review.}
\label{tab:overview-communication-settings}
\end{table*}

\section{Physical Realizations \label{sec:implementations}}

A central challenge in scaling quantum computing lies in connecting qubits across different hardware components.
This challenge takes different forms depending on the length scale involved:
at the on-chip level, the goal is to expand the connectivity of qubits beyond nearest neighbors;
at the multi-chip level, the focus shifts to integrating distinct quantum modules into a single logical system;
and at the inter-device level, the task becomes the coherent coupling of spatially separated quantum processors.
Distributed quantum architectures aim to meet these challenges by enabling quantum state transfer and entanglement generation across these physical scales.
The feasibility and implementation strategies for such architectures vary widely across platforms, shaped by both physical constraints and technology-specific capabilities.

In the following sections, we review experimental progress on quantum interconnects~\cite{kimble2008quantum, awschalom2021development} and distributed hardware across several leading quantum computing platforms.
We highlight how different platforms approach on-chip, multi-chip, and inter-device quantum communication, and how these capabilities inform their potential for distributed quantum information processing. 

\subsection{Modular architectures}

Modular architectures for quantum information processing aim to address the scaling challenge by connecting smaller subsystems into a larger, composite system, either to offload computational tasks, to achieve higher connectivity~\cite{beals2013efficient}, or to extend the quantum state space across modules~\cite{barral2025review}.
This modularity allows for dividing a large quantum computation across multiple physical nodes, each with fewer resources than the task might nominally require~\cite{haner2021distributed}.
Moreover, it facilitates hardware reusability and specialization, where certain nodes handle storage and others focus on distinct computational subroutines.

Different types of modularity can be distinguished by whether the nodes can exchange only classical data or also quantum information.
Classical-only approaches are arguably easier to realize and suffice for some tasks, but quantum links enable operations like state transfer, entanglement distribution, teleportation, and distributed gates.
Importantly, even if these quantum links are lossy, they can be useful~\cite{jiang2007distributed,nickerson2014freely,li2016hierarchical,covey2023quantum,ramette2024faulttolerant}.

The goal of complexity theory and, in particular, communication complexity is to study how much information must be exchanged between nodes to perform a certain task.
The early results of Holevo~\cite{holevo1973bounds} showed that, on average, the transmission of quantum bits does not allow more classical information to be transmitted than by sending classical bits, placing limits on the classical information capacity of quantum channels.
However, later work revealed tasks in which quantum protocols, especially those that leverage entanglement, can reduce the required communication, sometimes requiring exponentially fewer rounds of information exchange~\cite{buhrman2010nonlocality}.
These insights have led to a range of new algorithms that demonstrate the advantages of distributing quantum resources.
The different communication settings relevant to modular architectures are summarized in Table~\ref{tab:overview-communication-settings}, where we distinguish between classical and quantum communication for transmission of classical and quantum data, respectively.
In this review, while we also touch on hybrid approaches involving small quantum computers~\cite{bravyi2016trading,dunjko2018computational, piveteau2024circuit} connected by classical channels, we focus primarily on tasks that distribute quantum information across quantum networks.
We will now discuss protocols and realizations of quantum interconnects that enable such distributed quantum communication.

\subsection{Protocols for quantum interconnects \label{ssec:interconnects-protocols}}

We first summarize some of the key protocols for sharing quantum information across a network.
These protocols enable three major building blocks:
(i) quantum state transfer, (ii) transduction, and (iii) remote entanglement generation.

\subsubsection{Quantum state transfer \label{sssec:qst}}

\begin{figure}[b]
    \centering
    \includegraphics[width=\linewidth]{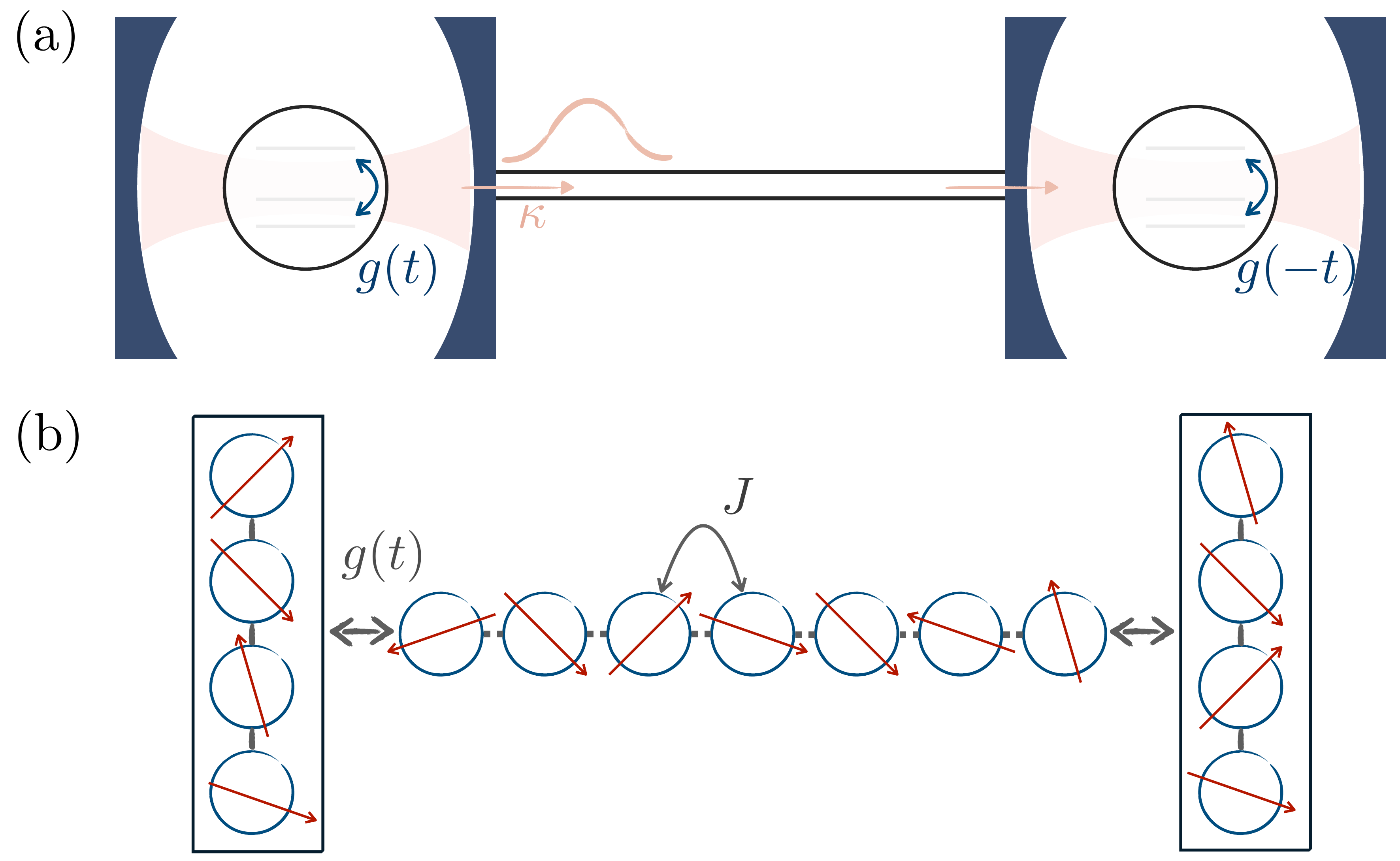}
    \caption{Deterministic quantum state transfer protocols.
    (a) Cavity quantum electrodynamics setup with a tunable coupling $g(t)$ and cavity decay rate $\kappa$.
    The photon decays into the quantum link and is reabsorbed at the second node.
    (b) Two registers connected by a quantum spin chain that is governed by a nearest-neighbor coupling at rate $J$.}
    \label{fig:state-transfer}
\end{figure}

Quantum state transfer refers to the transmission of a quantum state over a physical distance.
This can be accomplished in various ways, depending on the physical system and the degree of control available over its parameters.
The first basic scheme for quantum state transfer was proposed in~\cite{cirac1997quantum}.

A common approach is to convert stationary qubits into mobile carriers of information, so-called flying qubits, which can be realized using optical photons~\cite{ritter2012elementary}, microwave photons~\cite{axline2018ondemand,campagne-ibarcq2018deterministic}, or phonons~\cite{bienfait2019phononmediated}.
These flying qubits propagate through a transmission channel and may later be reabsorbed at a remote location.
Photonic flying qubits are commonly represented using a Fock-state encoding, defined by the presence or absence of a photon, a time-bin encoding, defined by orthogonal spatio-temporal modes of the quantum interconnect, dual-rail encodings or using the photon's polarization~\cite{northup2014quantum,beukers2024remoteentanglement}.

The emission and absorption of flying qubits are facilitated by coupling stationary qubits to optical or microwave cavities.  
By dynamically tuning the coupling between the qubit and the cavity mode, the timing and shape of the emitted photon wave packet can be controlled.  
Protocols that rely on such shaped emission followed by absorption at a distant node are referred to as \emph{pitch-and-catch} schemes; see Fig.~\ref{fig:state-transfer}(a) for a schematic depiction.

A key requirement for achieving high-fidelity transfer in pitch-and-catch protocols is the control of the temporal envelope of the emitted photon.  
This was formalized in~\cite{cirac1997quantum}, where it was demonstrated that deterministic state transfer between distant cavities is possible if the photon is emitted with a time-symmetric wave packet matched by the absorption dynamics of the receiver.
In this protocol, the qubit–cavity coupling is modulated in time to shape the emission profile, ensuring that the photon is absorbed without reflection or loss, in the ideal case. 
The approach provides a blueprint for coherent transfer of quantum states as well as intermodule two-qubit gates and has since been adapted to various physical platforms (Sec.~\ref{ssec:platforms}).

Faithful transmission of quantum states should be possible even in the presence of noisy channels~\cite{xiang2017intracity,vermersch2017quantum}.
However, there are several limitations of the idealized pitch-and-catch framework that recent works have studied.
Most protocols neglect distortions experienced by the photon wave packet during propagation, as well as non-Markovian effects arising from time-dependent control fields during emission and reabsorption.
These issues can be partially addressed with a correction strategy that improves the fidelity of quantum state transfer~\cite{penas2023improving}.
In scenarios where the spectral properties of the sender and receiver systems differ, direct transmission may lead to mismatched absorption.
This can be mitigated by applying a unitary transformation to the flying qubit, which aligns the photon's time-frequency profile with that of the receiver to enable higher-fidelity transfer~\cite{randles2023quantum}.
Building on this, the role of spectral overlap between the emitted and ideal wave packets was analyzed and connected to the success probability of quantum state transfer~\cite{randles2024success}.

A distinct paradigm for quantum state transfer arises in quantum spin networks, where qubits remain stationary and information is propagated through coherent many-body dynamics~\cite{nikolopoulos2014quantum}.
In such systems, spin chains with engineered Hamiltonians can serve as passive quantum wires, transferring quantum states without requiring time-dependent control once initialized, as is schematically depicted in Fig.~\ref{fig:state-transfer}(b).
A common model is a Heisenberg spin chain with nearest-neighbor couplings, where the Hamiltonian is tailored to optimize transmission fidelity~\cite{banchi2017pretty}.
Depending on the choice of couplings and boundary conditions, these setups can support perfect state transfer~\cite{bose2003quantum,christandl2004perfect,bernasconi2008quantum,kay2010perfect}, or approximate schemes achieving high fidelity over longer distances, often referred to as pretty good state transfer~\cite{camposvenuti2007qubit,camposvenuti2007longdistance,vinet2012almost,godsil2012numbertheoretic}.
In practice, today's spin-qubit experiments instead make use of spin-shuttling techniques, which have shown great promise and are described in Sec.~\ref{ssec:quantum-dots}

\subsubsection{Quantum transduction}

Physical systems that operate at incompatible frequency ranges may be connected using quantum transducers, which enable coherent signal conversion.
This is particularly important for converting signals between the microwave and optical domains~\cite{lauk2020perspectives}.
Microwave-to-optical photon transduction may enable information transfer between cryogenic superconducting quantum processors over optical fiber links used for long-distance communication~\cite{mirhosseini2020superconducting}, and phonon-photon transducers may be used to mediate couplings between a variety of different physical systems~\cite{kuzyk2018scaling,lemonde2018phonon,yue2025deterministic}.
Efficient quantum transduction is therefore a key ingredient in the deployment of quantum networks.
A key quantity to assess the utility of a quantum transducer is its efficiency, \emph{i.e.}, the probability of successful conversion.
The highest microwave-optical conversion efficiencies that have been achieved to date reach $50\%$~\cite{andrews2014bidirectional,higginbotham2018harnessing}.
Realizing quantum transducers comes with additional challenges as it requires not only high conversion efficiencies between signals with large potential frequency mismatch, but also low added noise to the input signal and a broad conversion bandwidth~\cite{wang2022quantum}.

Protocols for quantum transduction come in different flavors:
In one type of transducer, a qubit is directly converted from one frequency domain to another.
Quantum frequency conversion from an input beam of light to an output beam of a different frequency was first demonstrated more than three decades ago~\cite{huang1992observation}.
While frequency conversion between optical fields is relatively advanced today, a direct conversion of modes with very different frequencies is challenging, because their interaction is highly off-resonant.
The conversion between the optical and microwave domains requires a large frequency mismatch of five orders of magnitude to be bridged, which may be achieved on the basis of nonlinear physical processes such as three-wave mixing.
Usually both up-conversion (microwave-to-optical) and down-conversion (optical-to-microwave) processes are needed from quantum transducers~\cite{caleffi2025quantum}.

Instead of a direct conversion, transduction may also be achieved using an intermediate system that generates an effective coherent coupling between the different modes~\cite{zhong2020proposal,meesala2024quantum}.
A~variety of physical systems may be used for this purpose (Sec.~\ref{ssec:platforms}).
In these systems, it is possible to generate hybrid microwave-optical entanglement.
This can be achieved through a two-mode squeezing interaction and spontaneous parametric down-conversion processes~\cite{krastanov2021optically} or a beam-splitter-type interaction and an additional pump field~\cite{jiang2020efficient}.

\subsubsection{Multiplexing, parallelization and ultrafast state transfer}\label{ssec:multiplexing}

Basic quantum state transfer and entanglement distribution protocols have been experimentally demonstrated~\cite{hu2023progress, munro2015quantum}.
However, quantum communication in quantum local area networks still faces significant challenges, especially as these start to scale to hundreds or thousands of qubits.
In that regime, transferring qubit states one by one becomes too slow to keep up with limited coherence times, thus creating a major bottleneck for distributed quantum algorithms. Multiplexing strategies for parallelization of complex quantum state transfer are therefore key to accelerate quantum communication~\cite{lopiparo2019quantum, munro2010quantum, covey2023quantum, liu2020orbital, haldar2025reducing}.

\textit{Multi-mode quantum state transfer protocols}.\textemdash
In deterministic quantum state transfer, a quantum state is turned into a propagating wave packet and later absorbed at a distant second node by a control pulse~\cite{cirac1997quantum}.
These ideas were generalized to a scenario with $N$ bosonic modes that can be transferred to a distant quantum register containing the same number of modes
\cite{xiang2023universal}. 
In addition, the quantum state transfer can be designed in a way that an arbitrary unitary transformation $U$ between the modes can be implemented in a single time step that scales linearly with $N$, and does not require a network of $O(N^2)$ beam splitters~\cite{reck1994experimental}.  
Beyond quantum communication applications, unitary transformations between large numbers of modes are particularly appealing for boson sampling~\cite{brod2019photonic, hoch2025quantum}, linear quantum computing~\cite{knill2001scheme} and certain quantum chemistry applications~\cite{huh2015boson,shang2024boson}.

\textit{Parallel autonomous entanglement distribution}.\textemdash
Entanglement distribution is the creation and sharing of quantum correlations between distant particles or nodes, serving as a fundamental resource for quantum networks~\cite{munro2015quantum}. 
It often also requires generating multiple entangled pairs between two nodes and consuming them at nearly the same time.
Entanglement can be distributed by first generating it locally and then transferring one of the states to a distant party. 
These methods are deterministic but require costly controls at both nodes. 
An alternative is to distill entanglement from multiple weakly entangled pairs without requiring expensive operations, which is probabilistic and suffers from an exponentially increasing failure rate as the number of desired entangled pairs grows~\cite{duan2001longdistance, barrett2005efficient, matsukevich2006entanglement, moehring2007entanglement, riedinger2018remote}.
Inspired by the generation of highly entangled pairs using two‑mode squeezed states of light~\cite{kraus2004discrete}, a deterministic entanglement distribution protocol has recently been explored using a nondegenerate parametric amplifier, which produces a continuous Gaussian two‑mode squeezed state~\cite{agusti2022longdistance, agusti2023autonomous, agusti2025nonmarkovian}. 
In this approach, the parametric amplifier emits two quantum‑correlated photon beams that can be directed to different nodes in the network. 
By varying the detuning of each qubit relative to the resonance frequency of the parametric amplifier, different patterns of parallel, autonomous entanglement distribution can be achieved.

\textit{Ultrafast quantum state transfer}.\textemdash
Quantum state transfer protocols with ultrashort pulses constitute a promising route to significantly accelerate quantum communication. 
Here, ``ultrashort'' refers to pulse durations on the order of the inverse qubit transition frequency, which is approximately $100$ times shorter than the wave packets employed in conventional protocols, thus implying a potential two-orders-of-magnitude speedup. 
However, processing quantum information at such timescales requires careful treatment beyond the rotating wave approximation.
Recent work by~\cite{zhu2021quantum} demonstrated the feasibility of ultrafast single- and two-qubit gates in superconducting circuits under such conditions. 
Via numerical optimization, their work identified fundamental limits on gate speed that are imposed by the qubit's intrinsic nonlinearity, and that are independent of the specific hardware design. 
One can expect that, with highly anharmonic flux qubits, and commercially available electronics, high-fidelity operations (error below $10^{-4}$) can be executed in the $\sim\!100$\,ps regime. 
Notably, a compressed version of Shor's algorithm [factoring $15$~\cite{vandersypen2001experimental}] in just $1$\,ns was simulated, thus highlighting the viability of ultrafast gates and suggesting that a hundredfold speedup over current implementations may be within reach.

\subsubsection{Remote entanglement generation}

Remote entanglement generation (REG) plays a central role for quantum-network applications, as it enables teleportation-based protocols (Sec.~\ref{ssec:teleportation}) and underpins many approaches to distributed quantum operations.
Often REG is considered between two nodes;
the problem of how to simultaneously distribute entanglement between multiple node pairs is known as routing~\cite{pant2019routing,li2021effective}.

Remote entanglement and state transfer can be realized both probabilistically~\cite{chou2005measurementinduced,moehring2007entanglement,lee2011entangling} and deterministically~\cite{kurpiers2018deterministic,humphreys2018deterministic,almanakly2025deterministic}.
Deterministic REG relies on deterministic transfer protocols such as the one discussed in Sec.~\ref{sssec:qst}.
In many probabilistic protocols, entangled states between two network nodes are generated by heralding~\cite{barz2010heralded,bernien2013heralded,delteil2016generation}, which are then used for state transfer by means of teleportation~\cite{pfaff2014unconditional,tsurumoto2019quantum,li2022quantum,hermans2022qubit,lago-rivera2023long}, but entanglement can also be generated probabilistically without heralding~\cite{choi2010entanglement,ritter2012elementary}.
In principle, deterministic operations can be carried out even between probabilistically connected nodes, which is usually associated with an overhead in additional resource requirements~\cite{jiang2007distributed}.

A key building block of many probabilistic REG protocols is the cavity-assisted interaction between flying and stationary qubits~\cite{duan2004scalablea,waks2006dipole,childress2006faulttolerant,vanloock2006hybrid}.
Depending on the specific implementation, this interaction may enable controlled-$X$ or controlled phase gates on a photonic flying qubit, controlled by a stationary qubit.
The photon is then sent to a detector for measurement, where it is measured in a basis that preserves the entanglement between the stationary qubits.
Following~\cite{beukers2024remoteentanglement} we distinguish between three topologies for heralded remote-entanglement protocols:
(i) detector-in-midpoint configurations based on spontaneous emission,
(ii) detector-in-midpoint configurations based on conditional reflection, and
(iii) sender-receiver configurations based on conditional reflection.

\begin{figure}[t]
    \centering
    \includegraphics[width=\linewidth]{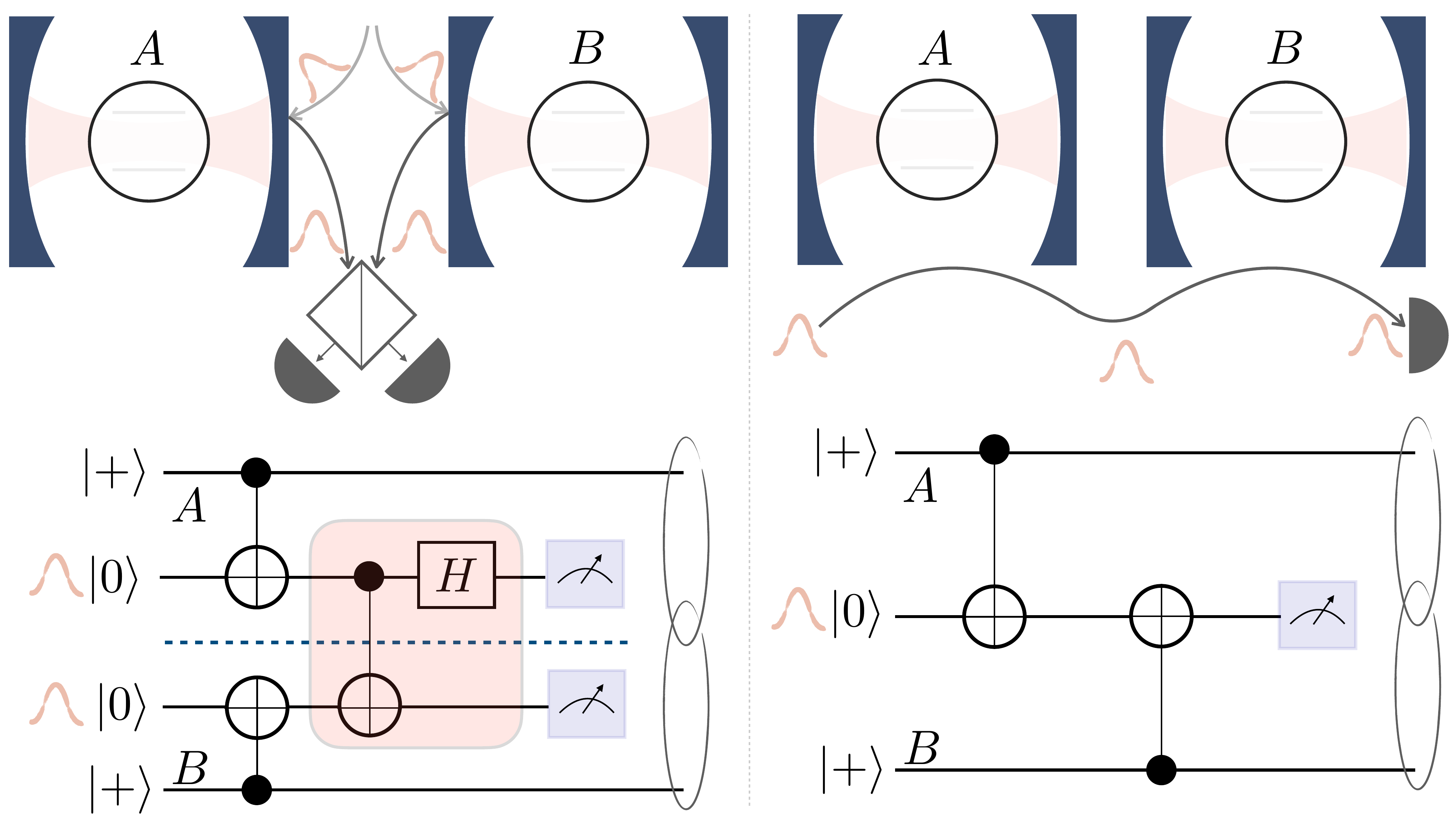}
    \caption{Two different types of remote entanglement protocols.
    (Left) Detector-in-midpoint configuration where photons are sent to a beam splitter, effectively realizing a Bell-basis measurement (red region in circuit).
    This enables heralded entanglement generation between the systems at nodes $A$ and $B$ (indicated by gray oval shapes at the end of the circuit).
    (Right) Sender-receiver protocols where a conditional gate between a single photon and each of the systems at nodes $A$ and $B$ is realized.
    Figure adapted from Ref.~\cite{beukers2024remoteentanglement}.}
    \label{fig:remote-entanglement}
\end{figure}

In (i) and (ii) detector-in-midpoint topologies, photons arrive from the nodes at a detector in the middle and are then measured in the Bell basis (Sec.~\ref{ssec:swap-test}).
This measurement projects the systems at both nodes in an entangled state.
In (iii) sender-receiver topologies, a single photon successively interacts with both nodes, after which the photon is measured.
These different topologies are schematically depicted in Fig.~\ref{fig:remote-entanglement}.
Previous implementations of these protocols are discussed in the following Section.

\subsection{Platforms for quantum interconnects \label{ssec:platforms}}

In the following, we discuss implementations of modular quantum computing.
Many are based on strong light-matter coupling in cavity~\cite{walther2006cavity} and circuit~\cite{blais2021circuit} quantum electrodynamics, which has been demonstrated in many physical systems, including trapped atoms~\cite{volz2006observation,ritter2012elementary} and ions~\cite{blinov2004observation,herskind2009realization,stute2012tunable}, superconducting circuits~\cite{wallraff2004strong,majer2007coupling}, nitrogen-vacancy centers in diamond~\cite{togan2010quantum}, and spin qubits in quantum dots~\cite{mi2018coherent,landig2018coherent,samkharadze2018strong,dijkema2025cavitymediated}.
For more detailed discussions of physical realizations, we refer the reader to other review articles that are more focused on the challenges and history of quantum-network implementations~\cite{wehner2018quantum,wei2022realworld,reiserer2022colloquium,azuma2023quantum}.

\subsubsection{Neutral atoms \label{ssec:neutral-atoms}}

Neutral atoms in optical-tweezer or lattice arrays are a versatile platform for quantum information.
Entanglement between encoded qubits can be created using controlled collisions or spin-exchange dynamics~\cite{calarco2000quantum, brennen1999quantum,sorensen1999spinspin,weitenberg2011quantum}.
The most prominent mechanism for entanglement employs Rydberg excitation to generate strong interactions.
Rydberg atoms and molecules occupy highly electronically excited states, which endows them with characteristics such as large sizes, large dipole moments, high sensitivity to electric fields and large interaction strengths~\cite{gallagher1988rydberg,sibalic2018rydberg}.
Since the early days of Rydberg physics, the center of attention has shifted from atomic physics and cavity quantum electrodynamics~\cite{haroche2013nobel} to quantum many-body physics~\cite{browaeys2020manybody} and technology~\cite{adams2020rydberg}.
While the first observation of interactions between Rydberg atoms~\cite{raimond1981spectral} and their proposed use as building blocks of quantum gates~\cite{jaksch2000fast,lukin2001dipole} is a little while ago, it has taken some time for neutral-atom quantum processors to become a powerful platform for realizing quantum computers~\cite{saffman2010quantum,saffman2016quantum,wu2021concise} and simulators~\cite{scholl2021quantum}.

Single atoms can now be held in place by tightly focused beams of light that are called optical tweezers, allowing flexible geometries and precise control of individual qubits~\cite{kaufman2021quantum} as well as entanglement generation between them~\cite{levine2019parallel,graham2019rydbergmediated,madjarov2020highfidelity,dordevic2021entanglement}.
Atomic qubits in tweezer arrays provide a path towards large-scale quantum processors~\cite{ebadi2021quantum} and can be coherently transported through different zones of a module, enabling dynamic and nonlocal connectivities between qubits~\cite{bluvstein2022quantum,bluvstein2024logical,reichardt2024faulttolerant,saffman2025quantum}.
To date, arrays of several thousands of coherent atomic qubits have been realized in two-dimensional geometries~\cite{manetsch2024tweezer} and smaller systems even in 3D~\cite{barredo2018synthetic}.
High-fidelity mid-circuit measurements were demonstrated in several experiments~\cite{deist2022midcircuit,singh2023midcircuit,graham2023midcircuit}.

More than a decade ago, pitch-and-catch state transfer (Sec.~\ref{ssec:interconnects-protocols}) has been demonstrated with atomic qubits in high-finesse optical cavities spatially separated by about 20~meters~\cite{ritter2012elementary}, and remote entanglement was generated between individual atoms~\cite{hofmann2012heralded}.
Improved REG protocols (Sec.~\ref{ssec:interconnects-protocols}) are still being explored and may lead to higher entanglement rates and fidelities in future experiments~\cite{li2024highrate}.

Large-scale networks of neutral atoms share many properties with trapped-ion networks (see following paragraphs) and will most likely be based on photonic links.
Efficient light-matter interfaces are enabled by optical cavities~\cite{covey2023quantum,kimble2008quantum,reiserer2015cavitybased}.
Atom-cavity coupled network nodes have been used to mediate heralded, nonlocal quantum gates~\cite{daiss2021quantumlogic} and perform Bell-basis measurements between distant atomic qubits~\cite{welte2021nondestructive}.
While most previous work was focused on networks with single-qubit nodes, first steps towards scaling the number of atoms per node have been taken~\cite{langenfeld2021quantum,thomas2024fusion}.
Future improvements will require progress with such multi-qubit nodes, direct telecom operation~\cite{vanleent2022entangling} and multiplexed networking.
Alternatively, exchange dynamics between photons and atoms could be leveraged for a cavity-free approach relying on Rydberg polaritons~\cite{thompson2017symmetryprotected, khazali2019polariton, zhang2025dualtype}.
Also, there have been several approaches to increasing the photon collection efficiency, \emph{e.g.}, using optical cavities~\cite{takahashi2017cavityinduced, takahashi2020strong,schupp2021interface, bochmann2010lossless,ramette2022anytoany, hu2025siteselective}, microcavity lens arrays~\cite{shaw2025cavity}, and recently cavity-free collection directly in the telecom band~\cite{li2025parallelized}.

\subsubsection{Trapped ions \label{ssec:trapped-ions}}

When ions are confined by suitable electromagnetic fields, they can be used as qubits with long coherence times~\cite{leibfried2003quantum}.
Two common types of ion traps are Penning traps~\cite{penning1936glimmentladung}, which employ static electric and magnetic fields, and radio-frequency or Paul traps~\cite{paul1953notizen} that use a combination of static and oscillating electric fields to confine charged particles.
Trapped ions may be used to implement universal gate sets and are a main contender for realizing quantum computing hardware~\cite{cirac1995quantum,monroe1995demonstration} and quantum simulators~\cite{porras2004effective,blatt2012quantum}.
While two-dimensional trapped-ion systems are actively being explored for quantum information processing tasks in both specifically designed Paul traps~\cite{kiesenhofer2023controlling,guo2024siteresolved} and Penning traps~\cite{hawaldar2024bilayer,jain2024penning}, most studies have been performed on linear strings confined in linear Paul traps.
One-dimensional linear arrays of trapped ions achieve exceptional performance and, in the past 15 years, have been used to produce relatively large entangled states~\cite{monz201114qubit} and demonstrate several quantum algorithms~\cite{bruzewicz2019trappedion}.
Single-qubit and two-qubit gates between arbitrary pairs of ions are implemented by appropriate laser beams, achieving all-to-all connectivity among the qubits in an ion chain~\cite{debnath2016demonstration}.
However, individual control over large numbers of qubits and high-fidelity entangling gates become impractical as these chains grow in length~\cite{zhu2006trapped,cetina2022control}.
This scaling challenge may be addressed using modularity at different length scales.

One proposed solution for short-range modularity is that of the quantum charge-coupled device (QCCD) architecture, which uses mobile ions as qubits that are transported between different processing zones using electric fields~\cite{wineland1998experimental,kielpinski2002architecture}.
The transport of ions in segmented Paul traps has been demonstrated in early experiments~\cite{rowe2002transport,hensinger2006tjunction,blakestad2009highfidelity} and high fidelities may be preserved during transport~\cite{kaufmann2018highfidelity} using efficient shuttling schedules~\cite{sterk2022closedloop,schoenberger2024shuttling}.
Recent advances have underlined the viability of this approach toward multi-core quantum processors~\cite{pino2021demonstration,moses2023racetrack}.
The thereby connected qubit registers may be addressed and operated simultaneously~\cite{kwon2024multisite,mordini2025multizone}.
The QCCD architecture has recently enabled progress in quantum error correction~\cite{ryan-anderson2021realization,paetznick2024demonstration}, the realization of non-Abelian topological order~\cite{iqbal2024nonabelian} and random circuit sampling~\cite{decross2025computational}.
It may be extended to multiple modules with fast qubit transport from one module to adjacent modules using electric fields~\cite{lekitsch2017blueprint}.
This has recently been demonstrated experimentally~\cite{akhtar2023highfidelity}.
Overall, the QCCD architecture allows for scaling up trapped-ion processors but suffers from an overhead due to the shuttling times between gate operations.
Photonic interconnects offer another route towards modular trapped-ion computing.

Modular ion-trap architectures based on photonic interconnects were already considered more than ten years ago and are under active investigation~\cite{moehring2007entanglement,monroe2014largescale,hucul2015modular,brown2016codesigning,stephenson2020highrate}.
Connecting multiple modules within one laboratory and also extending networks to larger distances benefit from optical links~\cite{monroe2013scaling,northup2014quantum}.
Recently, remote entanglement of two trapped ions separated by hundreds of meters~\cite{krutyanskiy2023entanglement} and long-distance quantum repeater nodes based on trapped ions were demonstrated~\cite{krutyanskiy2023telecomwavelength}, as well as ion-photon entanglement over a 101-km-long fiber channel~\cite{krutyanskiy2024multimode}.
High-fidelity remote entanglement generation over a distance of several meters has been demonstrated using time-bin photons~\cite{saha2025highfidelity}, which have also been used to entangle higher-dimensional qudit memories in separate ion traps at a distance of $2$~m~\cite{shalaev2025photonic}.
The first steps towards distributed quantum computations have been taken by executing Grover's algorithm between optically connected trapped-ion modules~\cite{main2025distributed}.
Many theoretical proposals that are geared towards trapped-ion architectures may also be implemented with neutral atoms, and vice versa~\cite{ramette2022anytoany}.

\subsubsection{Superconducting circuits \label{ssec:sc-qubits}}

Superconducting circuits are among the most advanced platforms for quantum computing, combining fast gate operations, high fidelities, and compatibility with microfabrication technologies~\cite{clarke2008superconducting,krantz2019quantum,kjaergaard2020superconducting}.
They support various qubit modalities, such as transmons~\cite{koch2007chargeinsensitive}, flux qubits~\cite{friedman2000quantum,vanderwal2000quantum} and fluxoniums~\cite{manucharyan2009fluxonium}, and can also be adapted to encode higher-dimensional quantum information using qudits~\cite{goss2022highfidelity}.
With processors now surpassing the 100-qubit scale~\cite{kim2023evidence,googlequantumaiandcollaborators2025quantum,gao2025establishing}, superconducting devices have become a central platform for implementing quantum algorithms and exploring near-term applications.

Further scaling is challenged by qubit connectivity, crosstalk, fabrication yield and the complexity of wiring in cryogenic environments.
To address these challenges, several modular approaches are currently being explored, based on low-loss quantum interconnects~\cite{awschalom2021development}, ranging from short-range to metropolitan-area networks.
Different types of modularity are enabled by on-chip nonlocal couplers, short-range chip-to-chip as well as meter-range and beyond meter-range microwave interconnects, and microwave as well as optical longer-distance links~
\cite{bravyi2022future}.
Note that, while detectors for single itinerant microwave photons are still more challenging to realize than in the optical domain, they have been subject of several theoretical works~\cite{romero2009microwave,sathyamoorthy2014quantum,royer2018itinerant} as well as experimental demonstrations~\cite{opremcak2018measurement,besse2018singleshot,kono2018quantum,lescanne2020irreversible}.

Short-range modular architectures based on three-dimensional (3D) integration techniques, such as flip-chip bonding~\cite{rosenberg20173d,foxen2018qubit,gold2021entanglement,kosen2022building}, are emerging as a promising path to scaling up superconducting quantum processors, as reviewed in~\cite{rosenberg2020solidstate}.
With flip-chip technology, separate quantum modules can be fabricated on different chips and stacked together.
This technique enables compact packaging, increased qubit count, and higher fabrication yield, while maintaining or even improving coherence times~\cite{field2024modular,putterman2025hardwareefficient,norris2025performance}.
Experiments have demonstrated high-fidelity two-qubit gates between qubits on different chips, comparable to the best intra-chip gates~\cite{gold2021entanglement,conner2021superconducting,field2024modular,dalton2025resourceefficient}, indicating that modular designs need not compromise performance.
Modular setups also support long-range connectivity between qubits, a feature that could simplify the layout of more complex quantum circuits~\cite{wu2024modular,wang2025demonstration}.

Deterministic pitch-and-catch state transfer and entanglement generation between qubits on separate chips at the same cryogenic node were demonstrated several years ago~\cite{kurpiers2018deterministic,axline2018ondemand,campagne-ibarcq2018deterministic}.
Submeter and meter-scale quantum interconnects can be realized as low-loss transmission lines using coaxial cables and may be used to connect multiple modules within a single cryostat~\cite{burkhart2021errordetected,niu2023lowloss}, enabling remote entanglement generation~\cite{chang2020remote,zhong2021deterministic,song2025realization,mollenhauer2025highefficiency}, remote state preparation~\cite{pogorzalek2019secure} as well as state~\cite{fedorov2021experimental} and gate teleportation~\cite{chou2018deterministic}.
Other implementations have utilized qubits that interact with standing-wave modes of meter-scale interconnects~\cite{zhong2019violating,leung2019deterministic,qiu2025deterministic,heya2025randomized}.
To date, most works have focused on networks with two nodes, and it remains an important goal to build and characterize multi-node networks.
In a recent experiment, five quantum modules were connected with low-loss coaxial cables~\cite{niu2023lowloss}.
Quantum secret sharing of classical information was demonstrated in a network with three nodes, connected through long coplanar waveguides, at a pairwise distance of about 1~m~\cite{yan2025quantum}.

Microwave quantum links were also built between superconducting circuits housed in separate cryogenic systems~\cite{magnard2020microwave,yam2025cryogenic}.
Physical node distances of up to 30 meters have been demonstrated using this approach, enabling
fundamental Bell tests~\cite{storz2023loopholefree}, proofs of concept of randomness amplification~\cite{kulikov2024deviceindependent} and self-testing~\cite{storz2024complete}.

Metropolitan-scale networks and beyond may require microwave-to-optical transduction to connect quantum processors in distant dilution refrigerators.
Yet operating and addressing superconducting qubits optically still poses several challenges, such as a large desired bandwidth, low added noise and high conversion efficiencies~\cite{delaney2022superconductingqubit}.
Microwave-optical quantum transducers may be realized in different ways, \emph{e.g.}, optomechanically with an intermediary mechanical mode that can interact with microwave and optical fields~\cite{aspelmeyer2014cavity}, with Rydberg atoms~\cite{han2018coherent,vogt2019efficient} or electro-optically~\cite{tsang2010cavity}.
Different realizations bring specific advantages, for example, electro-optical transducers currently achieve only low efficiencies of about ten percent, but they also add less noise to the signal~\cite{sahu2022quantumenabled}.
More information about physical realizations of quantum transduction can be found in~\cite{lauk2020perspectives}, references therein, and~\cite{mirhosseini2020superconducting,brubaker2022optomechanical,kumar2023quantumenabled,weaver2025scalable}.

Setups for up-conversion and down-conversion between microwave and optical frequencies have recently been used for all-optical readout~\cite{lecocq2021control,arnold2025alloptical,vanthiel2025optical} and coherent optical control of superconducting qubits~\cite{warner2025coherent}.
Future improvements of the total loss between transducer and qubits will be essential to realize high-fidelity state transfer based on optical photons.
A recent analysis showed that the performance of microwave-to-optical quantum links must be significantly improved before they can become practically useful~\cite{ang2024arquin}.

\subsubsection{Color centers \label{ssec:color-centers}}

Color centers are optically active point defects in diamond and suitable candidates for quantum-network nodes~\cite{awschalom2018quantum, ruf2021quantum}.
The first single-defect observation of the nitrogen-vacancy (NV) center almost thirty years ago~\cite{gruber1997scanning} kickstarted efforts to realize solid-state qubits at room-temperature.
Several other point defects have been investigated in the past decade. 
One group is referred to as Group-IV defects and includes SiV~\cite{neu2011single, sipahigil2014indistinguishable, sipahigil2016integrated}, GeV~\cite{iwasaki2015germaniumvacancy} and SnV~\cite{iwasaki2017tinvacancy} centers.
The vacancy centers generally differ in their physical properties such as their sensitivity to electric fields~\cite{bradac2019quantum}, with relevance to nanophotonic integration of the vacancy centers.
Different host materials are also being investigated, such as a vacancy centers in silicon carbide (SiC)~\cite{christle2017isolated}, which is a promising candidate for integrated photonics by leveraging state-of-the-art wafer technology~\cite{liu2024silicon}.
Another promising recent direction studies so-called T-centers in silicon, which are telecom single-photon emitters~\cite{bergeron2020siliconintegrated, zhang2025laserinduced}, or the carbon-based G-center also in the telecom band~\cite{redjem2020single, prabhu2023individually}.

In the following part, we focus our discussion on the NV center, which is currently among the most advanced platforms for realizing quantum networks~\cite{ruf2021quantum, pompili2021realization,nemoto2016photonic,bradley2022robust}.
NV centers are color defects that occur when two neighboring sites in a diamond lattice host a nitrogen atom and a vacancy defect, respectively~\cite{doherty2013nitrogenvacancy}.
The energy levels of NV centers are sensitive to magnetic~\cite{maze2008nanoscale} and electric~\cite{dolde2011electricfield} fields, temperature~\cite{kucsko2013nanometrescale} and strain~\cite{maze2011properties}, making them highly suitable for quantum sensing applications~\cite{zhou2020quantum}.
It is also appealing that NV center spins can be coherently operated even well above room temperature~\cite{liu2019coherent}.
However, the sensitivity of NV centers to electric fields makes nanophotonic integration challenging, and current experiments have been realized mostly with bulk diamond.

Spin-photon entanglement is a fundamental building block for quantum networks based on NV centers and was demonstrated in 2010~\cite{togan2010quantum}.
The first experimental demonstration of a loophole-free Bell inequality violation was demonstrated using NV centers in diamond~\cite{hensen2015loopholefree}.
This landmark experiment highlighted the potential of NV centers as reliable carriers of quantum information.
In distributed quantum computing architectures, NV centers may serve as versatile nodes capable of local processing and long-distance entanglement distribution.
Their spin-photon interface enables remote entanglement generation through heralded protocols that exploit spin-photon entanglement and subsequent photon interference at beam splitters~\cite{bernien2013heralded}.

Recent advances have focused on improving the entanglement generation rate and fidelity by integrating color centers with nanophotonic structures such as waveguides and optical cavities~\cite{riedel2017deterministic, knaut2024entanglement}.
These structures enhance photon-collection efficiency and enable Purcell-enhanced emission, addressing one of the major bottlenecks in realizing scalable NV-based quantum networks.
At the same time, the development of quantum memories based on nuclear spins in the vicinity of the NV center provides a valuable resource for buffering entanglement and implementing temporal multiplexing schemes~\cite{kalb2017entanglement}.

NV centers have also been instrumental in pioneering quantum repeater protocols and fault-tolerant link-layer architectures.
These efforts are further supported by error-correction techniques tailored to the specific error channels of NV systems, including those arising from photon loss and spin decoherence~\cite{nemoto2014photonic}.
A quantum network of three nodes~\cite{pompili2021realization} and entanglement distribution across non-neighboring nodes has been demonstrated in recent years~\cite{hermans2022qubit}, also over metropolitan distances~\cite{stolk2024metropolitanscale}.
Together, color centers are regarded as a leading platform for realizing quantum networks, but the field is still dynamic and defect types beyond the NV center are actively explored.
For example, a quantum network node based on SiV centers was demonstrated~\cite{stas2022robust} and used for implementing a small-scale blind quantum computing protocol~\cite{wei2025universal}.

\subsubsection{Semiconductor spin qubits \label{ssec:quantum-dots}}

Quantum dots (QDs) are semiconductor-based artificial atoms which have been established as a promising platform for quantum information processing several decades ago~\cite{loss1998quantum}.
They can be fabricated, for example, by electrical gating in semiconductor heterostructures or by self-assembly techniques.
Self-assembled QDs are optically active and serve as efficient light–matter interfaces.
When embedded in photonic nanostructures, they can achieve near-unity coupling to optical modes and function as high-quality single-photon sources~\cite{lodahl2018quantumdot}.
Gate-defined QDs, in contrast, confine single electrons or holes but are not optically active, and represent a promising platform for scalable spin qubits due to their long coherence times and compatibility with established semiconductor technologies; see ~\cite{burkard2023semiconductor} for more detailed information about such platforms and creating long-range coupling between distant spins.
The discussion in the following paragraphs focuses on gate-defined QDs.

Two-qubit entangling gates are commonly based on exchange interactions, which enable universal quantum computation~\cite{divincenzo2000universal}.
Spin-spin exchange interactions have a relatively short physical range, which is determined by the region in which electronic wavefunctions of neighboring dots significantly overlap.
They are routinely used for qubit control and constitute a crucial resource for spin qubits.
To scale up to larger systems, it can also be beneficial to bridge larger distances.
This may be achieved with modular devices, in which all modules are hosted on the same chip, and with on-chip quantum links to connect the individual modules.
Two conceptually different approaches toward on-chip quantum links may be distinguished.

One type of approach is to extend the coupling range to larger distances than is possible by exchange coupling.
This may presumably be achieved through capacitive couplings, which arise from Coulomb interactions and allow for longer interaction ranges.
That has been theoretically investigated~\cite{taylor2005faulttolerant,stepanenko2007quantum,calderon-vargas2015directly,srinivasa2015capacitively}, and demonstrated for the case of the singlet-triplet qubit~\cite{shulman2012demonstration}.
Another promising strategy is the use of a mediator, \emph{i.e.}, an additional dot or short chain of QDs which is placed between the spins of interest, enabling an effective interaction through virtual processes, often described as superexchange.
Although the occupation of the intermediate dot does not change, the outer dots can be indirectly coupled in this way, enabling long-range charge transfer~\cite{braakman2013longdistance,busl2013bipolar}.
Intermediary systems have also been used to induce coherent spin-spin coupling at a distance~\cite{baart2017coherent,martins2017negative,malinowski2019fast}.
Finally, yet another strategy toward long-range coupling is the integration of spin qubits with on-chip superconducting resonators within the circuit-QED framework~\cite{blais2021circuit,burkard2020superconductor}.
Strong spin-photon coupling~\cite{samkharadze2018strong,mi2018coherent,landig2018coherent} and resonant spin-photon-spin~\cite{borjans2020resonant} coupling have been achieved with such hybrid QD-cavity devices.
Recently, the control of a dot-resonator-dot system enabled the realization of two-qubit iSWAP oscillations between spin qubits that were more than 200~$\mu$m apart~\cite{dijkema2025cavitymediated}.

A different approach to realizing on-chip quantum links is via the transport of qubits over a distance.
Here, we can again distinguish between two basic types of strategies.
The first encompasses schemes where the electrons, or generally the charge carriers whose spin degree of freedom is used to define a qubit, are not moved themselves, but the qubit transport is realized via an exchange-based spin bus~\cite{friesen2007efficient}.
Early work has suggested that spin chains can be used to transfer quantum states from one end of the chain to the other~\cite{bose2003quantum}.
Suitable spin couplings potentially allow for perfect state transfer~\cite{christandl2004perfect}.
In experiment, coherent SWAP operations have been realized to transfer single-spin and entangled states back and forth in an array without moving any electrons~\cite{kandel2019coherent}.
Recently, the propagation of single-spin and two-spin excitations has been observed in exchange-coupled QDs~\cite{farina2025siteresolved}.
The second strategy, which is experimentally more advanced, is based on shuttling, where electrons are physically displaced while preserving their spin state.
Once brought sufficiently close to one another, formerly distant electron spins can couple through their exchange interaction.
There exist various shuttling-based physical mechanisms and architectures for spin-qubit transport.

First, conveyor-mode shuttling refers to a technique where a spin is transported in a traveling-wave potential or moving QD.
This moving potential can be generated by surface acoustic waves~\cite{delsing20192019,wang2024electron}, which have been analyzed for their potential to realize universal quantum transducers~\cite{schuetz2015universal}, or by phase-shifted sinusoidal signals applied to successive gate electrodes~\cite{taylor2005faulttolerant}.
With sinusoidal gate voltages, charge transport was recently demonstrated~\cite{xue2024si}, as well as spin-coherent shuttling of individual electrons~\cite{struck2024spineprpair} and high-fidelity spin transport~\cite{desmet2025highfidelity}.
With surface acoustic waves, the key idea is that in piezoelectric devices the sound waves are accompanied by a traveling electric potential, which can be used to form a train of QDs moving along a transport channel.
Based on this approach, first charge transport was demonstrated~\cite{mcneil2011ondemand,hermelin2011electrons}, followed by experiments that demonstrated preservation of spin coherence during the shuttling process~\cite{bertrand2016fast,jadot2021distant}.
In platforms that are not piezo-electric, such as silicon or germanium, this approach is more challenging to realize.

Second, bucket-brigade shuttling refers to spin transport through an array of QDs by means of electron hopping between neighboring dots.
This can be achieved by appropriately adjusting the electrochemical potentials within the array~\cite{taylor2005faulttolerant}.
With this approach, successful charge hopping was demonstrated~\cite{mills2019shuttling}, as well as preservation of spin projection~\cite{baart2016singlespin} and some level of spin coherence in GaAs~\cite{fujita2017coherent,flentje2017coherent} and silicon~\cite{noiri2022shuttlingbased,yoneda2021coherent,zwerver2023shuttling}.
It has already enabled hopping-based quantum logic operations with relatively high fidelities~\cite{wang2024operating,desmet2025highfidelity}.

Several architectural proposals for scalable spin-qubit hardware are based on some of the previously mentioned shuttling processes; see, for example,~\cite{vandersypen2017interfacing,boter2022spiderweb,langrock2023blueprint,kunne2024spinbus}.
Note that all strategies discussed so far target on-chip quantum links between semiconductor spin qubits.
Some aspects of off-chip quantum links between optically active QDs, which we have not yet addressed, will be mentioned in the next section.

\subsubsection{Photonic systems \label{ssec:photonic-systems}}

Photons play a central role in distributed quantum computing architectures due to their natural suitability for long-distance quantum communication~\cite{obrien2009photonic,wang2025scalable}.
As such, photons are the prevalent choice for information carriers to realize quantum interconnects.
Depending on the physical system, photonic information carriers can have vastly different carrier frequencies.
For example, optical photons usually have little interaction with their environment, enabling low-loss transmission through free-space links or optical fibers, without the need for quantum transduction processes~\cite{psiquantumteam2025manufacturable}, and can be manipulated with high precision using integrated optical components~\cite{wang2020integrated};
microwave photons form the basis of circuit quantum electrodynamics~\cite{blais2021circuit} and superconducting quantum information processing~\cite{gu2017microwave}, and they usually require cryogenic environments to suppress thermal fluctuations.

A major direction in photonic quantum information processing is measurement-based quantum computation using cluster states~\cite{raussendorf2001oneway}.
These entangled resource states can be prepared offline in constant circuit depth and consumed by local projective measurements to perform quantum computation~\cite{briegel2009measurementbased}.
This paradigm is particularly relevant in distributed scenarios, where a server can generate and distribute cluster states to clients who then perform only local measurements, enabling blind or delegated quantum computing (see Sec.~\ref{ssec:verification}).
Early works analyzed how to prepare matrix-product states, which include GHZ, W and cluster states, with atomic~\cite{schon2005sequential,schon2007sequential} and solid-state~\cite{lindner2009proposal} emitters in cavities undergoing pumping and decay.
In these schemes, which have been realized experimentally in the optical domain with trapped atoms~\cite{yang2022sequential,thomas2022efficient} and optically active quantum dots~\cite{schwartz2016deterministic,istrati2020sequential,cogan2023deterministic,coste2023highrate,meng2024deterministic,huet2025deterministic} and in the microwave domain with superconducting circuits~\cite{besse2020realizing,ferreira2024deterministic,osullivan2025deterministic}, unitaries are applied to the emitter between engineered emission steps, where each emission acts as a two-qubit gate between emitter and photon, such that repeating this sequence produces entangled photon states.
Although some of these works have tapped into the generation of higher-than-one-dimensional states, it remains a challenge to create large entangled resource states beyond 1D.
Yet even with one-dimensional cluster states and fusion measurements~\cite{browne2005resourceefficient,bartolucci2023fusionbased} only, it is possible to realize fault-tolerant quantum computations~\cite{paesani2023highthreshold,dessertaine2024enhanced,pettersson2025deterministic,chan2025tailoring}.

In the previous section, we had mentioned that off-chip quantum links based on optical photons can be realized with self-assembled quantum dots, which are used to build single-photon emitters~\cite{arakawa2020progress}.
In these systems, heralded entanglement generation between distant spin qubits has been demonstrated using electron~\cite{stockill2017phasetuned} and hole~\cite{delteil2016generation} spins, hosted several meters apart in separate cryostats.
Scaling up to larger networks, QD-based light and single-photon sources can be used for quantum key distribution and other quantum-network applications~\cite{schimpf2021quantum,schimpf2021quantuma,lu2021quantumdot}.
Resonantly driven QDs can produce streams of photons suitable for time-bin or polarization entanglement schemes, and recent developments have pushed their operation wavelengths into the telecom bands~\cite{yu2023telecomband}.

In the continuous-variable (CV) setting, universal quantum computation can be implemented using Gaussian cluster states with additional non-Gaussian operations or measurements~\cite{menicucci2006universal}.
CV systems have been employed to generate two-dimensional~\cite{larsen2019deterministic,asavanant2019generation} and three-dimensional~\cite{roh2025generation} cluster states.
Recent experiments have also demonstrated the generation of optical non-Gaussian states, such as GKP states~\cite{gottesman2001encoding}, with free-space optical components~\cite{konno2024logical} and integrated photonic chips~\cite{larsen2025integrated}, respectively.

Miniaturized footprints of quantum photonic chips can help scaling up modular systems, as is discussed in detail in the review article~\cite{luo2023recent}.
While fully chip-based implementations of quantum algorithms remain challenging,
the past years have shown tremendous progress with integrated photonic chips, as they have been used to demonstrate, for example, chip-based quantum key distribution~\cite{sibson2017chipbased}, quantum teleportation~\cite{llewellyn2020chiptochip}, genuine multipartite entanglement~\cite{wang2018multidimensionalb,reimer2019highdimensional,bao2023verylargescale} as well as small instances of selected quantum algorithms~\cite{paesani2019generation,arrazola2021quantum}.
Recent experiments have demonstrated the generation of large-scale photonic cluster states across spatially separated chips~\cite{aghaeerad2025scaling} and atomic systems~\cite{thomas2024fusion}.
There exists also a cloud-accessible quantum processor based on single photons~\cite{maring2024versatile}.

While photonic systems excel in communication protocols and measurement-based quantum computation, they also find applications in several protocols discussed in Sec.~\ref{sec:algorithms-applications}.
For instance, Bell-basis measurements and programmable interferometry have been used to experimentally estimate overlaps between photonic quantum states~\cite{zhan2025experimental}.
Continued progress with scaling up photonic chips, improvements of photon sources and detectors, as well as integration with quantum memories will be essential in the next years.
Their inherent compatibility with fiber networks and modular architectures ensures that photonic setups will remain a vital component of distributed quantum computing platforms.

\section{Fundamental Concepts \label{sec:concepts}}

Distributed QIP is a vast but not clearly defined subfield of quantum information science.
The compilation of relevant background material covered in this review must therefore be selective.
To maintain clarity, we follow the notation in Table~\ref{tab:notation}, distinguishing between operators (hats), matrices (bold font), vectors (arrows), scalars and functions (italics), and using bra-ket notation for quantum states.

\begin{figure}[t]
    \centering
    \includegraphics[width=\linewidth]{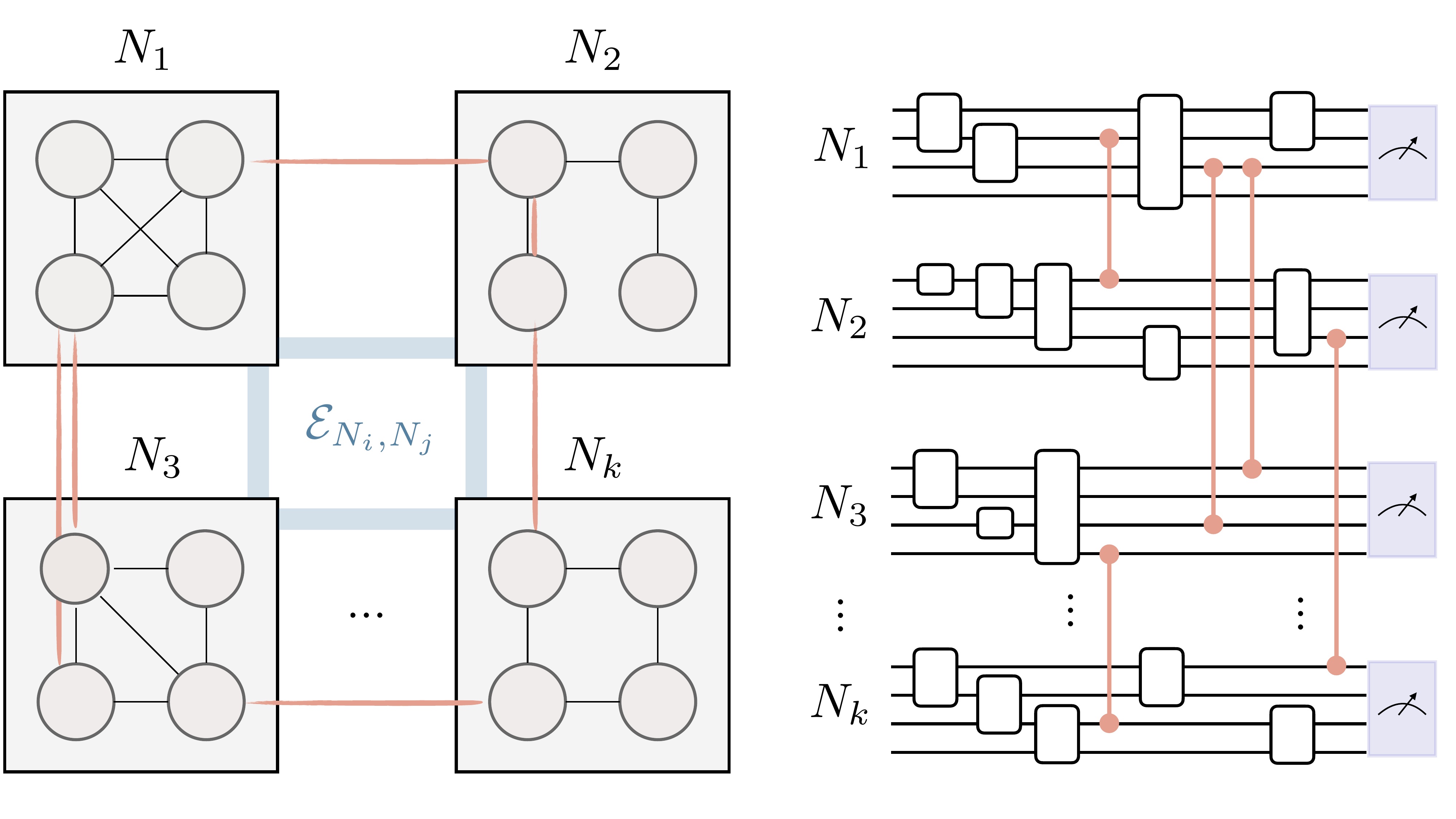}
    \caption{Schematic overview of distributed quantum computation with network architecture (left) and distributed quantum circuit (right).
    Several nodes ($N_1, \cdots,N_k$) of a quantum network are connected through quantum channels ($\mathcal{E}_{N_i,N_j}$ between nodes $N_i$ and $N_j$), \emph{i.e.}, communication channels which can transmit quantum information (Sec.~\ref{ssec:quantum-channels}).
    The individual nodes may be realized using different physical platforms (Sec.~\ref{ssec:platforms}) and have different connectivities (as indicated by black solid lines).
    The quantum channels allow the execution of nonlocal gates (orange lines both in the left and right subplots) between different nodes.
    }
    \label{fig:schematic-figure}
\end{figure}

Fig.~\ref{fig:schematic-figure} is a schematic depiction of a distributed architecture with several nodes $N_1, \cdots, N_k$, that are connected by quantum interconnects.
These enable the realization of nonlocal gates between the different nodes, or modules.
Since the realization of quantum interconnects still poses experimental challenges (see Sec.~\ref{sec:implementations}), we will cover both scenarios where quantum communication is possible, and those where only classical communication is available (see Table~\ref{tab:overview-communication-settings}).

\begin{table}[t!]
\centering
\begin{tabular}{ll}
\toprule
\textbf{Symbol} & \textbf{Meaning} \\
$\rho, \sigma$ & Quantum states (density operators) \\
$N$ & System size (number of qubits) \\
$d$ & Dimension ($d=2^N$ if not stated otherwise) \\
$K$ & Number of (coherently accessible) state copies \\
$\ket{\psi}, \bra{\psi}$ & Braket notation for pure quantum states \\
$\mathcal{H}$ & Hilbert space \\
$\mathcal{E}$ & Quantum channels (CPTP maps) \\
$\{K_i\}$ & Kraus operators for quantum channels \\
$U$ & Unitary operator \\
$\tr, \tr_1$ & Trace and partial trace over subsystem 1 \\
$\mathcal{D}(\rho, \sigma)$ & Trace distance between quantum states \\
$\mathcal{F}(\rho, \sigma)$ & Fidelity between quantum states \\
$\mathcal{F}_p$ & Process fidelity \\
$\mathbb{I}$ & Identity operator \\
$\otimes$ & Tensor product \\
$O$, $\hat{O}$ & Operator and its estimate \\
$\vec{v}$ & Vector \\
$\mathbf{M}$ & Matrix \\
$\mathbb{S}$ & SWAP operator \\
$\mathcal{P}(s)$ & Power set of $s$ \\
$\{I,X,Y,Z\}$ & Pauli basis: identity and Pauli matrices \\
\toprule
\end{tabular}
\caption{Notation used throughout this work.}
\label{tab:notation}
\end{table}

\subsection{Quantum channels \label{ssec:quantum-channels}}

Quantum channels represent changes in quantum systems.
Such changes include desired operations, such as those from gates or state transmission, as well as undesired losses from decoherence.
A quantum channel is given by a completely positive trace-preserving (CPTP) map which represents the transformation of a density operator describing the state of a system to another density operator describing the state after a physical process.
For any quantum state $\rho$, a CPTP map $\mathcal{E}$ can be written in Kraus representation as
\begin{equation}
\mathcal{E}(\rho) = \sum_i K_i \rho K_i^\dagger,
\end{equation}
with the Kraus operators $K_i$ that satisfy
\begin{equation}
\sum_{i=1}^r K_i^\dagger K_i = \mathbb{I}.
\end{equation}
The smallest possible number $r$ that achieves this is called the Kraus rank.

Examples of quantum channels include, but are not limited to:
\begin{enumerate}
\item Identity: $\mathcal{E}(\rho) = \rho$.
\item Time evolution: $\mathcal{E}(\rho) = U \rho U^\dagger$.
\item Depolarizing noise: $\mathcal{E}(\rho) = (1-\lambda)\rho + \frac{\lambda}{d} \mathbb{I}, \ \lambda\in[0,1]$.
\end{enumerate}
To describe a \emph{quantum state transfer} process using the channel $\mathcal{E}_t$, where Alice wishes to transmit a quantum state $\rho_{A}$ to Bob who initially holds a state $\rho_{B_0}$, the final state that Bob receives can be written as
\begin{equation}
\rho_B = \mathrm{tr}_A (\mathcal{E}_t (\rho_A \otimes \rho_{B_0})).
\end{equation}
The quantum state transfer process is perfect if $\rho_A$ can be recovered from $\rho_B$ by a decoding unitary operation.
Perfect and imperfect state transfer has been studied in a variety of physical systems both theoretically~\cite{christandl2004perfect,yao2011robust,lorenzo2013quantumstate,huang2021benchmarking,penas2023improving,penas2024multiplexed} and experimentally~\cite{chapman2016experimental,kurpiers2018deterministic,kandel2021adiabatic}.
A standard metric for validating the success of a state-transfer implementation is the average quantum state fidelity (Sec.~\ref{ssec:metrics}) between the state that Alice prepares and the state that Bob receives.

Another important concept is the \emph{Choi-Jamiołkowski isomorphism}, sometimes also known as the channel-state duality.
It allows to represent quantum channels as quantum states in an enlarged Hilbert space~\cite{jamiolkowski1972linear,choi1975completely}.
In more detail, the Choi-Jamiołkowski isomorphism constructs a one-to-one correspondence between a channel  $\mathcal{E}: \mathcal{L}(\mathcal{H}_A) \to \mathcal{L}(\mathcal{H}_B)$ and a positive semidefinite operator  $\rho_\mathcal{E}$  on  $\mathcal{H}_A \otimes \mathcal{H}_B$. 
Specifically, let $\{|i\rangle\}$ be an orthonormal basis of $\mathcal{H}_A$.
Then define: 
\begin{align}
    \rho_\mathcal{E} = \sum_{i,j} |i\rangle\langle j|\otimes \mathcal{E}\bigl(|i\rangle\langle j|\bigr).
\end{align}
When $\mathcal{E}$ is a completely positive map, 
$\rho_\mathcal{E}$ is positive semidefinite, and when $\mathcal{E}$ is trace-preserving,  $\mathrm{tr}_B(\rho_\mathcal{E}) = \mathbb{I}_A.$ 
From  $\rho_\mathcal{E}$, 
one can also recover $\mathcal{E}$ by performing partial traces and rearrangements. 

\subsection{Metrics \label{ssec:metrics}}

Quantum metrics are essential for evaluating the accuracy, reliability, and efficiency of quantum states, operations, and processes.
Metrics like fidelities quantify how closely a quantum state, operation, or process aligns with its intended or ideal counterpart, providing a direct measure of quality.
On the other hand, metrics such as coherence and gate times indicate how many operations can be performed within the lifetime or dephasing time of a qubit, reflecting the speed and robustness of the system.
Additionally, the number of qubits in a device sets a fundamental limit on the scale of computations it can support, influencing the complexity and depth of algorithms that can be executed.

Together, these metrics define three crucial pillars for assessing quantum device performance: quality, speed, and scale.
In this section, we summarize key metrics that will be used throughout the review, and that are among the quantities commonly employed to benchmark and compare various quantum platforms and protocols.

\subsubsection{Trace Distance}

Classically, to measure how close two probability distributions $\{P_x\}$ and $\{P'_x\}$ are over the same set $x$, the Kolmogorov distance~\cite{kolmogorov1965approximation, nielsen2012quantum}, also known as the $L_1$ distance, is used and defined as $\mathcal{D}(P_x, P'_x) = \frac{1}{2}\sum_{x}|P_x - P'_x|$. By extending this metric to the quantum information setting, the trace distance between two quantum states $\rho$ and $\sigma$ is defined as:
\begin{equation}
    \mathcal{D}(\rho, \sigma) \coloneq \frac{1}{2}\|\rho - \sigma\|_1 = \frac{1}{2} \tr |\rho - \sigma|,
\end{equation}
where $|A| \equiv \sqrt{A^{\dagger}A}$.
The trace distance plays an important role in evaluating the performance of various quantum algorithms~\cite{khatri2020principles,bittel2025optimal}.
Several important properties of trace distance are listed below~\cite{nielsen2012quantum}:

\begin{enumerate}
    \item Symmetric: $\mathcal{D}(\rho,\sigma) = \mathcal{D}(\sigma,\rho)$.
    \item Bounded: $0 \leq \mathcal{D}(\rho,\sigma) \leq 1$; $\mathcal{D}(\rho, \sigma) = 0 \Leftrightarrow \rho = \sigma$.
    \item Invariant: $\mathcal{D}(\rho,\sigma) = \mathcal{D}(U\rho U^\dagger, U\sigma U^\dagger)$.
    \item For two pure states $\rho=\ket{\psi}\bra{\psi}$ and $\sigma=\ket{\phi}\bra{\phi}$: $\mathcal{D}(\ket{\psi},\ket{\phi}) = \sqrt{1 - |\langle \psi | \phi \rangle|^2}$.
\end{enumerate}

\subsubsection{Quantum state fidelity}

Quantum state fidelity, commonly denoted by $\mathcal{F}(\rho, \sigma)$, measures closeness of two arbitrary quantum states $\rho$ and $\sigma$.
Some of its key properties are~\cite{jozsa1994fidelity}:
\begin{enumerate}
    \item Symmetric: $\mathcal{F}(\rho,\sigma) = \mathcal{F}(\sigma,\rho)$.
    \item Bounded: $0 \leq \mathcal{F}(\rho,\sigma) \leq 1$; $\mathcal{F}(\rho, \sigma) = 1 \Leftrightarrow \rho = \sigma$.
    \item Invariant: $\mathcal{F}(\rho,\sigma) = \mathcal{F}(U\rho U^\dagger, U\sigma U^\dagger)$.
    \item If either state is pure: $\mathcal{F}(\rho, \sigma) = \mathrm{tr}(\rho \sigma)$.
\end{enumerate}
Besides, an important inequality relation between trace distance and fidelity is:
\begin{equation}
1-\mathcal{F}(\rho,\sigma)\leqslant\mathcal{D}(\rho,\sigma)\leqslant\sqrt{1-\mathcal{F}^2(\rho, \sigma)}.
\end{equation}
These basic requirements comply with an infinite class of norm-based fidelity measures~\cite{liang2019quantum}.
The most commonly employed one is the Uhlmann-Jozsa fidelity:
\begin{equation}\label{eq:fidelity-eq1}
    \mathcal{F}_1(\rho, \sigma) \coloneq \underset{|\phi\rangle,|\psi\rangle}{\mathrm{max}} |\langle\phi | \psi\rangle|^2 = \left ( \mathrm{tr} \sqrt{\sqrt{\rho} \sigma \sqrt{\rho}} \right )^2.
\end{equation}
In Eq.~\eqref{eq:fidelity-eq1} the states $|\psi\rangle$ and $|\phi\rangle$ are called \emph{purifications} of $\rho$ and $\sigma$, respectively.
A purification is a pure state in an enlarged Hilbert space that, when the auxiliary subsystem is traced out, yields the mixed state $\rho$.
The Uhlmann-Jozsa fidelity is thus a natural mixed-state extension of the fidelity between pure states, because the fidelity between two pure states, $\rho = |\psi_\rho\rangle\langle \psi_\rho|$ and $\sigma = |\psi_\sigma\rangle\langle \psi_\sigma|$, is simply:
\begin{equation}
    \mathcal{F}(\rho, \sigma) = |\langle \psi_\rho | \psi_\sigma \rangle |^2.
\end{equation}

Yet another definition that meets all the desired criteria is:
\begin{equation}\label{eq:fidelity-eq2}
    \mathcal{F}_2(\rho, \sigma) \coloneq \frac{\mathrm{tr}(\rho \sigma)}{\mathrm{max}\{\mathrm{tr}(\rho^2), \mathrm{tr}(\sigma^2)\}}.
\end{equation}
The numerator in Eq.~\eqref{eq:fidelity-eq2} is the \emph{inner product} between $\rho$ and $\sigma$.
If one of the states is known and pure, \textit{e.g.}, the target state is $\sigma = |\psi\rangle\langle\psi|$, methods such as direct fidelity estimation~\cite{flammia2011direct} may be used to obtain an accurate estimate of $\mathrm{tr}(\rho \sigma) = \langle \psi | \rho | \psi \rangle$, see Sec.~\ref{ssec:randomized-measurements}.
Estimating the inner product between two \emph{unknown} quantum states can be achieved with a SWAP test which is reviewed in Sec.~\ref{ssec:swap-test}.
In the literature, the task of estimating it using a distributed architecture is also referred to as \emph{distributed inner product estimation}, which is discussed in Sec.~\ref{ssec:two-copies}.

\subsubsection{Process and gate fidelities}

The notion of mixed-state fidelity can be extended to quantum processes~\cite{raginsky2001fidelity}, \textit{i.e.}, mathematical descriptions of the time evolution of quantum systems.
This can be useful for comparing and benchmarking different quantum devices against each other~\cite{greganti2021crossverification,knorzer2023crossplatform,zheng2024crossplatform}.
A distance measure to compare different quantum experiments should adhere to a number of plausible criteria, similar to the properties listed above for quantum state fidelities~\cite{gilchrist2005distance}.
Two different quantum channels may be compared by evaluating the inner product between their respective Choi states.
For two unitary processes $U_1$ and $U_2$ described by $d\times d$ unitary matrices, this yields a quantum process fidelity defined as:
\begin{equation}
    \mathcal{F}_\mathrm{p}(U_1, U_2) = \frac{|\mathrm{tr}(U_1^\dagger U_2)|^2}{d^2}.
\end{equation}
This can be generalized to the average case for two general quantum processes. For example, for two channels $\mathcal{E}_1$ and $\mathcal{E}_{2}$, the fidelity-based metrics for estimating the distance between them can be naturally defined as:
\begin{equation}
    \mathcal{F}_\mathrm{p}(\mathcal{E}_1, \mathcal{E}_2) = \int  \mathcal{F}(\mathcal{E}_1(\psi), \mathcal{E}_2(\psi)) \, d\psi,
\end{equation}
where $\mathcal{E}(\psi)$ denotes the output state after applying the channel $\mathcal{E}$ to the input state $\ket{\psi}$, and $\mathcal{F}$ is the fidelity between the output states from the two channels. When one of the channels is a unitary operation $U$, this definition yields a specific measure for how well the (implemented) quantum process $\mathcal{E}$ approximates the (desired) unitary $U$~\cite{bowdrey2002fidelity, nielsen2002simple}:
\begin{equation}
    \mathcal{F}_\mathrm{p}(\mathcal{E}, U) = \int \bra{\psi} U^{\dagger} \mathcal{E}(\psi) U \ket{\psi} \, d\psi,
\end{equation}
where the integral is over the uniform Haar measure $d\psi$ on state space.

To estimate quantum process fidelity in practice, many approaches have been developed~\cite{lu2015experimental,emerson2007symmetrized,dankert2009exact,moussa2012practical,flammia2011direct,dasilva2011practical}. Among them, randomized benchmarking~\cite{knill2008randomized, helsen2022general, magesan2011scalable, hashagen2018real, gambetta2012characterization, cross2016scalable, helsen2019new, wood2018quantification, harper2017estimating, sheldon2016characterizing, gaebler2012randomized, onorati2019randomized, erhard2019characterizing, mckay2019threequbit, huang2019fidelity} has emerged as one of the most widely used methods.
It has also been experimentally implemented across various platforms, including ion traps~\cite{erhard2019characterizing, gaebler2012randomized, knill2008randomized}, superconducting qubits~\cite{mckay2019threequbit, gambetta2012characterization}, and quantum dots~\cite{huang2019fidelity} to characterize the performance of quantum devices. The basic protocol of randomized benchmarking proceeds as follows: the system is first initialized in the all-$\ket{0}$ state. Then, $N$ random sequences of $m$ Clifford gates $C_1,C_2,\cdots,C_m$ are generated from the Clifford group. Each sequence is followed by its inverse and applied to the system.
After that, all qubits are measured. By recording the average survival probability $P(m)$—\textit{i.e.}, the probability of measuring the system in state $\ket{0}$—over $N$ repetitions and varying the sequence length $m$, the data is fitted to the function $P(m) = Ap^m + B$ to extract the decay parameter $p$.
The average gate error is then given by $(1-p)(1-\frac{1}{d})$, where $d=2^N$ is the Hilbert space dimension for $N$ qubits.
Consequently, the average gate fidelity is estimated as:
\begin{equation}
    \hat{\mathcal{F}_{p}}(C) =  \frac{(d-1)p+1}{d}.
\end{equation}

\subsubsection{Coherence times}

Qubits are fragile and typically have limited coherence time to maintain superpositions of quantum states. The decoherence of qubits is commonly characterized by the $T_1$ and $T_2$ times~\cite{houck2008controlling, kubica2023erasure, you2007lowdecoherence, cywinski2008how, rigetti2012superconducting, fraval2005dynamic, paik2011observation, nguyen2019highcoherence, somoroff2023millisecond, ithier2005decoherence}, originally from nuclear magnetic resonance (NMR)~\cite{hahn1950spin, bloch1946nuclear}.
The $T_1$ time refers to the energy relaxation time of a qubit, indicating how long it takes for a qubit to decay exponentially from the excited state (\textit{e.g.}, $\ket{1}$) to the ground state (\textit{e.g.}, $\ket{0}$), \textit{i.e.}, the longitudinal relaxation on the Bloch sphere.
This process, also known as energy dissipation, arises due to coupling between the quantum system and the outer environment.
To measure $T_1$ in practice, the qubit can be prepared in the excited state $\ket{1}$ by applying a $\pi$-pulse (\textit{i.e.}, a Pauli $X$ gate).
After waiting for a variable delay time $t$, the probability of the qubit remaining in the $\ket{1}$ state is measured.
By repeating this procedure for various values of $t$ and observing the exponential decay in the measured probability, the $T_1$ time can be extracted.
Specifically, $T_1$ corresponds to the time at which the probability of finding the qubit in the $\ket{1}$ state falls to $1/e$.
The $T_2$ time is the phase coherence time.
It is the time over which a qubit maintains phase coherence between $\ket{0}$ and $\ket{1}$.
This process can cause the loss of off-diagonal elements in the density matrix of the state.
The $T_2$ time can be determined by preparing the qubit in a superposition state, allowing it to evolve, and then measuring in the same basis.
Without further pulses, this Ramsey-type sequence yields the dephasing time $T_2^*$.
To separate out slow fluctuations, a refocusing $\pi$-pulse can be inserted halfway through the evolution, forming a Hahn echo sequence.
The decay of this signal defines $T_2$, which is typically longer than $T_2^*$.

\subsection{Quantum teleportation \label{ssec:teleportation}}

\subsubsection{Quantum state teleportation}

\begin{figure}[b]
    \centering
    \includegraphics[width=\linewidth]{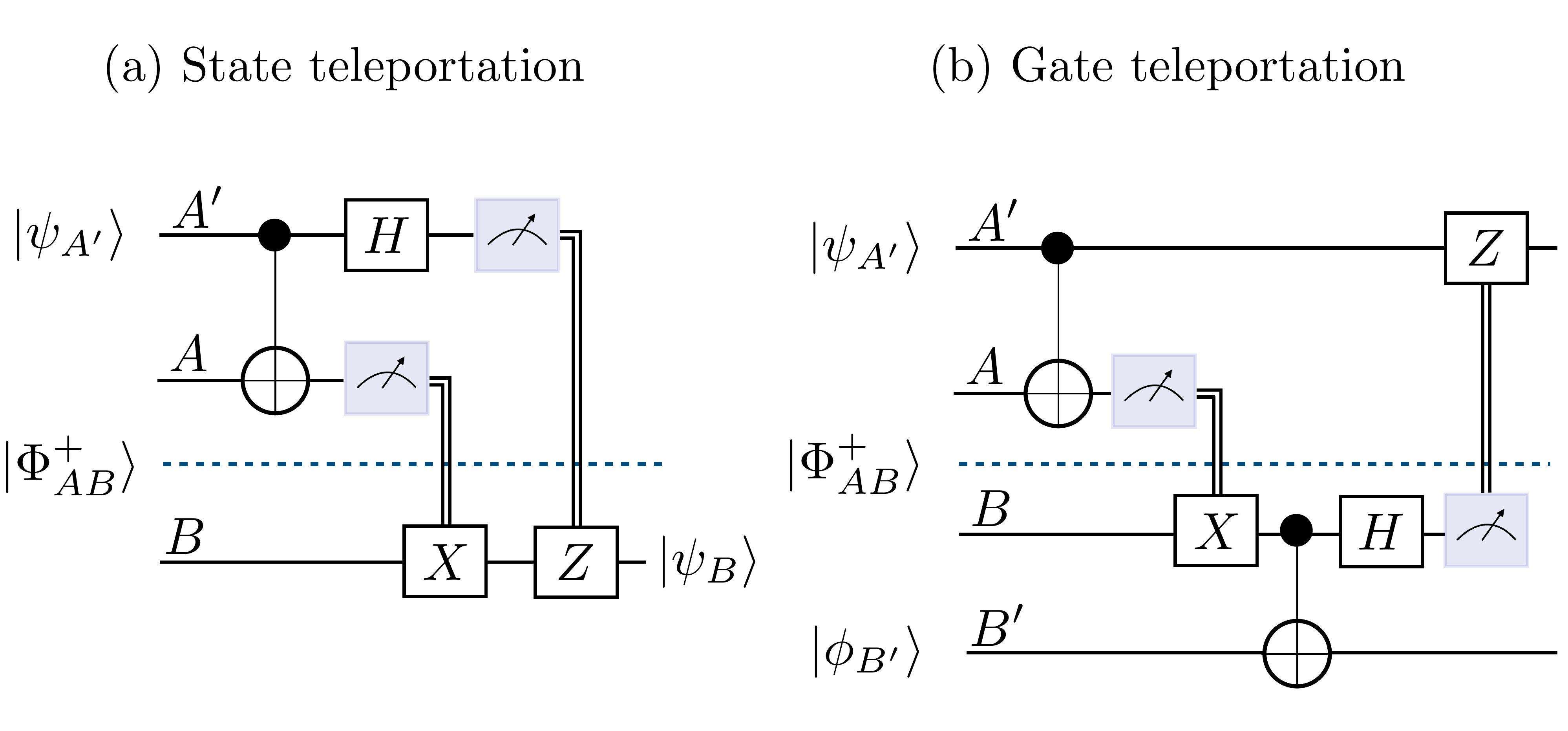}
    \caption{Quantum teleportation protocols.
    (a) State teleportation of a single qubit and (b) teleportation of a controlled-NOT gate.
    The latter implements a CNOT operation between registers $A^\prime$ and $B^\prime$, consuming the Bell pair $\ket{\Phi_{AB}^+}$.
    }
    \label{fig:teleportation}
\end{figure}

Quantum teleportation allows the transfer of a quantum state using shared entanglement and classical communication~\cite{bennett1993teleporting, boschi1998experimental} and is a key ingredient of quantum networks~\cite{gottesman1999demonstrating}.
The basic idea behind state teleportation can be shown from a single-qubit pure state.
Suppose Alice ($A$) and Bob ($B$) share a maximally entangled Bell state:
\begin{equation}\label{eq:bell-phi-plus}
    \ket{\Phi^{+}_{AB}} = \frac{1}{\sqrt{2}}(\ket{0_A 0_B} + \ket{1_A 1_B}),
\end{equation}
and Alice wants to send the unknown state $\ket{\psi_{A'}} = \alpha\ket{0_{A'}} + \beta\ket{1_{A'}}$ to Bob.
The combined state of the three qubits can be written as:
\begin{equation}
\begin{split}
    &\ket{\psi_{A'}} \otimes \ket{\Phi^{+}_{AB}} \\
    =\; &\frac{1}{2}\ket{\Phi^{+}_{A'A}} \otimes \ket{\psi_B}
    + \frac{1}{2}\ket{\Phi^{-}_{A'A}} \otimes Z_B \ket{\psi_B} + \\
    & \frac{1}{2}\ket{\Psi^{+}_{A'A}} \otimes X_B \ket{\psi_B}
    + \frac{1}{2}\ket{\Psi^{-}_{A'A}} \otimes Z_B X_B \ket{\psi_B},
\end{split}
\end{equation}
where $\{\ket{\Phi^{+}}\bra{\Phi^{+}}, \ket{\Phi^{-}}\bra{\Phi^{-}}, \ket{\Psi^{+}}\bra{\Psi^{+}}, \ket{\Psi^{-}}\bra{\Psi^{-}}\}$ form the Bell basis and
\begin{align}
    \ket{\Phi^{-}} &= \frac{1}{\sqrt{2}}(\ket{0 0} - \ket{1 1}), \\
    \ket{\Psi^{+}} &= \frac{1}{\sqrt{2}}(\ket{0 1} + \ket{1 0}), \\
    \ket{\Psi^{-}} &= \frac{1}{\sqrt{2}}(\ket{0 1} - \ket{1 0}).
\end{align}
Alice performs a local Bell measurement on qubits $A'$ and $A$, then sends the measurement result to Bob.
Based on this result, Bob applies the corresponding Pauli correction to recover the original state on his qubit: $\ket{\psi_{B}} = \alpha\ket{0_B} + \beta\ket{1_B}$.
The corresponding quantum circuit is shown in Fig.~\ref{fig:teleportation}(a).

In practice, creating Bell states between distant nodes is difficult due to fiber loss, decoherence, and other noise sources.
To enable long-distance quantum teleportation, long-range entanglement is required.
Entanglement swapping is a protocol that extends entanglement by connecting two shorter entangled links to form a longer one~\cite{pan1998experimental}.
This technique is essential for building large-scale quantum repeaters and the quantum internet~\cite{munro2015quantum}.
The basic idea is as follows.
Suppose Alice and Bob share a Bell pair $\ket{\Phi_{AB_1}^{+}}$, and Bob and Charlie ($C$) share another pair $\ket{\Phi_{B_2C}^{+}}$.
The joint state can be written as:
\begin{equation}
\begin{split}
& \ket{\Phi^{+}_{AB_1}} \otimes \ket{\Phi^{+}_{B_2C}} \\
=  & \frac{1}{2} \ket{\Phi^{+}_{B_1B_2}} \otimes \ket{\Phi^{+}_{AC}} 
+ \frac{1}{2} \ket{\Phi^{-}_{B_1B_2}} \otimes \ket{\Phi^{-}_{AC}} + \\
& \frac{1}{2} \ket{\Psi^{+}_{B_1B_2}} \otimes \ket{\Psi^{+}_{AC}} 
+ \frac{1}{2} \ket{\Psi^{-}_{B_1B_2}} \otimes \ket{\Psi^{-}_{AC}}.
\end{split}
\end{equation}

By performing a Bell-basis measurement on Bob’s local qubits $B_1$ and $B_2$, the state of Alice and Charlie collapses into one of the four Bell states.
This effectively creates an entangled link between them, even though they never interacted directly.\\

\implementations
Quantum teleportation is a central building block for many QIP applications, ranging from quantum cryptography~\cite{ekert1991quantum}, communication~\cite{duan2001longdistance, aspelmeyer2003longdistance} and the experimental implementation of loophole-free Bell tests~\cite{hensen2015loopholefree}, to scalable quantum networks, underpinning advances in distributed quantum computing~\cite{buhrman2003distributed} and quantum internet architectures~\cite{kimble2008quantum, wehner2018quantum}.
Moreover, although earlier works on entanglement-based approaches to quantum key distribution (QKD) did not explicitly require quantum teleportation~\cite{ekert1991quantum}, they highlighted the advantages of entanglement over previous protocols~\cite{bennett1983quantum, bennett1992quantum, bennett2014quantum}. 
Particularly, to ensure the security of quantum protocols remains a central goal, device-independent quantum key distribution (DI-QKD), based on the violation of Bell inequalities~\cite{ekert1991quantum}, eliminates the need to trust the internal workings of any involved devices~\cite{acin2006efficient,nadlinger2022experimental,acin2007deviceindependent,acin2006bells,masanes2011secure,vazirani2019fully,miller2016robust}.

Soon after the first theoretical proposal on quantum teleportation~\cite{bennett1993teleporting}, proof-of-concept experiments followed~\cite{bouwmeester1997experimental}, and quantum teleportation had eventually been achieved in laboratories with all kinds of technological platforms~\cite{pirandola2015advances}. 
Such platforms include photonic qubits encoded in polarization~\cite{bouwmeester1997experimental, ursin2004quantum, boschi1998experimental, jin2010experimental, kim2001quantum}, time-bin~\cite{landry2007quantum, deriedmatten2004long, marcikic2003longdistance}, spin-orbit~\cite{wang2015quantum}, single-~\cite{lombardi2002teleportation, giacomini2002active} and dual-rail~\cite{fattal2004quantum, metcalf2014quantum}, nuclear magnetic resonance~\cite{nielsen1998complete}, optical modes~\cite{furusawa1998unconditional, takei2005highfidelity, yonezawa2007experimental, lee2011teleportation, takeda2013deterministic,bowen2003experimental, zhang2003quantum, takei2005experimental, yukawa2008highfidelity}, 
atomic ensembles~\cite{sherson2006quantum, krauter2013deterministic, chen2008memorybuiltin, bao2012quantum}, trapped atoms~\cite{nolleke2013efficient, boozer2007reversible}, ions~\cite{riebe2004deterministic, barrett2004deterministic, riebe2007quantum, olmschenk2009quantum} and solid-state systems such as NV centers~\cite{pfaff2014unconditional}, quantum dots~\cite{gao2013quantum}, doped crystals~\cite{bussieres2014quantum, saglamyurek2011broadband, clausen2011quantum} and superconducting qubits~\cite{steffen2013deterministic}.

In the last decade, experimental quantum teleportation has seen continuous development~\cite{krishnan2025deterministic,laneve2024quantum,ren2017groundtosatellite,fedorov2021experimental,sun2016quantum,barasinski2019demonstration,daurelio2025boosted,strobel2024quantum,shen2023hertzrate,hoke2023measurementinduced,kucera2024demonstration}, and also shifted from simple to complex quantum states. These include multiple degrees of freedom~\cite{wang2015quantum, graham2015superdense,chapman2020timebin,luo2016teleportation,ru2021quantum,liu2024deterministic} and high-dimensional~\cite{hu2020experimental, luo2019quantum,martin2017quantifying,wang2018multidimensional,cervera-lierta2022experimental,lv2024demonstration} quantum states, continuous-variable quantum states~\cite{shi2023continuous, ulanov2017quantum}, and from proof-of-principle demonstrations to real-world applications~\cite{hu2023progress}. Moreover, long-distance quantum teleportation records have been consistently broken in the past years~\cite{hu2023progress, ren2017groundtosatellite}.
Photon polarization has allowed for quantum teleportation over distances of 144km~\cite{ursin2007entanglementbased} and 248km~\cite{neumann2022continuous} with optical fibres, but also in free-space links~\cite{yin2012quantum, ma2012quantum}, allowing for QKD implementations in metropolitan scales~\cite{krzic2023metropolitan}. Satellite-based approaches have allowed to reach distances over 1200 km~\cite{yin2017satellitebased, deforgesdeparny2023satellitebased} and space-to-ground quantum communication networks over 4600km have been established~\cite{chen2021integrated}.
To enable such long-distance quantum teleportation, many of the above-mentioned works employ quantum multiplexing (cf. Sec.~\ref{ssec:multiplexing}), which significantly boosts the transmission capacity of optical communication systems by combining multiple channels into a single link~\cite{covey2023quantum, lopiparo2019quantum, liu2020orbital, munro2010quantum, haldar2025reducing}.

\subsubsection{Quantum gate teleportation}

Quantum teleportation is not limited to quantum states, but can also be applied to quantum gates~\cite{gottesman1999demonstrating}.
Gate teleportation is a variant of quantum teleportation that distributes local gate operations between spatially separated particles. 
Similar to state teleportation, gate teleportation also requires shared entanglement, local measurements and classical feedback.
It is at the core of measurement-based quantum computing based on cluster states~\cite{raussendorf2001oneway, thomas2022efficient} and finitely correlated or projected entangled pair states~\cite{gross2007novel}, and also a foundational concept in distributed quantum computing~\cite{buhrman2003distributed}.
Moreover, the concept of gate teleportation is widely used in fault-tolerant quantum computations with quantum error correction, particularly for non-Clifford gates~\cite{sahay2025error, hillmann2022performance, knill1997theory, grimsmo2020quantum, noh2022lowoverhead, walshe2020continuousvariable, larsen2021faulttolerant}.
A quantum circuit for the teleportation of a CNOT gate, as proposed in \cite{eisert2000optimal}, is shown in Fig.~\ref{fig:teleportation}(b).\\

\implementations
Quantum gate teleportation has been realized experimentally across various platforms.
The first demonstrations were performed using photonic systems~\cite{huang2004experimental}, followed by more advanced realizations~\cite{gao2010teleportationbased, dong2024experimental,liu2024nonlocal, feng2025chiptochip}.
Subsequent experiments extended gate teleportation to other architectures, including superconducting qubits~\cite{qiu2025deterministic,chou2018deterministic}, trapped ions~\cite{wan2019quantum, main2025distributed}, QED cavities~\cite{daiss2021quantumlogic}, and spin qubits in quantum dots~\cite{qiao2020conditional, kojima2021probabilistic}. 
Notably, in 2018, gate teleportation was demonstrated using a microwave cavity~\cite{reagor2016quantum} coupled to a superconducting circuit~\cite{chou2018deterministic}, achieving a coherence time of approximately 1~ms and a process fidelity of around 79\%. 
In terms of spatial separation, gate teleportation has been demonstrated over distances of up to 0.42~m~\cite{fedorov2021experimental}, 2~m~\cite{main2025distributed}, and 5~m/1~km~\cite{feng2025chiptochip} in laboratory settings.
Moreover, gate teleportation over fiber links exceeding 7~km has been demonstrated using photonic qubits~\cite{liu2024nonlocal}, marking a significant step toward non-local quantum processing.

\subsection{Bell nonlocality and the CHSH inequality}\label{ssec:bell-nonlocality}

One of the most profound developments in the foundations of physics is Bell's theorem \cite{bell1964einstein}. Since the seminal Einstein-Podolsky-Rosen paper \cite{einstein1935can} and culminating in the 2022 Nobel Prize in Physics, the counterintuitive behavior of quantum mechanics has fascinated scientists. Bell's theorem lies at the heart of a precise formulation of this phenomenon. It states that Nature does not admit a description compatible with a local hidden-variable model. Formally, such a model would describe the correlations between distant observers as 
\begin{equation}
    p(a,b|x,y) = \int_\Lambda d\lambda ~\pi(\lambda) p_\mathrm{A}(a|x,\lambda)p_\mathrm{B}(b|y,\lambda).
    \label{eq:LHVM},
\end{equation}
with the probability density $\pi$, and the probability $p_\mathrm{A}(a|x,\lambda)$ ($p_\mathrm{B}(b|y,\lambda)$) that Alice obtains a result $a$ ($b$) if she conducts an experiment $x$ ($y$) and a local variable $\lambda$ describing the experiment.

Bell nonlocality \cite{brunner2014bell}, measured through the violation of a so-called Bell inequality, shows that some quantum correlations cannot be explained within this framework. The most celebrated Bell inequality is without a doubt the Clauser-Horne-Shimony-Holt \cite{clauser1969proposed} inequality, given by 
\begin{equation}\label{eq:chsh}
    S = \langle A_0 B_0 \rangle + \langle A_1 B_0 \rangle + \langle A_0 B_1 \rangle - \langle A_1 B_1 \rangle,
\end{equation}
with the measurement observables $A_0$, $A_1$, $B_0$ and $B_1$, stating that $S\leq 2$ under Eq.~(\ref{eq:LHVM}), whereas certain measurements on a maximally entangled pair of qubits can yield $S=2\sqrt{2}$.

Besides its philosophical implications, Bell nonlocality is now established as a resource that enables device-independent quantum information processing tasks \cite{acin2007deviceindependent}, allowing certification of quantum technologies from minimal assumptions, excluding even those on the internal working of the devices. For instance, Bell nonlocality certifies the presence of entanglement, and the former is strictly stronger than the latter, as the converse implication is false in general \cite{werner1989quantum, augusiak2014local, augusiak2015entanglement, augusiak2018constructing, bowles2016genuinely}.

Bell nonlocality finds practical applications in quantum information processing via self-testing \cite{supic2020selftesting}, which allows to infer the underlying physics of a quantum experiment in a black box scenario. Self-testing statements have been developed for important classes of quantum states \cite{bamps2015sumofsquares, salavrakos2017bell, baccari2020scalable, augusiak2019bell, kaniewski2019maximal, coladangelo2017all, wang2018multidimensional} and entangled subspaces \cite{baccari2020deviceindependent, makuta2021selftesting}.
Bell nonlocality is by now established as a resource enabling novel device-independent (DI) quantum information tasks, including DI randomness generation \cite{colbeck2011quantum, dhara2013maximal, li2025necessary, zhang2025randomness} and expansion \cite{colbeck2011private, colbeck2012free, augusiak2014elemental}; DI quantum cryptography \cite{ekert2014ultimate}, including quantum key distribution \cite{pironio2009deviceindependent, vazirani2019fully} , bit commitment \cite{aharon2016deviceindependent}, weak string erasure \cite{kaniewski2016deviceindependent} and position verification \cite{buhrman2014positionbased, ribeiro2018device}; and delegated quantum computing \cite{reichardt2013classical}.

In the deep multipartite regime, loophole-free Bell tests remain technologically too demanding and one is forced to relax the stringent requirements for Bell nonlocality. So-called Bell correlations can be revealed by measuring collective observables in experiments where individual addressing is out of reach, to effectively evaluate few-body symmetric Bell inequalities \cite{tura2014detecting, tura2015nonlocality, guo2023detecting, fadel2018bell, marconi2025symmetric, muller-rigat2021inferring, wagner2017bell, aloy2025integrability, aloy2024deriving, muller-rigat2024threeoutcome}. So-called Bell-operator correlations are stronger, requiring access to individual addressing \cite{tura2017energy, wang2025probing}. However, spacelike separation is often not enough to close the locality loophole. In the multipartite case, the depth of genuinely nonlocal correlations \cite{liang2015family, aloy2019deviceindependent, tura2019optimization, baccari2019bell} becomes a good figure of merit for benchmarking the intrinsic quantum behavior a device can produce.\\

\implementations
The first experimental confirmations of violations of Bell inequalities were in \cite{freedman1972experimental} and then more convincingly in \cite{aspect1982experimental}. Many experiments have followed since, aiming at closing experimental loopholes that would prevent a strictly conclusive falsification of Eq.~(\ref{eq:LHVM}), especially the so-called locality and detection loopholes, although even the freedom of choice loophole has been tackled \cite{abellan2018challenging}. The first loophole-free Bell tests were achieved simultaneously in \cite{hensen2015loopholefree} using NV centers and \cite{giustina2015significantloopholefree, shalm2015strong} using photons. Subsequently, in \cite{rosenfeld2017eventready} using entangled atoms, \cite{li2018test} using photons and randomness from distant stars, and \cite{storz2023loopholefree} using superconducting qubits.

Bell correlations have been  demonstrated in systems such as Bose-Einstein condensates \cite{schmied2016bell} and thermal ensembles \cite{engelsen2017bell}, whereas Bell-operator correlations have been recently demonstrated in superconducting devices \cite{wang2025probing}.

\subsection{Entanglement distillation} \label{ssec:ent-dist}

Many applications of distributed quantum processing, such as quantum teleportation (Sec.~\ref{ssec:teleportation}), require remote entanglement as a resource.
Because quantum interconnects and matter-photon interfaces are lossy and noisy, the raw Bell pairs available between the nodes often have fidelity below what the applications can tolerate.
An essential concept is therefore entanglement distillation, which allows to convert many low-fidelity entangled pairs shared across two devices into fewer, high-fidelity entangled pairs.
The terms \textit{entanglement distillation} and \textit{entanglement purification} are used interchangeably in the literature; to avoid confusion with other uses of purification (\textit{e.g.}, the quantum state purification in Sec.~\ref{ssec:purity_amp}) we use the term distillation throughout.
Conversely, a high-fidelity Bell pair can be diluted into a larger number of lower-fidelity pairs (entanglement dilution). 
For a concise and pedagogical introduction to entanglement distillation and dilution, we refer to the textbook~\cite{nielsen2012quantum}. 
We also highlight several foundational works~\cite{briegel1998quantum, bennett1996concentrating, deutsch1996quantum, dur1999quantum}.
More recently, the influence of noisy gates and finite memory storage times on the performance of distillation in a quantum network was studied in~\cite{victora2023entanglement}. 
We discuss fault-tolerant protocols for the distillation of Bell pairs in Sec.~\ref{ssec:qec}.\\

\implementations
Entanglement distillation has been demonstrated on several platforms. 
For discrete-variable photonics, single-photon entanglement and polarization entanglement were distilled using linear optics~\cite{pan2003experimental, salart2010purification}.
Also, several experiments considered the continuous-variable regime~\cite{hage2008preparation, takahashi2010entanglement, dong2010continuousvariable, kurochkin2014distillation}.
Beyond photonic systems, two-atom Bell pairs were distilled in trapped ions~\cite{reichle2006experimental}, and remote solid-state nodes demonstrated distillation with stored memory qubits~\cite{kalb2017entanglement}.

\subsection{Quantum circuits \label{ssec:quantum-circuits}}

Implementing a quantum algorithm across multiple nodes requires adapting quantum circuits to the realities of hardware: limited qubit connectivity, restricted communication bandwidth, and other architectural constraints.
To address these challenges, various circuit partitioning techniques are used to distribute tasks when operating under tight hardware limitations.
In this section, we introduce the key methods that enable the practical implementation and analysis of quantum circuits in distributed and hybrid quantum–classical settings.

\subsubsection{Circuit knitting}

Circuit knitting comprises a collection of techniques designed to overcome the limitations of current quantum hardware.
Its primary motivation is to increase the number of available qubits by executing computations on devices that are smaller than the original circuit nominally requires, thereby circumventing memory limitations.
By partitioning a large quantum circuit into smaller sub-circuits, each can be executed on separate quantum processors or sequentially on a single device, effectively enabling the simulation of larger quantum systems than the hardware would typically allow.

\begin{figure}[b]
    \centering
    \includegraphics[width=\linewidth]{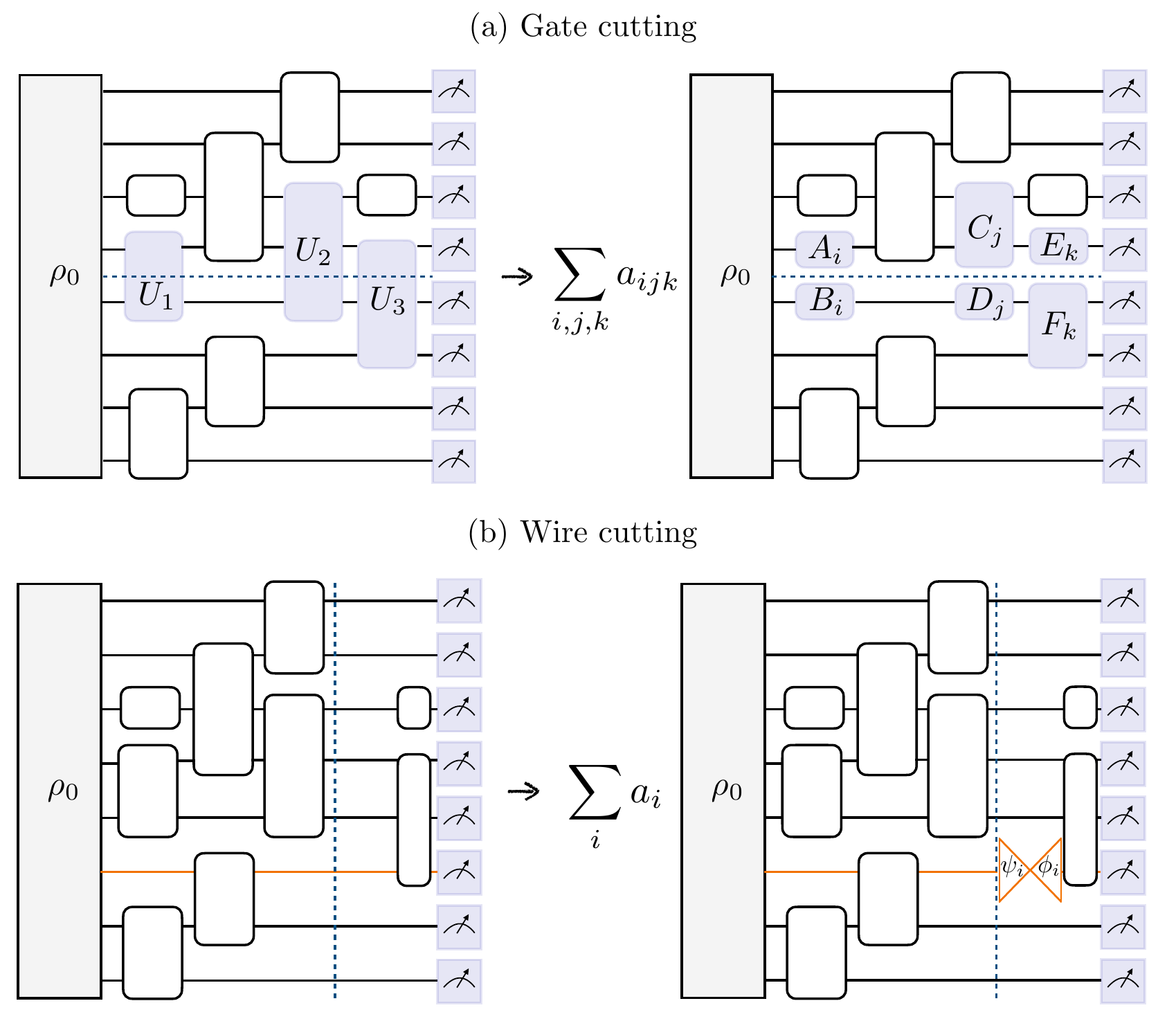}
    \caption{The width and depth of a quantum circuit may be reduced using circuit-knitting techniques.
    In (a) gate cutting, a nonlocal circuit may be simulated by cutting it in two halves and running the resulting local circuits separately.
    A (b) wire cut is a technique where a qubit wire is split, and the resulting segments are simulated independently, followed by classical post-processing to recover the full circuit behavior.
    Circuit cutting comes at the cost of a sampling overhead.
    }
    \label{fig:circuit_knitting}
\end{figure}

We distinguish between \emph{wire cuts} and \emph{gate cuts} when partitioning a quantum circuit into smaller circuits.
Wire cuts are vertical cuts through a circuit~\cite{perez-salinas2023shallow, pira2023invitation, fuller2025improved}, sometimes referred to as time-like cuts.
They cut a \emph{wire} of a quantum circuit, which belongs to a specific qubit, at a given location to divide it into shorter subcircuits~\cite{peng2020simulating}.
While circuit-knitting procedures reduce the hardware requirements to execute each of the more manageable circuits, they come at the cost of a sampling overhead that scales exponentially in the number of wires and gates involved in a cut~\cite{piveteau2024circuit}.
Gate cuts are horizontal or space-like splits of quantum circuits, to reduce the width of a circuit and make it fit on smaller devices~\cite{chen201864qubit}.
Gate cutting is used to simulate nonlocal operations by sampling a set of local operations, \textit{e.g.}, in order to construct virtual two-qubit gates by sampling single-qubit operations~\cite{mitarai2021constructing,mitarai2021overhead}.
This idea is schematically shown in Fig.~\ref{fig:circuit_knitting}.
The sampling overhead of any circuit-knitting technique scales exponentially in the number of cuts in the circuit.

Circuit knitting has been studied using quasiprobability decompositions of quantum channels~\cite{pashayan2015estimating,brenner2023optimal,harrow2025optimal}.
In this framework, a unitary quantum channel $\mathcal{U}$ is decomposed in a form
\begin{equation}
\mathcal{U} = \sum_{i=1}^m a_i \mathcal{E}_i,
\end{equation}
with real numbers $a_i$ and operations $\mathcal{E}_i$ that may be realized by the hardware.
Quasiprobability simulation is based on executing these operations randomly, yielding a sampling overhead that is related to the coefficients $a_i$~\cite{piveteau2024circuit,marshall2023all}.
Considering multiple gate cuts at once can further reduce overhead when allowing for classical communication and randomized measurements~\cite{lowe2023fast}.

Optimal partitioning of quantum circuits has been investigated using graph-based approaches~\cite{tomesh2023divide,brandhofer2024optimal,burt2025multilevel} and it has been shown that a machine learning model can approximate the outputs of a large quantum circuit using evaluations from significantly fewer, smaller partitioned circuit samples~\cite{marshall2023high}.
Circuit knitting approaches have also been widely applied to specific problems, such as Quantum Approximate Optimization Algorithm (QAOA)~\cite{bechtold2023investigating, soloviev2025scaling, yang2024understanding} and Hamiltonian simulations~\cite{yang2024understanding, harrow2025optimal}.
Furthermore, quantum circuits have been mapped to distributed architectures using a Quadratic Unconstrained Binary Optimization (QUBO)~\cite{bandic2023mapping} approach and the Overall Extreme Exchange (OEE) algorithm~\cite{baker2020timesliced}.

Another circuit-knitting technique, known as entanglement forging, uses the Schmidt decomposition to reduce quantum hardware requirements through classical postprocessing~\cite{eddins2022doubling}.
Any pure state $|\psi\rangle$ of $2N$ qubits, split into subsystems $A$ and $B$, can be written as:
\begin{equation}
|\psi\rangle = \sum_{n=1}^R \lambda_n\, U |b_n\rangle_A \otimes V |b_n\rangle_B
\end{equation}
where $\{\lambda_n\}$ are non-negative Schmidt coefficients, $U$ and $V$ are unitaries on each $N$-qubit subsystem, $R$ is the Schmidt rank, and $\{|b_n\rangle\}$ is a fixed orthonormal basis.
In practice, $U$ and $V$ are implemented as parameterized quantum circuits, optimized to minimize a target observable such as the energy, $E = \langle \psi | \hat H | \psi \rangle$.

To estimate expectation values, the quantum computer prepares $N$-qubit states $U|b_n\rangle$, $V|b_n\rangle$, and superpositions
\begin{equation}
|\phi^{p}_{b_n b_m}\rangle = \frac{1}{\sqrt{2}} \left( |b_n\rangle + i^p |b_m\rangle \right)
\end{equation}
with $p \in \{0, 1, 2, 3\}$, allowing efficient extraction of diagonal and off-diagonal matrix elements.
Measurement results are combined with the $\lambda_n$ coefficients in classical post-processing to reconstruct global observables.

The result is passed to a classical optimizer, which updates the parameters in $U$, $V$, and the $\lambda_n$.
This loop is repeated until convergence.
While the method reduces the number of required qubits from $2N$ to $N$, it introduces an overhead that depends on the entanglement in the system—smaller when the state can be partitioned into weakly entangled halves, such as in some molecular ground-state problems.\\

\subsubsection{Dynamic circuits and mid-circuit measurements}

Dynamic circuits combine quantum computation with classical information processing. 
They involve mid-circuit measurements, classical processing of the outcomes, and optionally, conditional quantum operations based on those outcomes.
Because dynamic circuits replace parts of the quantum evolution with measurement and classical logic, they are generally non-unitary, offering new tools for quantum algorithms and quantum information tasks.
A well-known example is quantum teleportation, but dynamic circuits have broader applications, including quantum error correction~\cite{preskill1997faulttolerant,terhal2015quantum,dur2007entanglement}, phase estimation~\cite{dobsicek2007arbitrary,corcoles2021exploiting}, long-range entanglement generation~\cite{baumer2024efficient,lu2022measurement,tantivasadakarn2024longrange}, and quantum state preparation~\cite{foss-feig2023experimental,raussendorf2003measurementbased,buhrman2024state,malz2024preparation,piroli2021quantum,piroli2024approximating}.
For instance, in quantum error correction, mid-circuit measurements extract error syndromes~\cite{dur2007entanglement}, which can then be used to apply appropriate corrections. 
In systems with limited qubit connectivity, generating long-range entanglement typically requires deep circuits due to many SWAP gates. 
Dynamic circuits can reduce this depth by using parallelized classical feedforward, enabling more efficient implementations~\cite{baumer2024efficient}.
However, they face hardware limitations. 
Conditional operations depend on accurate measurement outcomes, but current hardware still suffers from readout errors, which can lead to incorrect operations. 
Additionally, measurements and the associated classical processing can be time-consuming, requiring long coherence times to maintain quantum information throughout the circuit.\\

\implementations
Recent experiments across diverse quantum platforms have demonstrated the feasibility of mid-circuit measurements and dynamic circuits.
In superconducting circuits, dynamic feedback has been integrated with repeated mid-circuit measurements to realize teleportation, feedforward logic, long-range entanglement and quantum error correction protocols~\cite{baumer2024efficient,carreravazquez2024combining,song2024realization, corcoles2021exploiting,baumer2024quantum}.
In trapped-ion systems, high-fidelity mid-circuit measurements have been combined with qubit reset and reuse, enabling multiple rounds of measurement, real-time feedback, and scalable implementations of quantum error correction and verifiable quantum computation~\cite{foss-feig2023experimental, wan2019quantum, decross2023qubitreuse, zhu2023interactive, ringbauer2025verifiable}.
Neutral-atom platforms have demonstrated projective mid-circuit measurements in tweezer arrays~\cite{muniz2025repeated, deist2022midcircuit}, with corrections of atom losses by conditionally refilling atomic sites~\cite{norcia2023midcircuit}.
Coherent control after mid-circuit measurement has been shown using two-qubit gate sequences, atom rearrangement, and measurement-based branching~\cite{graham2023midcircuit}, enabling conditional logic within programmable atom arrays.
These developments establish measurement and feedback operations as viable primitives for quantum control across different platforms.

\subsubsection{Software and tools}
Automatically compiling quantum circuits for distributed architectures~\cite{ferrari2021compiler,cuomo2023optimized,promponas2024compiler,zhou2025optimizing, liu2025ecdqc} is important for optimal resource allocation~\cite{liu2025codesign,bahrani2024resource,mao2023qubit}.
Several programming tools and libraries have been developed to facilitate the deployment of circuit-partitioning techniques in numerical and actual experiments with early-generation hardware.
IBM’s Qiskit add-on \emph{Circuit Cutting} (formerly known as \emph{Circuit Knitting Toolbox}) provides both gate-cutting and wire-cutting capabilities within the Qiskit ecosystem, supporting automated partitioning and reconstruction workflows via Python APIs~\cite{javadi-abhari2024quantum}.
The package \emph{QCut}~\cite{nivala2024qcut}, built on top of Qiskit, enables the decomposition of large quantum circuits into smaller subcircuits that eliminate the need for reset gates or mid-circuit measurements.
More recently, \emph{Qdislib}~\cite{tejedor2025distributed} has introduced a unified framework for both gate and wire cutting within hybrid quantum-classical HPC environments.
It features graph-based circuit partitioning, employs PyCOMPS, a tool for managing simulations on high-performance computing clusters, for orchestration, and supports circuit formats from both Qiskit and Qibo, allowing flexible execution across CPUs, GPUs, and QPUs.
\emph{InterPlace} is also implemented in Qiskit and supports hardware-aware circuit partitioning~\cite{du2025optimizing}.
Other approaches have also contributed significantly to the evolution of circuit-knitting tools.
\emph{CutQC}~\cite{tang2021cutqc}, one of the earlier implementations from the Qiskit ecosystem, used mixed-integer programming to optimize cut placement and provided an early demonstration of efficient circuit reconstruction.
In parallel, hardware-aware gate-cutting methods have emerged to align cutting strategies with device-specific properties such as gate fidelity, topology, and noise profiles, thus minimizing sampling overhead.
Adaptive circuit knitting frameworks further introduce dynamic cut placement and resource allocation that respond in real time to hardware feedback and runtime constraints~\cite{mohseni2024how}.

Other libraries are designed to reduce distribution costs~\cite{mengoni2025efficient} and simulate quantum networking operations with remote entanglement generation~\cite{wu2022collcomm}.
For example, \emph{NetQASM}~\cite{dahlberg2022netqasm} is a low-level instruction set architecture.
\mbox{\emph{SimulaQron}}~\cite{dahlberg2018simulaqron} is an application-level simulator for quantum internet applications.
\emph{QuNet} is a quantum network simulator to design and test network protocols~\cite{diadamo2021qunetsim}.
\emph{SeQUeNCe} is a discrete-event simulator~\cite{wu2021sequence}.
\emph{QuISP}~\cite{satoh2022quisp} is an event-based simulation package for large-scale quantum networks.
\emph{QNodeOS} is designed as a quantum network operating system to delegate computations from a client to a server~\cite{delledonne2025operating}.
\emph{SwitchQNet} is a compiler that optimizes the scheduling of quantum communication for  quantum computers connected through switch networks~\cite{zhang2025switchqnet}.

Similarly to parallel computing models of classical machines that assess performance based on network characteristics such as latency and the number of processing units~\cite{culler1993logp}, quantum network architectures can be optimized using performance metrics~\cite{haner2021distributed} and standardized based on a quantum message passing interface~\cite{shi2023reference}.
Altogether, these and many other tools reflect a maturing ecosystem for distributed quantum circuit and quantum network simulation, increasingly optimized for diverse hardware targets, and integrated into scalable quantum-classical workflows.

\subsection{Quantum state tomography and multi-copy learning\label{ssec:state_tomography}}

Quantum state tomography is the process by which a quantum state is reconstructed from repeated measurements on an ensemble of nominally identical copies of said state.
Generally, the state of a $N$-qubit quantum system can be expressed in the Pauli basis as
\begin{equation}
    \rho = \frac{1}{2^N} \sum_{i_1,\cdots, i_N = 0}^3 c_{i_1,\cdots, i_N} P_{i_1,\cdots i_N},
\end{equation}
with $4^N-1$ independent coefficients $c_{i_1, \cdots, i_N}$ and $P_{i_1,\cdots, i_N}\in\{I,X,Y,Z\}^{\otimes N}$.
While full state tomography allows for a complete reconstruction of $\rho$, it requires a number of measurements which grows exponentially in the number of qubits.
In practice, it is thus currently limited to approximately $10$-qubit states~\cite{song201710qubit}.

To mitigate this exponential scaling, numerous approaches exploit structural assumptions on $\rho$, such as high purity or efficient matrix-product representations, to reduce the measurement overhead~\cite{cramer2010efficient,gross2010quantum,flammia2012quantum}, or overlapping tomography to measure subsystems~\cite{hu2025quantum, wang2025mitigating, araujo2022local, cotler2020quantum, hansenne2025optimal}.
Moreover, many tasks in quantum information processing do not even require full reconstruction of $\rho$, but rather the estimation of expectation values, fidelities, or entanglement measures.
In such cases, randomized measurements and the classical-shadow framework enable highly efficient inference from a restricted set of measurement outcomes~\cite{huang2020predicting,elben2022randomized} (cf. Sec.~\ref{ssec:randomized-measurements}).

\begin{figure*}[t]
    \centering
    \includegraphics[width=0.48\linewidth]{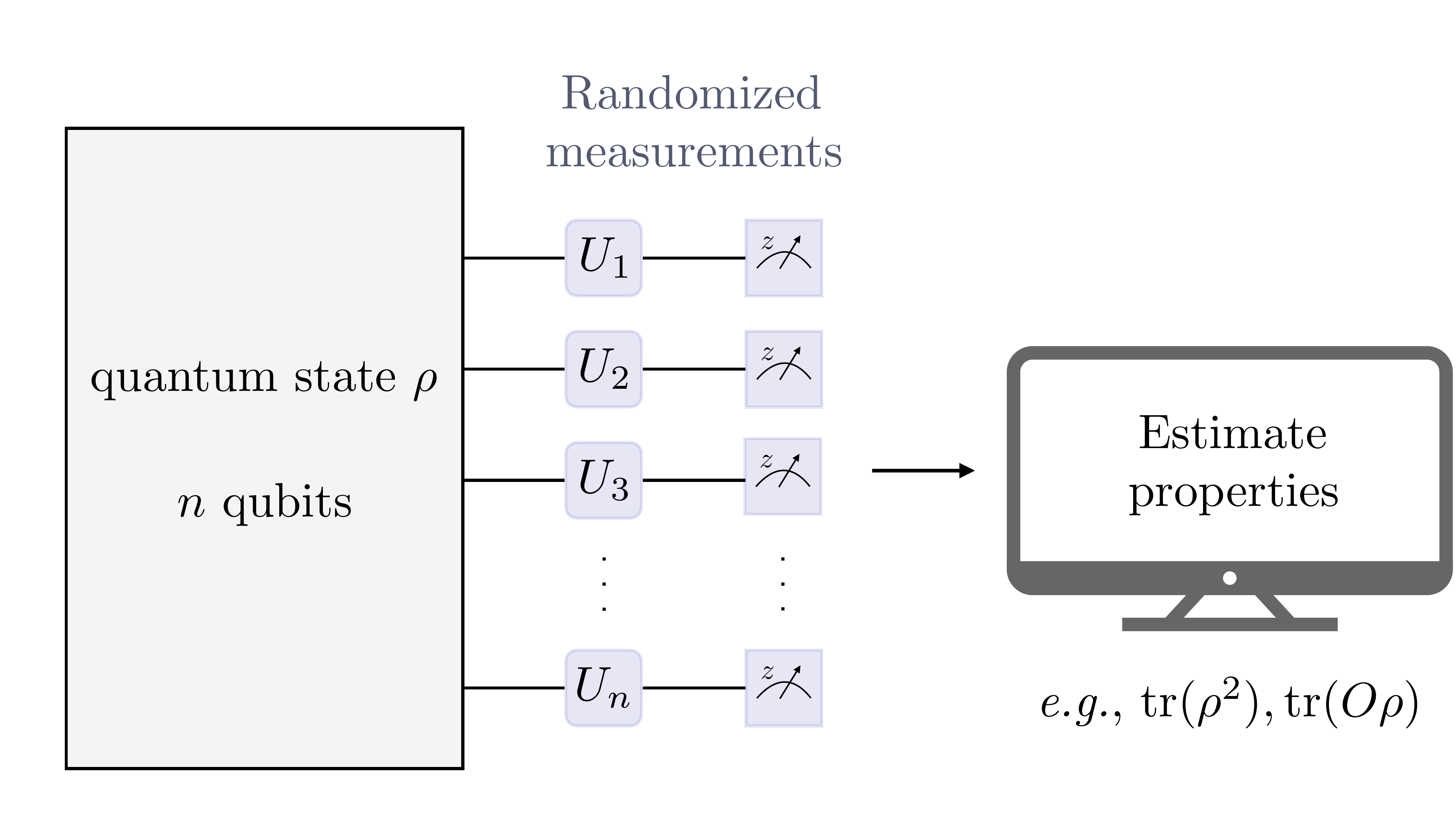}
    \includegraphics[width=0.48\linewidth]{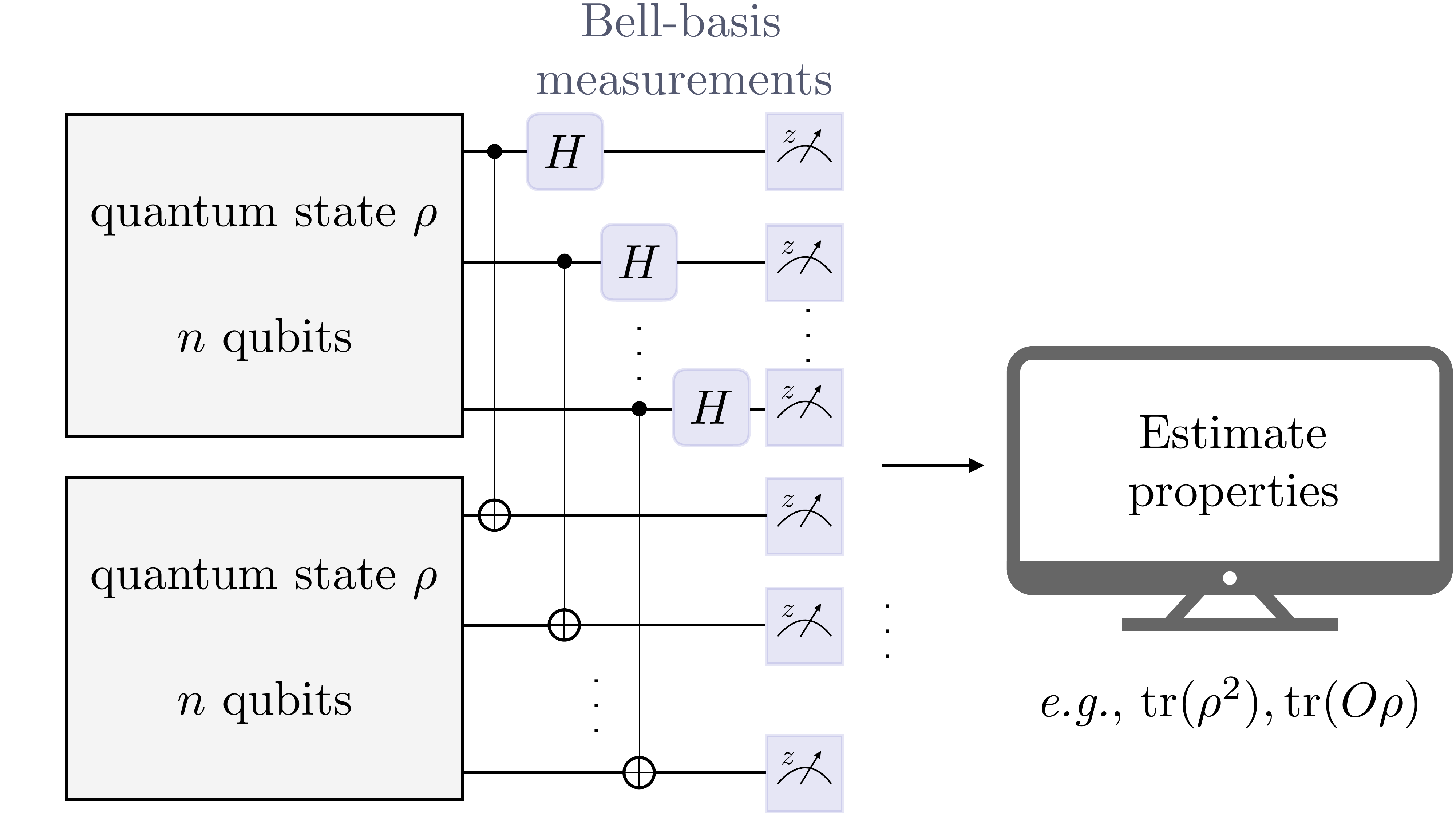}
    \caption{Procedures of (left) randomized measurements with local random unitaries $U_1, \ \ldots, U_n$ (adapted from Ref.~\cite{elben2022randomized}) and (right) Bell sampling, where two copies of $\rho$ are measured in the transversal Bell basis.
    Both approaches may be applied to the estimation of observables or state purity, but the hardware and sampling requirements differ in both cases as discussed in Secs.~\ref{ssec:randomized-measurements}, \ref{ssec:swap-test} and \ref{ssec:two-copies}.
    }
    \label{fig:randomized_measurements}
\end{figure*}

Often, tomographic reconstructions rely on repeated, independent measurements of single copies of a quantum state.
However, beyond single-copy strategies, entangled measurements can yield substantial improvements in the resource requirements of quantum state tomography~\cite{haah2016sampleoptimal,odonnell2016efficient}.
In fact, access to multiple identical copies of $\rho$ within the same coherent measurement setting benefits several quantum learning tasks~\cite{aharonov2022quantum}.
For example, coherent Bell-basis measurements on two copies of $\rho$ allow for efficient estimation of the absolute value of arbitrary Pauli-string expectation values~\cite{huang2020predicting} (cf.~Sec.~\ref{sssec:Bellsampling}), without the need to sample from exponentially  many measurement settings.
More generally, collective measurements on $K$ copies of a quantum state enable the estimation of nonlinear functions of $\rho$, such as $\tr(\rho^K)$, which are, \emph{e.g.}, relevant for entanglement quantification.
These learning strategies are representative of a broader paradigm in which information about quantum states is inferred not by reconstructing the state classically, but by performing joint measurements across multiple copies.
This will be discussed in more detail in Sec.~\ref{sec:algorithms-applications}.

\subsection{Classical shadows and randomized measurements \label{ssec:randomized-measurements}}

Our understanding of quantum mechanics dictates that it is too costly to fully characterize generic quantum systems with only classical data.
First and foremost, this can be seen as an invitation to build large-scale quantum machines to exploit all possible kinds of quantum advantage.
But it also seems to restrict what we can learn from a quantum experiment based on classical measurement data.

As discussed in Sec.~\ref{ssec:state_tomography}, full quantum state tomography becomes experimentally prohibitive already for systems composed of $\gtrsim 10$ qubits due to its exponential cost.
Fortunately, incomplete descriptions of quantum states are sufficient for many purposes~\cite{huang2020predicting,paini2021estimating}.
Such classical descriptions are known in the literature as \emph{classical shadows}; and earlier work introduced the term \emph{shadow tomography}, which refers to the task of predicting properties of a quantum state without fully characterizing it~\cite{aaronson2017shadow}.
Research on classical shadows has recently spurred several theoretical~\cite{chen2021robust,koh2022classical} and experimental~\cite{zhang2021experimental,struchalin2021experimental} developments, many of which resort to the framework of randomized measurements (RMs)~\cite{elben2022randomized}.
Classical shadows and RMs find applications in many areas of quantum information science and are useful for, \textit{e.g.}, detecting entanglement~\cite{elben2020mixedstate,vermersch2024manybody}, estimating the quantum Fisher information~\cite{rath2021quantum,yu2021experimental}, characterizing noise~\cite{stilckfranca2024efficient}, estimating fidelities~\cite{elben2020crossplatform,leone2023nonstabilizerness}, learning from experiments~\cite{huang2022quantum}, estimating gate-set properties~\cite{helsen2023shadow} and investigating measurement-induced phase transitions~\cite{ippoliti2024learnability}.

The philosophy behind this line of research has been succinctly expressed with the slogan: \emph{Measure first, ask questions later}.
It allows to construct a classical description of a quantum state using a number of measurements that is experimentally manageable and to decide later what this information should be used for.
To be more specific, consider a quantum state $\rho$ that is experimentally prepared and should be characterized.
For this, first a unitary operation $U$ is selected at random from a suitable ensemble of unitaries.
In Fig.~\ref{fig:randomized_measurements}(left), this unitary operation is given by a product of independent local random unitaries, \textit{i.e.}, $U = \bigotimes_{j=1}^N U_j$, that evenly cover the Bloch sphere.
Such random unitaries form \emph{unitary designs}~\cite{gross2007evenly,dankert2009exact}, and an example is the single-qubit Clifford group.
The unitary $U$ is applied to the state $\rho$ upon which it is measured in the computational basis $\{ \ket{\mathbf{s}} \}$.
The measurement outcomes, $\mathbf{s} \in \{0,1\}^N$, are recorded and the process is repeated $S$ times with a fixed unitary and for $M$ unitaries, requiring $M \cdot S$ experimental runs.

Next, the obtained classical data can be used to predict the expectation value of an observable $O$~\cite{elben2022randomized}. 
Denoting the $m^\text{th}$ unitary as $U^{(m)}$, and the $k^\text{th}$ sample as $\mathbf{s}^{(m,k)}$, an $m^\text{th}$ approximation $\hat{\rho}^{(m)}$ to the quantum state $\rho$ is given as
\begin{align}
    \hat{\rho}^{(m)}=\frac{1}{S} \sum_{k=1}^{S} \bigotimes^N_{j=1}\left(3(U_j^{(m)})^\dag \ket{s_j^{(m,k)}}\bra{s_j^{(m,k)}}U_j^{(m)}-\mathbb{I}\right),
\end{align}
and the collection $\{\hat{\rho}^{(m)}\}$ is referred to as a classical shadow of $\rho$.
The estimator $\hat{o}$ to the expectation value $o=\tr(O\rho)$ is then simply the average over $M$ independent results,
\begin{align}
    \hat{o} = \frac{1}{M} \sum_{m=1}^{M} \tr\big(O\hat{\rho}^{(m)}\big).
\end{align}
The power of classical shadows lies in the ability to estimate a large number of physical properties from the classical data.
An informal version of the theorem of classical shadows states that $M\propto\log(L)4^w/\varepsilon^2$ independent randomized measurements are sufficient to approximate a collection of $L$ expectation values of observables, each with a support bounded by $w$, to within an error $\varepsilon$ with high success probability.

Besides observables, classical shadows can be employed to estimate the purity $P_2=\tr(\rho^2)$~\cite{brydges2019probing,huang2020predicting} and nonlinear-properties of the form $\tr(O \rho^2)$~\cite{du2025optimal}. 
The estimator for the purity is given by
\begin{align}
    \hat{P}_2=\frac{1}{M(M-1)}\sum_{m\neq m'} \tr\big(\hat{\rho}^{(m)}\hat{\rho}^{(m')}\big).
\end{align}
Beyond the purity, this technique can be applied to general powers of density matrices $\tr(\rho^K)$~\cite{elben2020mixedstate,elben2019statistical,shin2025resourceefficient,zhou2024hybrid}.

Another application of randomized measurements is to estimate the inner product between two states $\rho$ and $\sigma$, which are prepared in separate experiments.
It is given by~\cite{elben2020crossplatform}
\begin{align}\label{eq:inner-product-weingarten}
    \tr(\rho \sigma) = 2^n \sum_{\mathbf{s},\mathbf{s}^\prime} (-2)^{-\mathcal{D}[\mathbf{s},\mathbf{s}^\prime]} \overline{P^{(1)}_U(\mathbf{s}) P^{(2)}_U(\mathbf{s}^\prime)},
\end{align}
where $\mathcal{D}[\mathbf{s},\mathbf{s}^\prime]$ denotes the Hamming distance between bitstrings $\mathbf{s}$ and $\mathbf{s}^\prime$, which are the measurement outcomes in the two separate experiments that are denoted by upper-case indices.
The notation $\overline{\cdots}$ refers to a statistical average over realizations of random unitaries $U$.
Together with a way to estimate purities, the inner product estimate yields the fidelity as defined in Eq.~\eqref{eq:fidelity-eq2}.

The corresponding sample complexity to estimate the fidelity for an $n$-qubit quantum state within additive error $\varepsilon$ is $O(2^n/\varepsilon^2)$. 
While this improves over full quantum tomography by a factor of $2^n$, it is not efficient. 
For certain practical applications where the support of the states concentrate on few low-weight Pauli operators, much fewer samples are needed; see also the discussion in Sec.~\ref{ssec:dist_inner_product}.

If a pure target state $\sigma = |\psi\rangle\langle\psi|$ is given, the fidelity between $\sigma$ and an experimental state $\rho$ can be estimated via importance sampling~\cite{flammia2011direct,barbera-rodriguez2025sampling}.
This technique, known as direct fidelity estimation, involves expanding both states in the Pauli basis $\rho=2^{-n/2}\sum_j a_j P_j$ and $\sigma=2^{-n/2}\sum_j b_j P_j$, such that the fidelity is given as:
\begin{align}
   \mathcal{F}(\rho, \sigma)=\tr(\rho \sigma)=\sum_j a_jb_j=\sum_j \frac{a_j}{b_j}b_j^2,
\end{align} 
Estimating $a_j$ empirically and the $b_j$ known, the probability distribution $b_j^2$ can be used with importance sampling to compute an estimate $\hat{\mathcal{F}}(\rho,\sigma)$ for the fidelity.
For some classes of states, this approach is resource-efficient~\cite{zhang2021direct,leone2023nonstabilizerness}.
More recently, it was shown that the fidelity between $\rho$ and any pure target state $\sigma$ can be certified efficiently with few single-qubit measurements~\cite{huang2024certifying,gupta2025few}.

\subsection{SWAP test, Bell measurements and permutation test \label{ssec:swap-test}}

The SWAP test is a primitive for comparing two quantum states.
In its original form, it involves an ancillary qubit and controlled-SWAP operations, which may be difficult to realize experimentally.
Alternatively, Bell-basis measurements may be used to accomplish the same task.

\begin{figure}[b]
    \centering
    \includegraphics[width=0.9\linewidth]{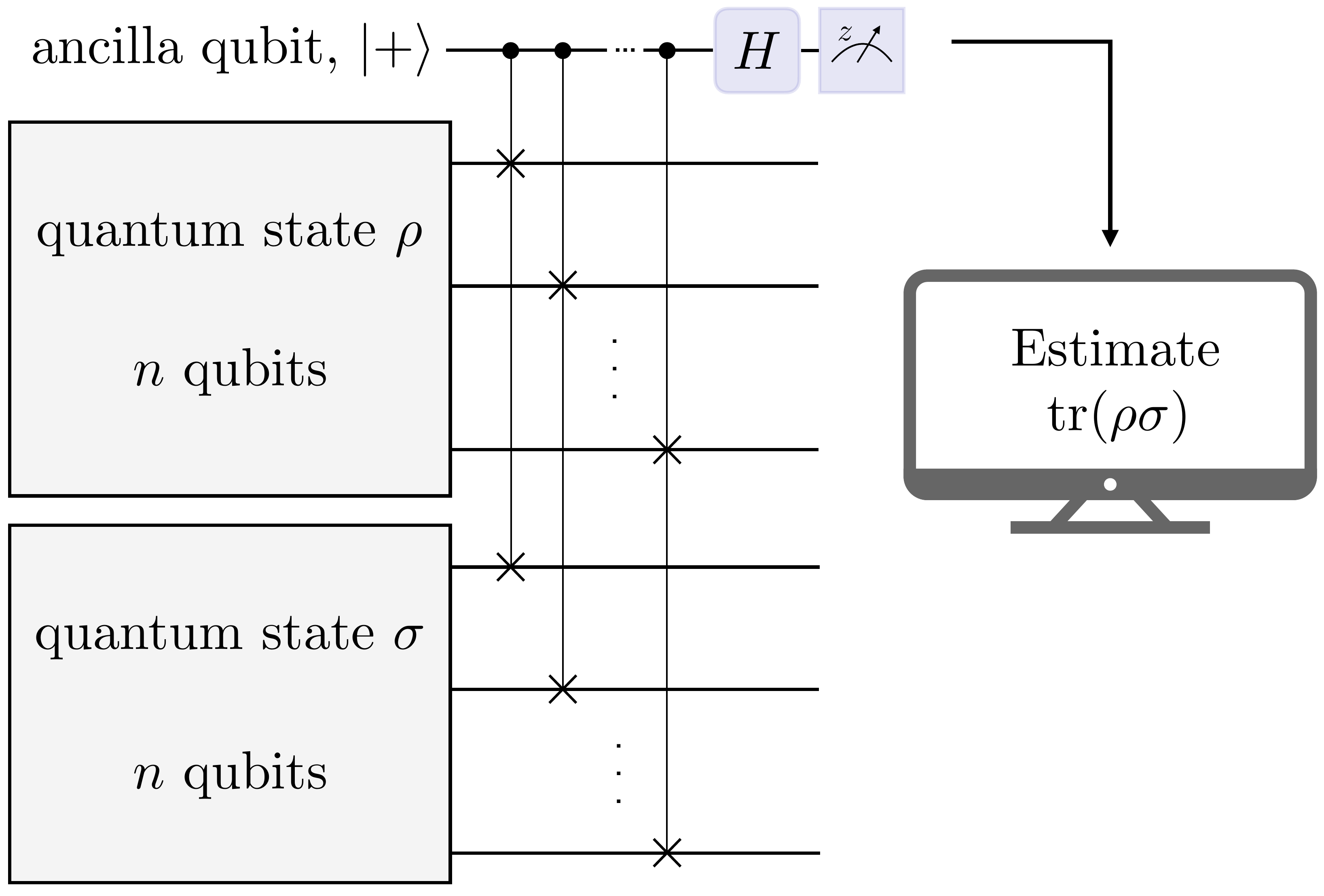}
    \caption{Circuit of the SWAP test which is commonly used as a subroutine in other algorithms and to estimate the trace overlap, $\tr(\rho \sigma)$, between two quantum states $\rho$ and $\sigma$.
    }
    \label{fig:swap_test}
\end{figure}

\subsubsection{SWAP test}
The SWAP test is a fundamental procedure~\cite{barenco1997stabilization,buhrman2001quantum} in quantum computation, commonly as a subroutine of several quantum algorithms and used for tasks like classification~\cite{rebentrost2014quantum}, clustering~\cite{wiebe2015quantum}, characterization of multipartite entanglement~\cite{beckey2021computable}, projective measurements~\cite{chabaud2018optimal} and generally in quantum machine learning~\cite{biamonte2017quantum, schuld2015introduction}.
The circuit implementing the SWAP test is shown in Fig.~\ref{fig:swap_test}.
Starting with an ancillary qubit initialized in $\ket{+}$ and two quantum states $\rho$ and $\sigma$, the SWAP test provides a way to efficiently estimate the inner product $\mathrm{tr}(\rho \sigma)$.
In case both states are pure, with $\rho = |\psi\rangle\langle\psi|$ and $\sigma = |\phi\rangle\langle\phi|$, controlled-SWAP gate acts on the total state of the system such that:
\begin{equation*}
    \ket{+}\ket{\psi}\ket{\phi} \rightarrow \frac{1}{\sqrt{2}}\ket{0} \ket{\psi}\ket{\phi} + \frac{1}{\sqrt{2}}\ket{1} \ket{\phi}\ket{\psi},
\end{equation*}
and after applying a Hadamard gate onto the ancilla qubit, the pre-measurement state is:
\begin{equation*}
    \frac{1}{2} \ket{0} \left ( \ket{\psi}\ket{\phi} + \ket{\phi}\ket{\psi} \right )
    + \frac{1}{2} \ket{1} \left ( \ket{\psi}\ket{\phi} - \ket{\phi}\ket{\psi} \right ).
\end{equation*}
Hence the ancilla qubit is finally measured in $0$ with probability
\begin{eqnarray}
    p_0 & = & \frac{1}{4} \braket{0|0} \left ( \bra{\psi}\bra{\phi} + \bra{\phi}\bra{\psi} \right ) \left ( \ket{\psi}\ket{\phi} + \ket{\phi}\ket{\psi} \right ) \nonumber \\
    & = & \frac{1+|\braket{\psi|\phi}|^2}{2}. \nonumber
\end{eqnarray}
By repeating the experiment, one may estimate the key component in fidelity estimations, $|\braket{\psi|\phi}|^2$ (or generally $\mathrm{tr}(\rho\sigma)$ if the states are mixed), up to an arbitrarily small additive error $\varepsilon$ using $k = O(1/\varepsilon^2)$ copies, which thereby achieves the optimal sample complexity for this task~\cite{anshu2022distributed}.
Moreover, note that a recent work has presented an optimal SWAP-based quantum algorithm for fidelity estimation between two quantum states, where one of them is pure, achieving $\varepsilon$-precision with $\Theta(1/\varepsilon)$ state preparation queries~\cite{fang2025optimal}. 
This technique can also be adapted to estimate the quantity $\sqrt{\mathrm{tr}(\rho\sigma^2)}$.


\subsubsection{Bell-basis measurements}\label{sssec:BBM}

The quantum circuit from Fig.~\ref{fig:swap_test} may be rewritten in a form that is more suitable for many implementations~\cite{garcia-escartin2013swap,cincio2018learning}.
It does not require an ancilla qubit, but instead resorts to Bell-basis measurements (BBM) between $\rho$ and $\sigma$, which is equivalent to the circuit depicted in Fig.~\ref{fig:randomized_measurements} (right).
There are two ways to demonstrate the equivalence between the SWAP test and BBM.

Firstly, the SWAP operator can be expressed in the Bell basis upon which it is easy to demonstrate how the inner product $\mathrm{tr}(\rho \sigma)$ can be obtained from post-processing BBM data~\cite{bandyopadhyay2023efficient}.
The SWAP operator generally takes the form:
\begin{equation}
    \mathbb{S} = \sum_{i,j} |i\rangle\langle j| \otimes |j\rangle\langle i|,
\end{equation}
and for qubits it can be rewritten in the form:
\begin{equation}\label{eq:swap-proj}
    \mathbb{S} = \sum_{i,j \in \{ 0, 1\}} (-1)^{i j} \Phi^{ij},
\end{equation}
where the four $\Phi^{ij}$ are the elements of the Bell basis,
\begin{equation}
    \mathcal{S} = \{ \ket{\Phi^+}\bra{\Phi^+},  \ket{\Phi^-}\bra{\Phi^-},  \ket{\Psi^+}\bra{\Psi^+},  \ket{\Psi^-}\bra{\Psi^-} \},\nonumber
\end{equation}
with the definitions of the Bell states as in Eq.~\eqref{eq:bell-phi-plus} and below, and we identify $\Phi^{00} = |\Phi^+\rangle\langle\Phi^+|$, $\Phi^{01} = |\Phi^-\rangle\langle\Phi^-|$, $\Phi^{10} = |\Psi^+\rangle\langle\Psi^+|$, $\Phi^{11} = |\Psi^-\rangle\langle\Psi^-|$.
The eigenspace spanned by the triplet (singlet) states is called the symmetric (antisymmetric) subspace.
Eq.~\eqref{eq:swap-proj} states that the SWAP operator can be expressed as the difference of the two projectors onto these subspaces.

The SWAP operator on $N$-qubit states can be written as:
\begin{eqnarray}\label{eq:swap-n-bell}
    & & \mathbb{S}^{(N)}  = \mathbb{S}^{\otimes N} \\
    & = & \left ( \sum_{k_1, l_1} (-1)^{k_1 l_1} \Phi^{k_1 l_1}_{1, 1} \right ) \otimes \cdots \otimes \left ( \sum_{k_N, l_N} (-1)^{k_N l_N} \Phi^{k_N l_N}_{N, N} \right ), \nonumber
\end{eqnarray}
where the subscript labels the qubits of the input states, \textit{e.g.}, the index \enquote{1,1} denotes a BBM between the first qubit in $\rho$ and the first qubit in $\sigma$.
More compactly, Eq.~\eqref{eq:swap-n-bell} can be written as:
\begin{equation}\label{eq:swap-n-bell-compact}
    \mathbb{S}^{(N)} = \sum_{\vec{k}, \vec{l} \in \{0, 1\}^N} (-1)^{\vec{k} \cdot \vec{l}} \Phi^{\vec{k}\vec{l}}
\end{equation}
with $\vec{k} = (k_1, \cdots, k_N)$, $\vec{l} = (l_1, \cdots, l_N)$ and:
\begin{equation}
    \Phi^{\vec{k}\vec{l}} = \Phi^{k_1 l_1}_{1,1} \otimes \cdots \otimes \Phi^{k_N l_N}_{N,N}.
\end{equation}
Next, we plug Eq.~\eqref{eq:swap-n-bell-compact} into the well-known identity:
\begin{equation}
\label{eq:swap-trick}
    \mathrm{tr}(\rho \sigma) = \mathrm{tr}\left ( \mathbb{S}^{(N)} (\rho \otimes \sigma) \right ),
\end{equation}
which yields:
\begin{equation}\label{eq:inner-product-bell}
    \mathrm{tr}(\rho \sigma) = \sum_{\vec{k}, \vec{l} \in \{ 0, 1 \}^N} (-1)^{\vec{k} \cdot \vec{l}}~ \mathrm{tr} \left ( \Phi^{\vec{k}\vec{l}} ( \rho \otimes \sigma ) \right ).
\end{equation}
Finally, Eq.~\eqref{eq:inner-product-bell} has a simple interpretation in terms of Bell measurements.
By performing BBM, $\Phi^{\vec{k}\vec{l}}$, on all qubit pairs, we obtain $2N$ bits stored in $\vec{k}$ and $\vec{l}$.
The post-processing of this data happens in the form of calculating $Z_t = (-1)^{\vec{k}\cdot\vec{l}}$, and repeating the experiment several times.
Then the final output $\bar Z = \tfrac{1}{T}\sum_{t=1}^T Z_t$ can be used to estimate $\mathrm{tr}(\rho\sigma)$.
Using the Hoeffding's inequality it follows that:
\begin{equation}
    \mathrm{Pr}\left (|\bar Z - \mathrm{tr}(\rho\sigma)| \leq \varepsilon \right ) \geq 1 - \delta,
\end{equation}
if the number of required samples, $T$, is of the order of $O(\frac{1}{\varepsilon^2}\ln{\frac{1}{\delta}})$.

\begin{figure}[b]
    \centering
    \includegraphics[width=\linewidth]{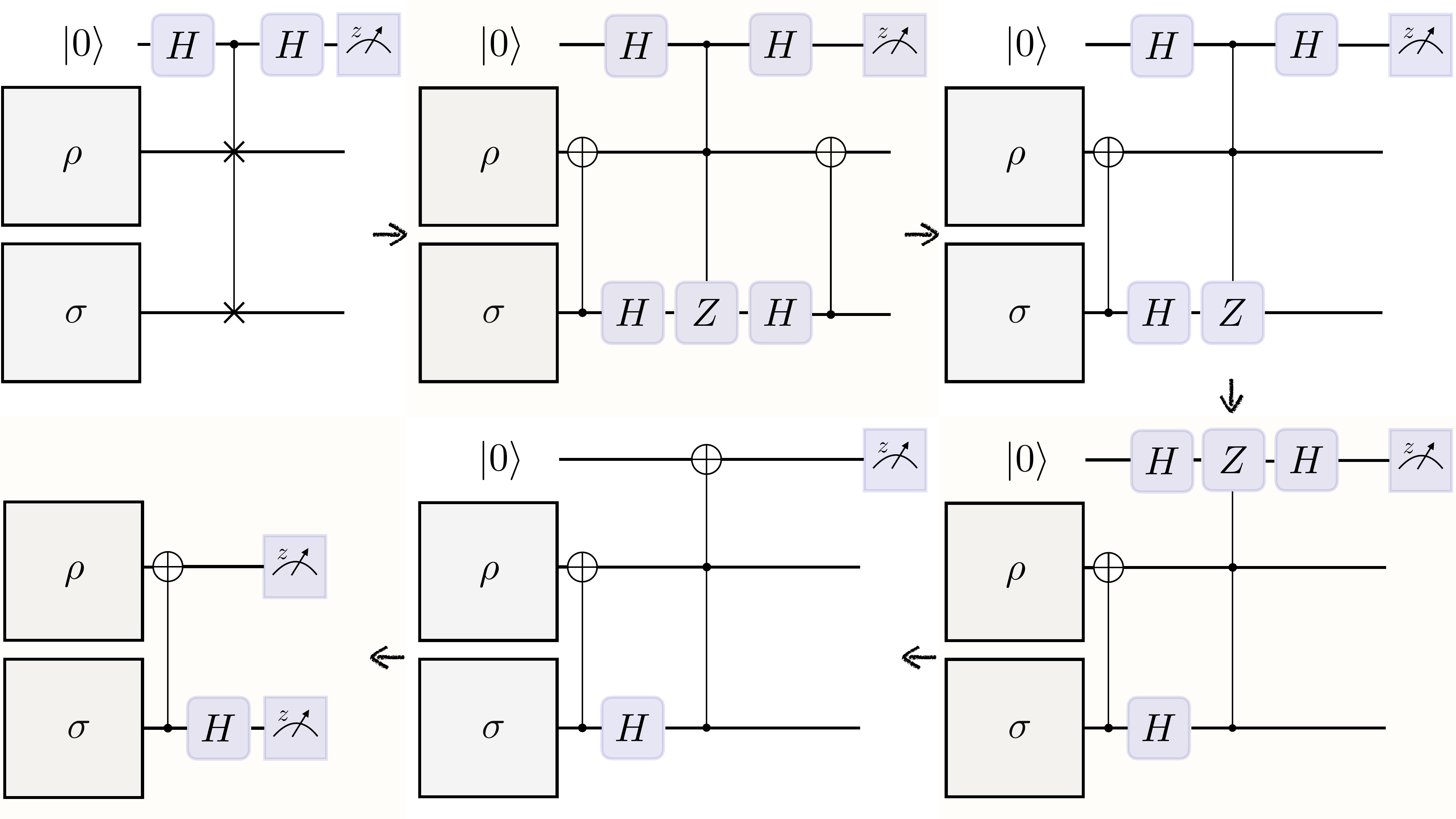}
    \caption{Starting from the original SWAP test (top left) which derives its name from the controlled-SWAP operation, circuit identities may be used to show that Bell-basis measurements (bottom left) may be equivalently employed to estimate inner products.
    This was originally discussed in Ref.~\cite{garcia-escartin2013swap}.
    }
    \label{fig:swap_bbm_equivalenfce}
\end{figure}

Another way to see the equivalence of the SWAP test and BBM is by using well-known quantum circuit identities and rewriting the circuit in Fig.~\ref{fig:swap_test}~\cite{garcia-escartin2013swap}.
The derivation is sketched in Fig.~\ref{fig:swap_bbm_equivalenfce}.

\subsubsection{Generalized SWAP tests}

There exist several measurement primitives that seek to take several copies of a quantum state as an input and project into the symmetric subspace between these copies~\cite{ekert2002direct,brun2004measuring,bacon2006efficient}.
In fact, this strategy was an early approach to dealing with noise before modern quantum error correction~\cite{berthiaume1994stabilisation,barenco1997stabilization,peres1999error}.
The SWAP test is merely a two-copy example of such a measurement primitive and can be used for inner product estimation~\cite{anshu2022distributed} or identity testing~\cite{kobayashi2001quantum,debeaudrap2004onequbit}.
Similar procedures have been used for measuring Rényi entanglement entropies and other polynomial functions of the density matrix~\cite{horodecki2002method,islam2015measuring,johri2017entanglement,subasi2019entanglementa,huggins2021virtual, yao2024nonlinear}.

\begin{figure}[b]
    \centering
    \includegraphics[width=0.99\linewidth]{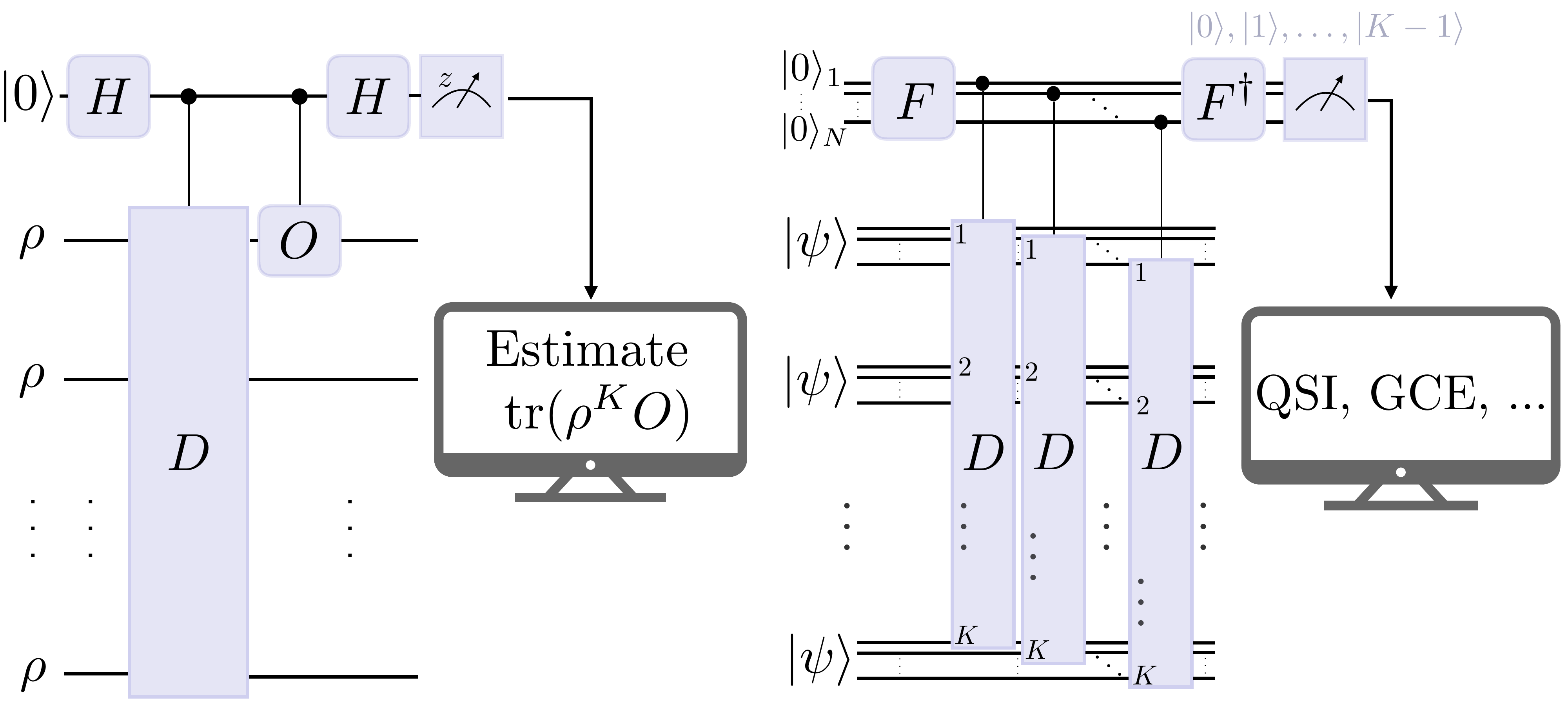}
    \caption{Circuits of the (left) circle and (right) permutation tests which are used for, \emph{e.g.}, quantum error mitigation, quantum state identity (QSI) testing and estimating generalized concentratable entanglement (GCE).
    }
    \label{fig:perm_test}
\end{figure}

In this spirit, the permutation test generalizes the SWAP test to multiple copies, as shown in Fig.~\ref{fig:perm_test} (\textit{right}).
It takes $K$ quantum states $\ket{\psi_1}, \cdots, \ket{\psi_K}$ as input, initializes a $d$-dimensional ancillary register in the state $\ket{0}$, and applies the quantum Fourier transform (QFT) $F_d$ over $d$ elements to it. 
Each level of the $d$-dimensional ancillary system coherently controls the target system $\ket{\psi}^{\otimes K}$ by applying a specific permutation to the $K$ quantum states, corresponding to a controlled-$D_K$ operation as depicted in Fig.~\ref{fig:perm_test} (\textit{right}).
Here, $2 \leq d \leq K!$, where $d$ denotes the number of different permutations applied across the $K$ quantum states. 
The Hadamard gate used in the SWAP test (cf.~Fig.~\ref{fig:swap_test}) can be seen as a special case of the QFT for $K=d=2$. 
Furthermore, different sets of permutations can be chosen depending on the application. 
For instance, cyclic (circle) and full $K!$ permutation tests are useful in quantum state identity testing~\cite{kada2008efficiency,buhrman2024permutation} and in the characterization of multipartite entanglement~\cite{liu2025generalized}.

\subsection{Quantum state purification} \label{ssec:purity_amp}

Any mixed state $\rho$ can be represented as the reduced state of a pure state $\ket{\psi}$ on a larger Hilbert space, $\emph{i.e.}$, $\rho = \tr_A(|\psi\rangle\langle\psi|)$ for some auxiliary system $A$.
This construction, though not unique, plays an important role in many protocols within quantum information processing.

Concurrently with early quantum error correction, work investigated the stabilization of noisy quantum computations via the preparation of approximately purified states~\cite{barenco1997stabilization,peres1999error}.
To overcome noise, several purification techniques, which generate high-purity qubits from low-purity qubits, have been proposed.
The optimal Cirac–Ekert–Macchiavello (CEM) purification procedure for a depolarized qubit is known~\cite{cirac1999optimal} but experimentally demanding~\cite{ricci2004experimental,hou2014experimental}.
Other works studied probabilistic qubit purification~\cite{keyl1999rate}, its optimization~\cite{fiurasek2004optimal}, and recently the purification of high-dimensional quantum states~\cite{li2024optimal}.

Quantum states may be recursively purified by using the SWAP test as a gadget of the protocol~\cite{childs2025streaming,grier2025streaming}.
Denoting the measurement outcome of the ancillary qubit in Fig.~\ref{fig:swap_test} with $a$, the post-selected, normalized state after the SWAP test can be written as
\begin{equation}
    \mathrm{SWAP}(\rho,\sigma) = \frac{1}{2} \frac{\rho + \sigma + (-1)^a (\rho \sigma + \sigma \rho)}{1 + (-1)^a \mathrm{tr} (\rho \sigma)}.
\end{equation}
Hence, post-selecting on $a=0$ and using two identical copies yields the state
\begin{equation}\label{eq:swap-purify}
    \mathrm{SWAP}(\rho,\rho)\vert_{a=0} = \frac{\rho + \rho^2}{1+\mathrm{tr}(\rho^2)}.
\end{equation}
This insight can be used to purify $\rho$.
For example, considering depolarizing noise and two copies of a state $\mathcal{E}(\rho, \lambda) = (1-\lambda)\rho + \frac{\lambda}{d} \mathbb{I}$,  with $\rho$ a pure state, the resulting state in Eq.~\eqref{eq:swap-purify} turns out as
\begin{equation}
    \mathrm{SWAP}(\mathcal{E}(\rho, \lambda),\mathcal{E}(\rho, \lambda))\vert_{a=0} = \mathcal{E}(\rho, \lambda^\prime),
\end{equation}
where
\begin{equation}
    \lambda^\prime = \frac{\lambda +\lambda^2/d}{2-2(1-1/d)\lambda+(1-1/d)\lambda^2} < \lambda,
\end{equation}
for $\lambda \in (0,1)$.
This analysis suggests a simple recursive purification algorithm based on the SWAP test, which is outlined in~\cite{childs2025streaming}.

In some settings, noisy states are shared between different parties and purification is realized using LOCC.
A well-known example is entanglement distillation, where the goal is to approach a Bell state with arbitrarily high fidelity (see Sec.~\ref{ssec:ent-dist}).
Recently, \cite{zhao2025power} showed that single-state two-to-one LOCC purification is possible when the target pure state is known in advance.
However, there is no non-trivial two-to-one LOCC purification protocol that improves fidelity for all pure states, all maximally entangled states, or all four Bell states.

\section{Algorithms and Applications \label{sec:algorithms-applications}}

This section explores how distributed quantum architectures enable new algorithmic capabilities, particularly when quantum information is shared or exchanged between spatially separated nodes.
We focus on tasks that benefit from entanglement, coordinated measurement, or access to multiple copies of quantum states.

Key applications include resource-efficient estimation of nonlinear observables (\emph{e.g.} the purity or fidelity), cross-platform state comparisons, and entanglement-assisted strategies for tomography and learning.
We also review complexity-theoretic results that clarify when distributed quantum setups provide computational advantages, and when they do not.
Our aim is to highlight how quantum networks not only extend the reach of quantum communication, but also serve as algorithmic platforms in their own right.

\subsection{Adaptations of known algorithms to distributed settings}\label{ssec:distributed-adaptations}

Quantum algorithms are used to solve problems in many areas, including cryptography, search, optimization, simulation of quantum systems, and linear algebra~\cite{montanaro2016quantum}.
Many of the foundational quantum algorithms were introduced in the 1990s and remain central to the field, including the Deutsch-Jozsa~\cite{deutsch1992rapid}, Bernstein-Vazirani~\cite{bernstein1997quantum}, Grover~\cite{grover1996fast}, Simon~\cite{simon1997power} and Shor's~\cite{shor1994algorithms} algorithms.
Since then, numerous refinements have been developed, which lend themselves more naturally to the capabilities and constraints of current and near-term quantum hardware.
In particular, recent work has extended many textbook quantum algorithms to distributed settings.

Shor's algorithm for factoring integers and computing discrete logarithms in polynomial time was one of the earliest to be considered in a distributed setting~\cite{yimsiriwattana2004distributed,meter2006architecture}.
The proposed approach partitions the quantum circuit across nodes using quantum teleportation.
This enables parallel execution of arithmetic subroutines, at the cost of $O(\log^2 N)$ communication overhead for factoring an integer~$N$.
A more recent distributed variant introduces a different model based on sequential execution across two quantum computers~\cite{xiao2022distributed}.
The resource costs of different distributed implementations of Shor's algorithm, in terms of qubit count and communication complexity, have been compared in~\cite{xiao2023distributed}.

The Deutsch-Jozsa algorithm, which determines whether a Boolean function is constant or balanced, has also been adapted to distributed architectures.
One approach is based on classical communication to collect and process the results of local operations across nodes~\cite{avron2021quantum}.
Another that relies on quantum communication, decomposes the global Boolean function into a set of local subfunctions, each of which is evaluated independently on a node~\cite{li2025distributed}.
The Bernstein-Vazirani algorithm, which identifies a hidden string with a single query, has also been tailored to networks~\cite{zhou2023distributed}.

Simon's algorithm, historically significant for demonstrating exponential quantum advantage, has seen multiple distributed adaptations.
One version executes independent instances of the algorithm across nodes without quantum communication between them~\cite{avron2021quantum}, while another has been tailored to multi-node networks using quantum teleportation~\cite{tan2022distributed}.

Grover's search algorithm has been distributed by splitting the search domain across quantum nodes and performing amplitude amplification in a synchronized fashion~\cite{qiu2022distributed}.
This preserves the quadratic speedup while distributing the computational workload and memory requirements.
Quantum sampling has been applied to distributed databases~\cite{chen2025optimal}.

\subsection{Two-copy measurements \label{ssec:two-copies}} 

\begin{figure}[tb]
\begin{overpic}[width=\linewidth]{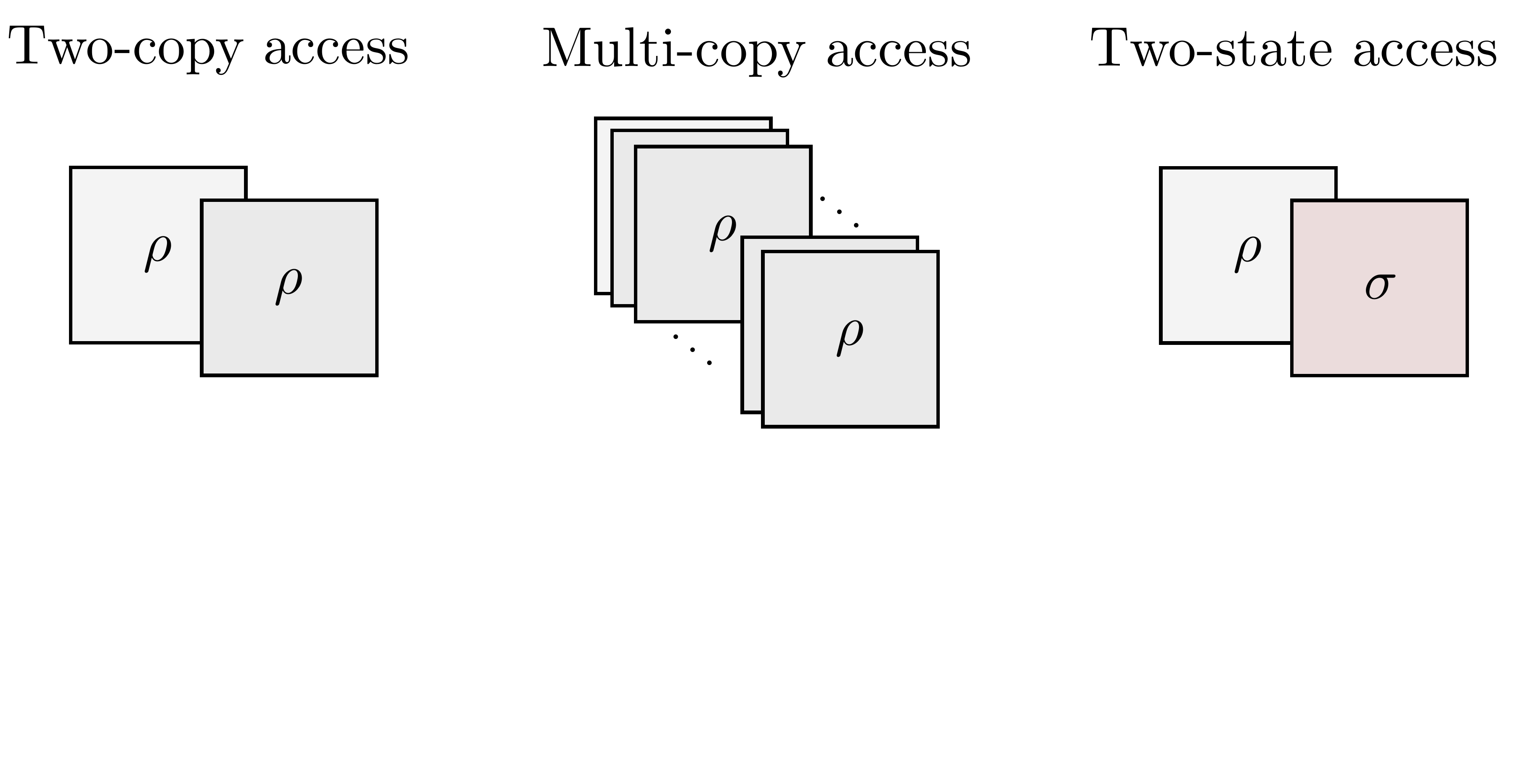}
\put(-40,16){\parbox{10cm}{$\bullet$ {\footnotesize Bell sampling [\ref{sssec:Bellsampling}]}}}

\put(-39.6,8){\parbox{10cm}{$\bullet$ {\footnotesize Entanglement [\ref{sssec:two-copy-ent}]}}}

\put(-6,16){\parbox{10cm}{$\bullet$ {\footnotesize Traces [\ref{sssec:multivariate-traces}]}}}

\put(-7,8){\parbox{10cm}{{\footnotesize and [\ref{sssec:multipartite-entanglement}]}}}

\put(25,16){\parbox{10cm}{$\bullet$ {\footnotesize DIPE [\ref{ssec:dist_inner_product}]}}}

\put(28.8,8){\parbox{10cm}{$\bullet$ {\footnotesize Conjugate [\ref{ssec:conjugate_states}]}}}

\put(25.5,0){\parbox{10cm}{$\bullet$ {\footnotesize Queries [\ref{ssec:memory-queries}]}}}

\end{overpic}
\caption{Overview of different scenarios underlying the distributed algorithms discussed in Sec.~\ref{sec:algorithms-applications}.
Applications within these settings are listed under each scenario.}
\label{fig:algorithms-overview}
\end{figure}

Coherent access to several copies of a quantum state can be highly advantageous for many applications in quantum information processing (see Fig.~\ref{fig:algorithms-overview} for a schematic overview of applications on the following pages).
While this may, in principle, be achieved in monolithic devices, it seems natural to consider this as an application of distributed architectures.

A relevant class of quantum algorithms concerns nonlinear observables that can be written as functionals which are quadratic in the state entries of one or two density matrices. 
By embedding the problem into a doubled Hilbert space, the problem can be written as a standard projective measurement of an observable $O$ on the joint two-copy system, $\tr[O(\rho \otimes \rho)]$.
This allows to efficiently perform Bell sampling (Sec.~\ref{sssec:Bellsampling}) or compute the state purity $\tr{(\rho^2)}$ (Sec.~\ref{sssec:two-copy-ent}), which may otherwise require less sample-efficient reconstruction methods such as full state tomography or the formula provided in Eq.~\eqref{eq:inner-product-weingarten}.
More generally, instead of two identical states, one can also consider two different quantum states $\rho$ and $\sigma$ to compute observables of the form $\tr[O(\rho \otimes \sigma)]$. 
Applications include the estimation of the inner product between both states (Sec.~\ref{ssec:dist_inner_product}), techniques for time-evolution (Sec.~\ref{ssec:Ham_evo}) and sample complexity gains when the latter state is the conjugate of the former (Sec.~\ref{ssec:conjugate_states}).

\subsubsection{Bell sampling} \label{sssec:Bellsampling}

In Section~\ref{sssec:BBM}, we have shown that transforming the SWAP test circuit into a Bell-basis measurement allows for estimating $\tr(\rho\sigma)$ without requiring an auxiliary qubit or a controlled-SWAP gate.
In fact, similar Bell-basis measurement circuits, such as those shown in Fig.~\ref{fig:randomized_measurements} and Fig.~\ref{fig:swap_bbm_equivalenfce}, can also be used to estimate $|\tr(P\rho)|$ for an $n$-qubit state $\rho$ and any $P \in \{I, X, Y, Z\}^{\otimes n}$, a technique known as Bell sampling~\cite{huang2021informationtheoretic}.
The key idea is that:
\begin{equation}
    \tr\left((P\otimes P) (\rho \otimes \rho)\right) = \left|\tr(P\rho)\right|^2,
\end{equation}
and for any $P = P_1 \otimes \cdots \otimes P_n$, where each $P_k \in \{I, X, Y, Z\}$, the Bell state is the eigenstate of the operator $P_k \otimes P_k$ with eigenvalues $\pm1$. Thus, Bell-basis measurements followed by classical postprocessing can be used to efficiently estimate $|\tr(P\rho)|$.
To estimate $|\tr(P\rho)|$ with $\varepsilon$ distance and with success probability at least $1 - \delta$, $\Theta(\log(1/\delta)/\varepsilon^4)$ pairs of $\rho$ are required. 
Note that determining the sign of $\tr(P\rho)$ requires majority voting based on projective measurements across extra copies of $\rho$, involving deeper coherent operations~\cite{huang2021informationtheoretic}.
Beyond that, Bell sampling is also useful for tasks such as learning stabilizer states~\cite{montanaro2017learning, allcock2024bell, leone2024learning}, estimating $T$-gate counts~\cite{leone2024learning}, preparing states and diagnosing circuits~\cite{hangleiter2024bell}.

\subsubsection{Entanglement measures}\label{sssec:two-copy-ent}

Entanglement measures are used to quantify the strength of quantum correlations between subsystems. 
Among these, entropy-based entanglement measures play central roles. 
The von Neumann entropy~\cite{vonneumann1996mathematische}, as a quantum generalization of Shannon entropy~\cite{shannon1948mathematical}, is one of the most widely used. 
However, it is often hard to compute for large quantum systems practically. 
To address this, alternative quantum entropies such as Rényi entropy~\cite{renyi1961measures} and Tsallis entropy~\cite{tsallis1988possible} are also used. 
These involve only traces of state powers, which are easier to estimate on quantum computers. 
Recent works have proposed estimating Rényi entropy using randomized measurements that require many single copies of the state~\cite{elben2018renyi, brydges2019probing}. 
Moreover, the second-order Rényi entropy, $S^{(2)}= - \log \tr(\rho^{2})$, can be computed from the state purity $\tr(\rho^2)$. 
Therefore, two-copy measurements have been shown to efficiently estimate this quantity~\cite{ekert2002direct, subasi2019entanglementa, johri2017entanglement, yirka2021qubitefficient}.
Two-copy measurements have also recently been applied beyond purity estimation, for example to efficiently detect non-Gaussianity in continuous-variable states via correlations generated by a beam splitter~\cite{hahn2025measuring}.
Similar methods have been used to analyze the entanglement properties of cold atoms in optical lattices~\cite{daley2012measuring,abanin2012measuring,pichler2016measurement}, and have been experimentally demonstrated~\cite{islam2015measuring,kaufman2016quantum,bluvstein2022quantum}.

Related methods allow generalization to other multipartite entanglement measures.
Particularly, a notable example is the family of multipartite entanglement measures known as concentratable entanglement~\cite{beckey2021computable,beckey2023multipartite}, which is able to recover several well-known entanglement measures mentioned above.
Concentratable entanglement for an $n$-qubit pure state $\ket{\psi}$, with labels $S=\{1,2,\cdots,n\}$, is defined as:
\begin{equation}
    \mathcal{C}^{(2)}_{\ket{\psi}}(s)\coloneq\frac{1}{2^{|s|}}\sum_{\alpha\in\mathcal{P}(s)} T_{\alpha}^{(2)}.
\end{equation}
Here, $T_{\alpha}^{(2)}$ denotes linear entropy (second-order Tsallis entropy) of the corresponding subsystem: $1-\tr(\rho_{\alpha}^2)$.
$s\in\mathcal{P}(S)\backslash\{\varnothing\}$ denotes the target subsystem of $S$ whose entanglement is being probed. 
$|s|$ is the cardinality of $s$ and $\rho_{\alpha}$ is the reduced density matrix corresponding to system $\alpha$ in the subset of $s$. 
Specifically, $\tr(\rho_{\varnothing}^2)=1$. 
Mathematically, the concentratable entanglement is defined as the average linear entropy over all subsystems of $s$. 
To compute this quantity in practice, one can utilize the parallelized SWAP test on two copies of the pure state $\ket{\psi}$, or parallelized Bell-basis measurements between pairs of qubits~\cite{beckey2023multipartite}. 
By executing the circuit sufficiently many times, one is able to easily compute $\tr(\rho_{\alpha}^2)$ for any subsystems $\alpha\in\mathcal{P}(s)$, thus computing concentratable entanglement from the probability distribution of the resulting bit strings on the auxiliary qubits.

\subsubsection{Distributed inner-product estimation} \label{ssec:dist_inner_product}

Two-copy measurements are provably more powerful than single-copy access, in some cases. 
Another application is distributed inner product estimation.
Here, the task is to estimate $\mathrm{tr}(\rho \sigma)$ for two experimentally available but unknown quantum states $\rho$ and $\sigma$, which are realized at physically separate locations.
This is a key step in estimating the fidelity between $\rho$ and $\sigma$ (cf.~Sec.~\ref{ssec:metrics}), which is useful for cross-platform verification (cf.~Sec.~\ref{ssec:verification}).
Three communication scenarios can be distinguished:
(\emph{a}) classical, (\emph{b}) unlimited quantum and (\emph{c}) limited quantum communication between the locations.

If only (\emph{a}) local operations and classical communication (LOCC) are allowed, it has been shown that, while much more efficient than full-state tomography~\cite{elben2020crossplatform}, the sampling complexity of this task is exponential in the system size across all communication and measurement settings, if arbitrary $\rho$ and $\sigma$ are considered.
It has been shown that $O(2^n)$ copies suffice for this task, relying on global random unitaries~\cite{anshu2022distributed}.
More experimentally feasible protocols relying on local random unitaries have recently been investigated~\cite{wu2025state,zheng2025distributed}.
If one does not consider generic states but instead focuses on quantum states with low magic and entanglement, the task may be efficiently solved via Pauli sampling~\cite{hinsche2024efficient}.
Provably efficient techniques also exist for quantum states with short-range correlations, based on learning their matrix-product operators~\cite{votto2025learning}.
Notably, complete information about one of the states makes this problem considerably more tractable~\cite{flammia2011direct}; see also Sec.~\ref{ssec:randomized-measurements}.

If (\emph{b}) unlimited quantum communication is allowed, the inner product can be efficiently estimated using a SWAP test or variants thereof~\cite{knorzer2023crossplatform}.
Experimentally, this is still challenging to realize for high-dimensional quantum states in most of the physical platforms.
Previously, the SWAP test was performed using photonic states~\cite{kang2019implementation,baldazzi2023swaptest,zhan2025experimental}, motional states of trapped ions~\cite{nguyen2021experimental} and also between three-qubit states in modular superconducting devices~\cite{dalton2025resourceefficient}.

If (\emph{c}) the amount of quantum information that can be transmitted is restricted, the presence of the quantum channel is still beneficial:
when both $\rho$ and $\sigma$ are $n$-qubit states and $q$ denotes the number of qubits that may be transmitted, the sampling complexity is $O(\sqrt{2^{n-q}})$~\cite{arunachalam2024distributed,gong2024sample}.
For $q=0$, this recovers the complexity within (\emph{a}) the LOCC setting and for $q=n$ the sampling complexity is $O(1)$, which can be achieved using (\emph{b}) the full SWAP test.

\subsubsection{Quantum simulation} \label{ssec:Ham_evo}

Time evolution under a Hamiltonian $H$, that is, the operation $e^{-iHt}$, is essential for simulating quantum many-body systems on quantum computers.
It appears in many algorithms, such as variational quantum algorithms~\cite{cerezo2021variational}, quantum phase estimation~\cite{kitaev1995quantum,nielsen2012quantum,gunther2025phase} and eigenstate preparation~\cite{schiffer2023quantum, liu2025preparing}.
For general $H$, exactly implementing $e^{-iHt}$ is hard, so it is usually approximated using Trotterization~\cite{suzuki1991general, berry2007efficient} or its variants~\cite{campbell2017shorter, campbell2019random, nakaji2024highorder, gunther2025phase}, which decomposes the time evolution into a sequence of simpler time evolutions. 
The cost depends on both the structure of $H$ and the evolution time $t$.

With distributed architectures, time evolution may be simulated more accurately and with lower gate complexity~\cite{feng2024distributed, huggins2021virtual}. 
In~\cite{huggins2021virtual}, the authors use QDrift~\cite{campbell2019random}, a randomized simulation method, together with two copies of the state and joint measurements, to simulate $e^{-iHt}\ket{\psi}$ more efficiently and accurately.
Recently, the authors of~\cite{feng2024distributed} introduced optimal communication-efficient protocols for distributed quantum simulation, enabling scalable many-body simulations across quantum networks.
Moreover, it is notable that if the goal is to compute $\bra{\psi}U_B^{\dagger} U_A\ket{\psi}$ but $U_B^{\dagger} U_A$ is too complex to implement, Alice and Bob can each apply local Hadamard tests to estimate $\bra{\psi}U_A\ket{\psi}$ and $\bra{\psi}U_B\ket{\psi}$. 
Then, by applying a SWAP test on their registers, they can compute $\bra{\psi}U_B^{\dagger} U_A\ket{\psi}$ from the measurement outcomes, combined with the earlier estimates. 
In particular, if $U_A = e^{-iHt}$ and $U_B = e^{iHt}$, this setup effectively computes the expectation value with doubled time evolution.
Note that this approach does not reduce the total number of gates to implement the longer time evolution, but results in a shorter circuit depth via parallelization.

In the context of quantum phase estimation~\cite{nielsen2012quantum}, several variants of the algorithm exist that rely on knowledge of the time series of the complex Loschmidt echo $\langle \psi|e^{-iHt}|\psi\rangle$~\cite{gardiner1997quantum, somma2019quantum, obrien2019quantum}. 
If the initial state $\ket{\psi}$ is prepared by a unitary circuit $U$, the projection can be realized by implementing $U^\dag$. 
In some applications, though, $U^\dag$ is not available, \emph{e.g.}, when $U$ describes (approximate) adiabatic state preparation on an analog quantum device, or when the state preparation is a dissipative process~\cite{mi2024stable, molpeceres2025quantuma}.
Then, the projection can be realized by a two-copy protocol~\cite{schiffer2025hardwareefficient}.

\subsubsection{Conjugate states} \label{ssec:conjugate_states}

Recent work has explored learning tasks that benefit from access to both an unknown quantum state $\rho$ and its conjugate $\rho^*$ instead of requiring access to $\rho^{\otimes K}$ for large $K$, thus providing a novel resource for specific learning tasks.
In Ref.~\cite{king2024exponential}, it has been shown that for a natural set of operators, measurements on $\rho \otimes \rho^*$ can provide exponential savings in sample complexity while using only a minimal quantum memory.
The setting considered there is a $d$-dimensional Hilbert space with discretized position and momentum operators and a natural limit of a continuous bosonic mode, and a learning task that is tailored to its bosonic displacement operator.
Since complex conjugation is not a physical operation, partly due to its basis dependence, it is generally difficult to produce $\rho^*$ from only copies of an unknown quantum state $\rho$.
Hence the authors of Ref.~\cite{king2024exponential} identify scenarios where experimental access to both $\rho$ and $\rho^*$ can naturally occur.
Besides applications in digital quantum simulation, they discuss how the learning task associated with bosonic displacement operators is directly connected to real-space arrays of quantum sensors.
In the context of quantum simulation, the task of measuring the distance of a state to an eigenstate of the Hamiltonian can also significantly benefit from access to a conjugate state~\cite{schiffer2022adiabatic}.

\subsection{Multi-copy measurements \label{ssec:multi-copy}}

Naturally, two-copy measurements can be extended to multi-copy measurements by applying coherent operations on multiple copies of a quantum state. 
In Section~\ref{sec:concepts}, we introduced several related fundamental concepts including generalized SWAP tests. 
These techniques are especially useful for estimating quantities such as $\tr(f(\rho))$, where $f(\rho)$ is a nonlinear function of a quantum state $\rho$.
Such functions are frequently used in quantum signal processing~\cite{Martyn2024parallel} and entanglement measures~\cite{beckey2021computable,liu2025generalized, johri2017entanglement, yirka2021qubitefficient}.
A key example is estimating $\tr(\rho^K)$, which underlies both polynomial functions of $\rho$ and Taylor approximations of non-polynomial functions. 
To estimate $\tr(f(\rho)) = \sum_{k=1}^{K} a_k \tr(\rho^k)$ within error $\varepsilon$, the generalized SWAP test requires $O(K^2/\varepsilon^2)$ samples of $\rho$~\cite{brun2004measuring}. 
With access to a purified oracle for $\rho$, this improves to $O(K/\varepsilon^2)$~\cite{Martyn2024parallel,wang2023quantum}.
Recently, similar scaling has been achieved without purified oracles using linear combination of unitaries, combined with parameterized quantum methods~\cite{yao2024nonlinear}.
Beyond function estimation, coherent operations on multiple copies of quantum states can also determine hidden subgroup structures of the states, including symmetry properties~\cite{hinsche2025abelian} and bipartition cuts~\cite{bouland2025state}.

\subsubsection{Quantum learning tasks}

Quantum learning theory is a recent and successful research program, which addresses tasks such as quantum state and process learning, or learning classical functions encoded as quantum states~\cite{anshu2023survey}.
Some problems, such as fully learning an unknown quantum state without coherent access to multiple state copies~\cite{kueng2014low,chen2024adaptive}, require a sample size that grows exponentially with system size, $N$.
When learning a quantum state, two ways of circumventing the need for an exponential sample size are the restriction to specific classes of states or the relaxation of the learning goal to certain observables~\cite{aaronson2007learnability}.
For example, if the unknown state is well approximated by a matrix-product state~\cite{cramer2010efficient} or known to be a stabilizer state~\cite{aaronson2004improved}, it can be efficiently identified.

Collective measurements on multiple copies can improve the sample efficiency, \textit{e.g}, via Bell-basis measurements across state copies~\cite{montanaro2017learning,gross2021schur,grewal2023improved}.
In fact, several learning tasks admit complexity-theoretic separations between single-copy strategies and multi-copy measurements~\cite{aharonov2022quantum}.
Even more intruigingly, recent works have established a fine-grained hierarchy of multi-copy advantage in quantum learning tasks~\cite{ye2025exponential,noller2025infinite}:
for example, estimating $\tr(O\rho^K)$ is exponentially hard for a large class of observables $O$ with access to only $K-1$ (or fewer) copies, while one additional copy reduces the complexity to a constant.

Frameworks that have been used to prove limits for sample complexity and time complexity of learning problems include probably approximately correct (PAC) learning, shadow tomography, and online learning~\cite{arunachalam2017guest}.
Shadow tomography of a $N$-qubit state $\rho$ with $M$ observables requires $\tilde \Omega(\min(M, 2^N))$ samples without quantum memory, whereas access to quantum memory enables more sample-efficient strategies, with exponential separations obtained for tasks such as purity testing and distinguishing dynamical processes~\cite{chen2022exponential}.
Entangled measurements are necessary to achieve optimal copy complexity in property testing, with unentangled strategies requiring asymptotically more samples~\cite{bubeck2020entanglement}. 
Access to purifications of low-rank mixed states further enables constant-sample algorithms for tasks such as purity estimation, virtual cooling, and quantum Fisher information estimation, while strategies without purification require exponentially many samples with bounded quantum memory~\cite{liu2024exponential}. 
A smooth tradeoff between entanglement and copy complexity in tomography has also been established, with collective measurements on batches of size $K$ reducing the sample complexity to $\tilde \Theta(d^3 / \sqrt{K}\epsilon^2)$, interpolating between the extremes of single-copy and fully entangled strategies~\cite{chen2024optimal}.
Similarly, in continuous-variable settings, estimating a random displacement channel exhibits an exponential separation in sample complexity between entanglement-assisted and entanglement-free schemes, with entanglement reducing the needed samples to be independent of mode number~\cite{oh2024entanglementenabled}.
These results provide a quantitative understanding of how coherent multi-copy access, quantum memory, and purification act as resources for quantum learning tasks.

From a broader perspective, quantum resource theories provide a unifying language for understanding such advantages. 
It has been shown that every quantum resource state provides an operational advantage, even in settings where the free set of states is non-convex, by linking generalized robustness to advantages in multi-copy and worst-case single-copy channel discrimination~\cite{kuroiwa2024every,kuroiwa2024robustness}. 
This highlights that separations between single-copy and multi-copy strategies are not isolated phenomena, but manifestations of a general principle: coherent access to quantum resources systematically enhances learning performance.

\subsubsection{Multivariate trace estimation \label{sssec:multivariate-traces}}

The multivariate trace denotes a specific nonlinear function of $K$ quantum states, $\tr(\rho_1\rho_2\cdots\rho_K)$, regardless of whether the quantum states are the same or different. 
For $K$ identical $\rho$, multivariate trace estimation boils down to estimating the quantity $\tr(\rho^K)$, which has been discussed previously, and for which many complexity-theoretic results exist~\cite{liu2024estimating,chen2025improved}. 
Multivariate trace estimation is widely used for fidelity estimation (cf. Section~\ref{ssec:two-copies}), quantum fingerprinting~\cite{buhrman2001quantum} and digital signatures~\cite{gottesman2001quantum}, quantum resource measurement~\cite{oszmaniec2024measuring}, linear independence of the state ensemble~\cite{oszmaniec2024measuring, buhrman2024permutation, kada2008efficiency}, among other applications. 

In this subsection, we focus primarily on non-identical states. 
In terms of distributed quantum computing architecture setup, a straightforward way of estimating multivariate traces is by utilizing the generalized SWAP test, favoring a generalization of the swap trick (Section~\ref{ssec:swap-test} and Eq.~\eqref{eq:swap-trick}) by replacing the swap operator $\mathbb{S}$ with a cyclic shift operator $\mathbb{D}$, as illustrated in~\cite{johri2017entanglement,ekert2002direct}. 
This method requires $O(K)$ circuit depth and $O(K)$ qubits.
Several improvements have been proposed to optimize circuit depth and reduce the number of required qubit registers in this approach. 
For qubit requirements,~\cite{yirka2021qubitefficient} reduced the number of required qubits to a constant using qubit resets. 
Regarding circuit depth,~\cite{subasi2019entanglementa} modified this method to achieve constant depth by replacing SWAP tests with Bell-basis measurements and doubling the number of prepared state copies per circuit execution. 
Additionally,~\cite{oszmaniec2024measuring} improved the linear dependence of circuit depth to a logarithmic one by decomposing the cyclic operators into $O(\log K)$ segments, each implemented using simultaneous SWAP operations.

Recently,~\cite{quek2024multivariate} developed an approach for multivariate trace estimation with constant circuit depth and $\lfloor{K/2}\rfloor$ auxiliary qubits, requiring only two-dimensional nearest-neighbor connectivity between registers, a typical feature in superconducting qubit quantum devices. 
First, the top $\lfloor{K/2}\rfloor$ auxiliary qubits are prepared in a $\lfloor{K/2}\rfloor$-party GHZ state with constant depth~\cite{quek2024multivariate,watts2019exponential} by connecting multiple Bell pairs with CNOT gates and performing measurement-conditional corrections. 
Second, since the cyclic shift $(1,2,\cdots,K)$ can be decomposed as
\begin{align}
    &\left\{\begin{array}{ll}
      \displaystyle \prod_{l=2}^{K/2} (l, K+2-l) \prod_{k=1}^{K/2} (k , K+1-k)  & : K \text{ even,} \\
      \displaystyle \prod_{l=2}^{\lceil K/2 \rceil} (l, K+2-l) \prod_{k=1}^{\lfloor K/2 \rfloor} (k , K+1-k) & : K \text{ odd,}\label{eq:decomposition}\\
\end{array} \right.
\end{align}
the cyclic shift can be divided into two independent parts, each of which can be implemented using controlled-SWAP operations acting on completely separate auxiliary qubits and target systems. 
In other words, each part can be executed within a single unit of circuit depth, ensuring the overall circuit depth remains constant. 
Finally, to estimate the multivariate trace, the auxiliary qubits must be repeatedly measured in either the $\ket{\pm}$ basis for the real part or the $\ket{\pm_Y}$ basis for the imaginary part. The expectation values of these measurements then serve as estimators for the multivariate trace. 
Notably, variants of this approach are also presented in~\cite{liang2023unified}, enabling a trade-off between circuit depth and the number of required qubit registers.

Multivariate traces are also known as Bargmann invariants~\cite{bargmann1964note} and can be used to study complex quantum theory and geometric phases.
Whether complex numbers are required in the quantum formalism or whether real numbers are sufficient has been heavily debated for many decades.
In a recent line of work, Bell-type experiments have been proposed~\cite{renou2021quantum} and realized~\cite{li2022testing} that show the necessity of a complex quantum theory.
In an alternative approach, quantum imaginarity can be tested by measuring multivariate traces~\cite{fernandes2024unitaryinvariant}, which can be estimated using cycle tests or several SWAP tests.

\subsubsection{Multipartite entanglement measures \label{sssec:multipartite-entanglement}}

Multipartite entanglement estimation is essential for certifying the quantumness of quantum systems and devices. 
However, as the size of the representation of quantum states scales exponentially with the number of their constituents, estimating entanglement in large-scale quantum devices becomes increasingly challenging. 
This necessitates the development of efficient methods for entanglement estimation.

In Section~\ref{sssec:two-copy-ent}, we discussed two-copy measurement approaches for estimating quantum Rényi entropy~\cite{ekert2002direct}. Similar to generalized SWAP tests, the two-copy measurement method can be extended to estimate Rényi entropies of any integer order by coherently placing multiple copies of the state and performing measurements~\cite{johri2017entanglement, subasi2019entanglementa, yirka2021qubitefficient}.
Moreover, the concentratable entanglement can also be generalized as $\mathcal{C}^{(K)}_{\ket{\psi}}(s)$ by replacing $T_{\alpha}^{(2)}$ with $T_{\alpha}^{(K)}$ for any $K>1$, named generalized concentratable entanglement~\cite{liu2025generalized}.
Here, $T_{\alpha}^{(K)}$ denotes quantum Tsallis entropy $\frac{1}{K-1}\left( 1-\tr(\rho_{\alpha}^K) \right)$~\cite{tsallis1988possible}.
It has been demonstrated that it remains a well-defined entanglement monotone and establishes a correspondence with quantum Tsallis entropies. 
Moreover, for $K\in\mathbb{Z}$, a parallelized cyclic permutation test applied across the distributed $K$ copies of state $\ket{\psi}$ is capable of efficiently estimating the generalized concentratable entanglement by postprocessing the probability distribution of the auxiliary qudit measurement results to compute $\tr(\rho_{\alpha}^K)$ for any $\alpha$.
It has further been shown that in the presence of noisy state copies $\ket{\psi}$, the estimation errors decrease as the number of noisy state copies $K$ increases, enhancing the robustness of this estimation method against errors.

\subsection{Memory-usage queries \label{ssec:memory-queries}}

Some distributed protocols treat quantum states as dynamic resources that generate operations, such as $\exp(-i \rho t)$.
These fit into a broader paradigm of circuits instructed by quantum states, enabling efficient, instruction-oblivious computation without reconstructing the state.

\subsubsection{Density matrix exponentiation}

Density matrix exponentiation (DME) takes as input the unknown density matrix $\rho$ and maps it onto the unitary operator $U(t) = \exp(-i \rho t)$ which is applied to a second density matrix, $\sigma$.
This mapping cannot be achieved perfectly with a single copy of $\rho$, as may be understood as an implication of the no-cloning theorem.
As shown in Ref.~\cite{lloyd2014quantum}, by using $n$ copies of $\rho$ one may implement it to desired precision with exponentially fewer copies than would be required via tomographically reconstructing $\rho$.
This protocol relies on the insight that
\begin{eqnarray}\label{eq:DME}
e^{-i\rho \delta} \sigma e^{i \rho \delta} & \approx &\mathrm{tr}_1 \left ( e^{-i\mathbb{S} \delta} \rho \otimes \sigma e^{i \mathbb{S} \delta} \right ) \\
& = &
\sigma \cos^2 \delta + \rho \sin^2 \delta - i \sin \delta \cos \delta \left [ \rho, \sigma \right ] \nonumber \\
& = &
\sigma - i \delta \left [ \rho, \sigma \right ] + O(\delta^2), \nonumber
\end{eqnarray}
where $\delta\mathbb{S} \equiv e^{-i\mathbb{S}\delta}$ is a partial SWAP operation.
By means of Trotterization, using $n$ copies of $\rho$ and partial SWAP operations $\exp(-i\mathbb{S}t/n)$, $U(t)$ can be applied to a state $\sigma$ up to precision of first order in $t^2/n$.
It has also been rigorously proved that the sample complexity of DME with $1-\varepsilon$ precision using multiple copies of $\rho$ is no larger than $4t^2/\varepsilon$~\cite{go2024density}.

Several algorithms may benefit from DME.
In quantum principal component analysis, it is used in combination with quantum phase estimation to extract the dominant eigenvalues and eigenvectors of a quantum state $\rho$~\cite{lloyd2014quantum}, and its exponential quantum advantage has been theoretically proven~\cite{huang2022quantum}.
Hamiltonian simulation is another application: for given $t$ and $\delta$, the minimum number of copies of $\rho$ necessary to implement $U(t)$ to a trace distance $\delta$, \textit{i.e.}, the sample complexity, can be optimized using DME~\cite{kimmel2017hamiltonian}.
A natural analogue of this approach has been shown capable of simulating Markovian dynamics under the Lindblad master equation using $O(t^2 / \varepsilon)$ samples of encoded states~\cite{patel2023wave}.
It can also be employed to access the spectral properties of a quantum many-body system~\cite{pichler2016measurement}, or as the means to efficiently exponentiate the kernel matrix to tackle machine-learning problems~\cite{rebentrost2014quantum}.

A two-qubit variant of DME has been experimentally demonstrated using superconducting circuits~\cite{kjaergaard2022demonstration}.
Using future implementations of DME, quantum algorithms and circuits may be encoded in quantum states, instead of compiling quantum circuits based on classical information.
Beyond the examples mentioned before, this approach paves the way for several promising applications in distributed quantum computing, such as quantum dynamic programming.

\subsubsection{Hermitian-preserving map exponentiation}

Hermitian-preserving map exponentiation can be regarded as a generalization of density-matrix exponentiation.
Beyond evolutions with the unitary $U_\mathrm{DME}=\exp(-i\rho t)$, it allows for evolutions of the form $U_\mathrm{HME} = \exp(-i\mathcal{N}(\rho) t)$ where $\mathcal{N}$ can be any Hermitian-preserving map~\cite{wei2024simulating}.
In comparison with the former, it replaces the SWAP operator $\mathbb{S}$ by the operator $N$, defined as the partial transpose of the Choi matrix of $\mathcal{N}$ (see Sec.~\ref{ssec:quantum-channels}).
In analogy with Eq.~\eqref{eq:DME}, the underlying trick used in this case is
\begin{equation}\label{eq:HME}
e^{-i\mathcal{N}(\rho) \delta} \sigma e^{i\mathcal{N}(\rho) \delta} \approx \mathrm{tr}_1 \left ( e^{-iN \delta} \rho \otimes \sigma e^{i N \delta} \right ),
\end{equation}
which can be trotterized in the same fashion as in the case of density-matrix exponentiation.

When combined with quantum phase estimation and Hadamard tests, Hermitian-preserving map exponentiation has several interesting applications, \emph{e.g.}, in the context of entanglement detection and quantification.
As a concrete example, negativities may be estimated by performing a controlled-$e^{-i\rho^{T_A} t}$ operation, where $\rho^{T_A}$ denotes the partial transpose of $\rho$ with respect to a subsystem $A$, and then running a Hadamard test, with which quantities like $\mathrm{tr}(\sin(\rho^{T_A} t))$ and $\mathrm{tr}(\cos(\rho^{T_A} t))$ may be estimated efficiently.
From these, the entanglement measure negativity may be estimated using Fourier analysis.

\subsubsection{Quantum dynamic programming}

Classical dynamic programming exploits a memory of intermediate results to solve recursive problems efficiently.
Quantum algorithms have traditionally lacked an analogous structure, relying instead on depth-intensive circuits that recompute subroutine outputs at every recursive level.
The proposal of quantum dynamic programming (QDP) fills this conceptual gap by introducing a memory-based scheme for quantum recursion that allows intermediate quantum states to be stored and coherently re-used~\cite{son2025quantum}.
A proof-of-concept experimental demonstration of a dynamic quantum algorithm was recently achieved with superconducting qubits for the thermodynamic task of cooling~\cite{alghadeer2025doublebracket}.

The core idea of QDP is to construct a sequence of quantum operations where each step of a recursive computation is implemented as a unitary transformation, and the resulting quantum state is retained in memory.
This stored state is then used as input to subsequent layers of the algorithm.
When applied appropriately, this paradigm yields an exponential reduction in circuit depth, since earlier computation steps no longer need to be repeated.
However, this gain in depth efficiency is traded against circuit width, as maintaining coherence over recursive branches requires additional qubits to store intermediate states.

In principle, the framework allows for a flexible interpolation between fully memoryless recursion and fully memory-based QDP.
With hybrid strategies, only a subset of recursive steps are stored, depending on the purity of the initial quantum states and the available hardware resources.
This tunable trade-off between depth and width makes QDP especially attractive in the context of near-term quantum devices, where circuit depth is often constrained by decoherence, but qubit count is becoming increasingly manageable.

\subsection{Complexity-theoretic results \label{ssec:complexity}}

Distributed quantum computing allows one to harness the capabilities of a collection of individual quantum processors.
As shown in this review, it offers multiple advantages when multiple copies of the same quantum state are available.
Beyond straightforward parallelization, where quantum states are processed in parallel, complexity-theoretic work has examined tasks that need to be solved jointly, without the players in the network necessarily having a complete view of the global system. 
In particular, such studies have explored how limitations on communication bandwidth, network topology or trust among participants can be efficiently obtained in a quantum setting.

A central question in this context is to what extent quantum communication can enhance the power of distributed algorithms~\cite{legall2025recent, arfaoui2014what, denchev2008distributed}.
An exponential separation between quantum and classical communication was shown in an early work by~\cite{raz1999exponential}.
Another branch of this research focuses on limited-trust scenarios, where some participants in the network may behave dishonestly. Early examples include quantum leader election~\cite{tani2012exact} and Byzantine agreement~\cite{ben-or2005fast}, with quantum money~\cite{aaronson2012quantum} serving as a notable illustrative application of cryptographic primitives.
For very recent studies, it has been demonstrated that a polynomial quantum advantage in tasks of both leader election and agreement can be achieved by applying quantum distributed algorithms~\cite{dufoulon2025quantum}.
Furthermore, for the tasks of routing information between two nodes in a network, there exists a quantum distributed algorithm that yields exponential quantum advantage over its classical counterparts~\cite{gall2025exponential}.

\subsubsection{Distributed quantum algorithms with limited communication}

One of the main avenues for studying the complexity of distributed quantum computing is to impose explicit limits on the amount of information that can be exchanged per round of communication. This setting captures realistic bandwidth constraints in quantum networks and leads to two widely studied abstract models of fault-free synchronous distributed computing: the CONGEST model, in which communication channels have limited bandwidth $O(\log n)$, with $n$ being the number of nodes in the network, and the LOCAL model, where bandwidth is unlimited and messages can be arbitrarily large~\cite{peleg2000distributed}.

Although early studies showed mostly negative results (both in quantum formulations of the LOCAL model~\cite{gavoille2009what} and in graph-theoretical problems pertaining to the CONGEST models~\cite{elkin2014can}), recent advances have shown how a quadratic speed-up exists in the CONGEST model to find the diameter of a network quadratically faster than any classical algorithm~\cite{legall2018sublineartime}, with an even larger quantum advantage with respect to the LOCAL model~\cite{legall2019quantum}. 
In the latter, two computational tasks are exhibited which require $\Omega(n)$ rounds in the classical (randomized) setting, with $n$ being the size of the network, whereas $O(1)$ rounds suffice in the quantum setting.

In the CONGEST model, the framework from~\cite{legall2018sublineartime} has been extended to weighted graphs~\cite{wu2022quantum} and the distributed setting framework has been generalized to any quantum query complexity~\cite{vanapeldoorn2022framework} which includes, for instance, quantum walks.
Further applications include determining the existence of a triangle subgraph in $\tilde O(n^{1/4})$~\cite{izumi2020quantum}, further improved to $\tilde O(n^{1/5})$~\cite{censor-hillel2022quantum}, both outperforming the best $\tilde O(n^{1/3})$ classical algorithm from~\cite{chang2019improved}.
This approach has been generalized to small-cycle~\cite{fraigniaud2024evencycle} and small-clique~\cite{censor-hillel2022quantum} finding.
If the CONGEST model, where messages can only be sent between adjacent nodes in each round, is extended to the CONGEST-CLIQUE model, which allows communication between any pair of nodes per round, the All-Pairs Shortest Path problem on weighted directed graphs can possibly be solved faster than with any classical algorithm ($\tilde O(n^{1/4})$ rounds~\cite{izumi2019quantum}, in contrast to $\tilde O(n^{1/3})$ classical rounds~\cite{censor-hillel2019algebraic}).

The quantum LOCAL model exhibits a vast separation between classical and quantum distributed algorithms~\cite{legall2019quantum}.
However, the problem used for this separation cannot be checked locally and $\Omega(n)$ rounds of communication are ultimately required.
Therefore, locally checkable problems (LCPs), for which output validity can be determined in a number of rounds that is independent of $n$, are of particular interest.
A central open question is whether there exists a quantum advantage in the LOCAL model for any LCP.
This question was considered in~\cite{arfaoui2014what} in the context of graph coloring (which can be checked by each node sending their color to its neighbors in constant time).
Unfortunately, recent results~\cite{akbari2025online, coiteux-roy2024no} are mostly negative, showing no advantage in networks of arbitrary topology. Nevertheless, finding a quantum advantage for LCP or proving that no quantum advantage exists remains open.

In the context of quantum verification, an all-powerful but untrusted party distributes a quantum state (also referred to as quantum proof in this setting) to the network elements, which also receive their respective inputs. The power of distributed quantum proofs investigated in~\cite{fraigniaud2021distributed} shows the existence of problems solvable by quantum protocols that require exponentially shorter quantum proofs compared to the classical setting. The potential and limitations of quantum proofs were further investigated in~\cite{hasegawa2024power}. The generation of quantum states using provers was investigated in~\cite{legall2023distributed} and the power of interactive quantum distributed proofs were considered in~\cite{legall2023distributeda}.

\subsubsection{Quantum consensus and Byzantine agreement}

As quantum networks approach maturity, an increasing number of networked tasks have emerged. Examples of such are quantum digital signatures~\cite{gottesman2001quantum, andersson2006experimentally, dunjko2014quantum}, quantum secret sharing~\cite{hillery1999quantum, gottesman2000theory}, secure anonymous protocols~\cite{hahn2020anonymous} and quantum Byzantine agreement~\cite{fitzi2001quantum, iblisdir2004byzantine, neigovzen2008multipartite, rahaman2015quantum, taherkhani2017resourceaware, sun2020multiparty}.

Experimental progress has closely followed theoretical developments, on quantum digital signatures
\cite{collins2014realization, clarke2012experimental, roberts2017experimental, pelet2022unconditionally, qin2022quantum, yin2023experimental}, quantum secret sharing~\cite{chen2005experimental, williams2019quantum}, secure anonymous protocols~\cite{huang2022experimental} and quantum Byzantine agreement~\cite{gaertner2008experimental, smania2016experimental, jing2024experimental}, with even the first demonstrations on quantum e-commerce having been recently demonstrated~\cite{cao2024experimental}.

Hence, achieving consensus in a quantum network becomes of paramount importance, not only in the adversarial setting~\cite{portmann2022security}, but also that of error tolerance in networks. Quantum Byzantine agreement leverages a quantum approach for handling the Byzantine generals problem, achieving consensus despite the presence of malicious players~\cite{ben-or2005fast}.

There are two fundamental differences between quantum and classical Byzantine agreement protocols.
The first one is the security loopholes of some public-key encryption methods used in the classical setting~\cite{rivest1978method, miller2016honey}, which can ultimately be broken with quantum computation~\cite{shor1994algorithms, fedorov2018quantum}. The second difference concerns the fault tolerance bound of the network. In the classical case, a minimum of $3f+1$ players is needed to tolerate $f$ malicious players~\cite{pease1980reaching, dolev1986possibility}.
In consequence, one cannot obtain classical Byzantine agreement in a three-partite setting. In~\cite{fitzi2001quantum} the first quantum solution to the three-party consensus was proposed. Experiments followed using entangled photons~\cite{gaertner2008experimental} and qudits~\cite{iblisdir2004byzantine, neigovzen2008multipartite, rahaman2015quantum, smania2016experimental}. In~\cite{fitzi2002generalized} an important impossibility theorem incorporated pairwise authenticated classical communication between the nodes, which restricted the existence of broadcast protocols to $f<n/3$.
However, in the past two decades, quantum-aided 'weak broadcast' protocols have been developed, building upon distributed entangled quantum states~\cite{fitzi2002unconditional}, surpassing the above mentioned classical impossibility theorem, thus offering a higher resilience, where $f<n/2$.
Further research, such as resource analysis for four-qubit singlet states followed~\cite{guba2024resource}, showcasing the suitability of near-term devices for the study of quantum networks.

The tasks concerning a multiparty evaluation of a classical function can be generalized to the quantum case~\cite{broadbent2010can}, where the inputs and outputs are quantum~\cite{crepeau2002secure, ben-or2006secure, dupuis2012actively, dulek2020secure, lipinska2020secure, alon2021efficient}, where the relation between inputs and outputs is most generally described by a CPTP map.
The framework in which one analyzes classical multiparty computation is so-called composable~\cite{cramer2015secure} but in the quantum case few works seem to maintain the composability property~\cite{ben-or2006secure}. There, it is proven that, assuming pairwise quantum channels and a classical broadcast channel between $n$ players, there exists a universally composable, statistically secure multiparty quantum computation protocol, that tolerates an adaptive adversary controlling up to $t<n/2$ faulty players.
Furthermore, the protocol has polynomial complexity in $n$ and the size of the circuit.

\subsubsection{Quantum money and quantum blockchain}
Two exemplary applications of quantum information processing in quantum networks are those of quantum money and proposals for a quantum blockchain.

In 1970 (only published in 1983), Wiesner proposed a scheme for quantum money that exploited the no-cloning theorem to achieve cryptographic tasks that would be otherwise impossible~\cite{wiesner1983conjugate}. There, an issuing bank would encode a serial number $s$ into a quantum state $\ket{\psi_s}$ consisting of $n$ qubits in a product state, each of them from the set $\{\ket{0}, \ket{1}, \ket{+}, \ket{-}\}$. The bank would maintain a classical description of $\ket{\psi_s}$ in a large database and the banknote could be verified in the bank by measuring each qubit in the appropriate $\sigma_z$ or $\sigma_x$ basis. This first scheme, though having some significant drawbacks (only the bank could verify, it is susceptible to online attacks~\cite{lutomirski2010online, aaronson2009quantum}, and requires a giant database), has been refined through the years~\cite{bennett1983quantuma, bennett1983quantum} though it lay dormant for more than two decades. After that, a series of works offered partial solutions to the verifiability problem:~\cite{mosca2009quantum} proposed to use blind quantum computing to delegate the verification to local merchants, while~\cite{gavinsky2012quantum} and subsequently~\cite{molina2013optimal, pastawski2012unforgeable} proposed variants that would only require classical communication between the merchant and the bank. Building on hidden subspaces,~\cite{aaronson2012quantum} proposed the first quantum money scheme that was public-key, allowing for anyone to verify a banknote as genuine, and cryptographically secure under well-established classical hardness assumptions.

Recent theoretical developments have provided new constructions and frameworks. For instance, how to make the Aaronson-Christiano-style public-key money robust to noise via a quantum error correction viewpoint~\cite{yuen2025noisetolerant}. Algorithms and verification refinements for group-action money, also proving security in the generic group-action model~\cite{doliskani2025publickey}, isogeny-based~\cite{zhandry2024quantum} and unifying group-action and non-collapsing-hash approaches~\cite{mutreja2025quantum} have been recently developed. Anonymous public-key quantum money, with and without traceability,~\cite{cakan2024anonymous} and public-key quantum money with a classical bank in a semi-quantum approach~\cite{shmueli2022publickey} are but some of the latest developments.

At the same time, recent work has also exposed limitations and vulnerabilities of existing proposals~\cite{liu2023another, bilyk2023cryptanalysis, ananth2023implausibility, kumar2019practically}; see also~\cite{sattath2023uncloneable} for a survey of unclonable primitives. 

Work on quantum blockchain is naturally influenced by developments in quantum money~\cite{zhandry2021quantum}. Quantum blockchain combines quantum-secured ledgers and intrinsically quantum encodings.
On the secured side, the use of metropolitan-fiber quantum key distribution schemes for information-theoretic authentication was demonstrated~\cite{kiktenko2018quantumsecured}.
On the encodings side, the use of temporal GHZ states has been considered for encoding the ledger directly using entanglement, as a chain in time~\cite{rajan2019quantum}, as well as using weighted hypergraph states~\cite{banerjee2020quantum, orts2023improving}.
Most recently, a proof-of-quantum-work prototype architectures have been proposed, where mining resources require a quantum computer~\cite{amin2025blockchain}, running consensus across four geographically distributed annealers. The increasing number of recent surveys highlight a growing interest in quantum blockchain technology~\cite{parida2023postquantum, gandhiS2024quantum, ghosh2025quantum}.

\subsection{Certification \& verification \label{ssec:verification}}

In addition to technological challenges, quantum information processing is confronted with at least two major conceptual questions:
First, how can quantum information be protected against environmental noise and unwanted errors, and second, how can one possibly verify the computational output of a $\text{BQP}$ machine without taking its reliability for granted?
On the theoretical level, the first question is addressed using quantum error correction~\cite{roffe2019quantum} which has seen fast-paced advances~\cite{campbell2024series} in recent years.
As discussed in Sec.~\ref{ssec:qem-qec}, distributed quantum error correction architectures are under active investigation.

The second major challenge is concerned with the verification, or certification, of the computational output of a quantum machine.
In a common scenario, verification protocols involve two parties, a \emph{verifier} and a \emph{prover}.
The verifier wants to verify the output that a prover provides~\cite{gheorghiu2019verification}.
A standard assumption of verification protocols is that the verifier is not herself in the possession of a mature quantum computer, and is thus confronted with the challenge of assessing the validity of a computation performed by a more powerful prover.
In some scenarios the verifier has limited quantum power, and in some they are assumed to possess only classical capabilities.

Classical computation can be verified by making predictions:
the result of a numerical calculation can be cross-checked on another computer.
By construction, this is not possible in the quantum setting if the verifier has access only to a classical computer and limited quantum resources.
Yet there exist many approaches to quantum verification~\cite{eisert2020quantum}, from commonly employed diagnostic methods like randomized benchmarking~\cite{knill2008randomized}, gate set tomography~\cite{nielsen2021gate}, direct fidelity estimation~\cite{flammia2011direct}, cross-platform verification~\cite{elben2020crossplatform} or compressed-sensing tomography~\cite{gross2010quantum}, to provably secure cryptographic techniques such as self-testing~\cite{supic2020selftesting}, verifiable blind quantum computing~\cite{fitzsimons2016private} or interactive proofs~\cite{aharonov2017interactive}.
In the following, we review various certification methods which are particularly relevant in modular architectures.

\subsubsection{Verifiable blind quantum computing}\label{sssec:bqc}

Blind quantum computing (BQC) addresses a key privacy aspect of commercial quantum devices:
ensuring that a potentially untrusted quantum server learns nothing about the computation it performs, while still allowing the client to check the correctness of the result~\cite{fitzsimons2016private,gheorghiu2015robustness}.
On grounds of the no-cloning theorem, BQC offers information-theoretic guarantees that allow a client with limited quantum capabilities to delegate quantum computations to powerful servers without revealing the delegated circuit or result of the computation~\cite{childs2005secure,arrighi2006blind}.
In the universal blind quantum computing protocol~\cite{broadbent2009universal,giovannetti2013efficient}, the verifier prepares single-qubit states chosen from a small discrete set and sends them to the prover, who embeds them into a larger entangled resource state and performs measurement-based quantum computation (MBQC)~\cite{raussendorf2001oneway}.
By applying secret, verifier-chosen rotations to the input states and adapting the measurement angles during the protocol, the computation is effectively hidden from the prover.

The fact that a verifier with limited or no quantum capabilities can verify the results of a quantum prover at all is made possible by the power of interaction~\cite{goldwasser1985knowledge,shamir1992ip}:
%
%
Basically, the verifier may \emph{repeatedly ask clever questions} to check the prover's answers for consistency~\cite{mahadev2018classical}.
Interactive verification protocols provide the verifier with a recipe for how to achieve this~\cite{aharonov2017interactive}.
A major axis of classification for interactive protocols concerns the quantum power of the verifier.
In several schemes, the verifier is assumed to possess a small quantum device capable of preparing simple quantum states or performing measurements.
These protocols fall into two primary families~\cite{gheorghiu2019verification}:
(i) prepare-and-send protocols and (ii) receive-and-measure protocols.
In prepare-and send protocols, the verifier prepares and sends simple quantum states, often single qubits, to the prover, who performs the computation~\cite{broadbent2018how}.
Hidden traps in the input allow the verifier to catch errors in a delegated computation within the universal BQC framework~\cite{fitzsimons2017unconditionally}.
In receive-and-measure protocols, the roles are reversed:
the prover sends quantum states, which need not necessarily be encrypted, and the verifier checks them by measurement.
Instead of relying on hidden traps, verifiability can be achieved if the client checks whether the state provided by the server is correct by testing stabilizers~\cite{hayashi2015verifiable}.

To eliminate the need for any quantum capability on the verifier's side, other protocols turn to either multiple entangled provers or cryptographic tools.
In the entanglement-based setting, a classical verifier interacts with two or more non-communicating quantum provers and uses non-local games, such as CHSH tests, to certify correct behavior~\cite{reichardt2013classical,coladangelo2017verifieronaleash}.
Underlying this scenario is a rigidity property of non-local quantum games:
if two entangled provers can pass a fraction $\omega^*-\varepsilon$ of the tests, where $\omega^* \approx 0.85$ denotes Tsirelson's bound, they must share an entangled state within distance $O(\sqrt{\varepsilon})$ of a Bell state, whose qubits are measured by pairs of anti-commuting observables.
Other protocols require multiple entangled provers~\cite{mckague2016interactive,natarajan2017quantum,fitzsimons2018post}.
Alternatively, cryptographic protocols like Mahadev's~\cite{mahadev2018classical} use classical encryption schemes to constrain a single quantum prover.

\subsubsection{Self-testing}

Self-testing comprises a set of methods for determining the underlying physics of a quantum experiment solely from measurement statistics.
Usually, it is aimed at determining whether some experimentally prepared unknown quantum state~\cite{coladangelo2017all,tavakoli2018selftesting} or observable~\cite{bowles2018selftesting} is close to a target state or observable.
An important goal of self-testing is to characterize the given state or observable with minimal assumptions.
We provide a brief review of self-testing results, as they fit into the topic of this article, and refer to~\cite{supic2020selftesting} for an in-depth account on the topic.

A central application is the device-independent certification of entangled states~\cite{bowles2018deviceindependent}.
In some black-box scenarios, the only accessible data is the observed violation $S$ of a Bell inequality, most prominently the Clauser–Horne–Shimony–Holt (CHSH) inequality~\eqref{eq:chsh}.
If $S$ approaches the maximal quantum value $S_\mathrm{max}=2\sqrt{2}$, the rigidity of the CHSH game ensures that the devices must share a state close to a maximally entangled pair, independent of their internal implementation~\cite{mayers2004self}.
Quantitative bounds relate $S$ to the distance between the actual and ideal Bell state~\cite{bardyn2009deviceindependent,mckague2012robust,yang2013robust}.
Different notions of closeness between states have been explored in this context, such as commutation-based measures~\cite{kaniewski2017selftesting,kaniewski2019maximal}, trace distance or fidelity, or usefulness for teleportation~\cite{cavalcanti2017all}.

Self-testing protocols must take into account experimental noise and statistical fluctuations due to finite sample sizes.
While the latter can be addressed using large-deviation bounds~\cite{yu2022statistical} and other results from statistical analysis~\cite{lin2018deviceindependent}, experimental imperfections are considered within the framework of \emph{robust} self-testing.
If noise is sufficiently low or the system admits exact solutions, analytical methods based on, \textit{e.g.}, Jordan's lemma~\cite{bardyn2009deviceindependent}, vector-norm or operator inequalities~\cite{mckague2014selftesting,bamps2015sumofsquares,kaniewski2016analytic} may allow statements about noise robustness.
Yet many results in the literature rely on numerical techniques based on the swap method~\cite{yang2014robust}, which provides a constructive way to prove self-testing results.
It introduces a local isometry that acts between each black-box device and a trusted ancilla, effectively swapping out the ideal target state into the ancillary registers while relegating all imperfections to a junk subsystem.
This allows one to certify that the observed correlations imply the existence of a state close to the ideal one, even without assumptions on the devices’ internal workings.

The simplest and historically most studied setting for self-testing is that of bipartite entangled states.
Achieving maximal violation of the CHSH inequality certifies (up to local isometries) the presence of a maximally entangled pair of qubits.
This seminal observation laid the foundation for the field of self-testing~\cite{mayers2004self}.
Subsequent works have generalized this paradigm to certify partially entangled two-qubit states~\cite{yang2013robust,coladangelo2017all}, higher-dimensional maximally entangled states~\cite{coladangelo2017all}, and states useful for teleportation~\cite{cavalcanti2017all}.
The key idea behind these results is to identify Bell inequalities or nonlocal games whose optimal quantum strategy uniquely characterizes a desired target state.
For instance, the tilted CHSH inequalities~\cite{coopmans2019robust} can be used to self-test the family of pure entangled two-qubit states.

While all bipartite pure states can be self-tested~\cite{coladangelo2017all}, only partial results exist for multipartite states. 
From a practical point of view, self-testing becomes more challenging in the case of multiple parties as it may require space-like separation between multiple devices~\cite{supic2020selftesting}.
The first result on self-testing for multipartite entangled state was derived for all graph states~\cite{mckague2014selftesting}
\begin{equation}
    \ket{G} = \Pi_{(i,j)\in E} \mathrm{CZ}_{i,j} \ket{+}^{\otimes N},
\end{equation}
where $\mathrm{CZ}_{i,j}$ is the controlled-$Z$ gate between qubits $i$ and $j$ and $E$ denotes a set of edges of a given graph $G$.
Other multipartite states that can be self-tested include the $N$-partite $W$~\cite{wu2016selftesting} and Dicke~\cite{fadel2017selftesting,supic2018selftesting} states.
Recently, a self-testing framework for any pure entangled state of an arbitrary number of subsystems has been developed for quantum networks~\cite{supic2023quantum}.

\subsubsection{Cross-platform verification}

Finally, there exist several certification methods with less stringent requirements and guarantees than self-testing and blind quantum computation, that can still be practically useful~\cite{eisert2020quantum}.
In modular architectures, one relevant method is cross-platform verification:
Consider two independent quantum devices that each prepare a state $\rho$ and $\sigma$, respectively.
One may wish to estimate the overlap between these states using only local operations and classical communication plus postprocessing. 
This can be achieved through estimation of the fidelity defined in Eq.~\eqref{eq:fidelity-eq2}~\cite{elben2020crossplatform,carrasco2021theoretical}.
This fidelity measure involves three components: $\tr(\rho^2)$, $\tr(\sigma^2)$, and $\tr(\rho\sigma)$.

The two purities can be estimated locally using randomized measurements, \textit{i.e.}, by applying random, completely separable unitaries and postprocessing the resulting measurement probabilities~\cite{vanenk2012measuring,elben2018renyi,elben2019statistical,elben2020mixedstate,elben2022randomized,huang2020predicting}.
To estimate $\tr(\rho\sigma)$, one can use second-order cross-correlations derived from the local measurement outcome distributions. 
The complexity is, though exponential, still significantly lower than that of full state tomography.
Specifically, estimating $\tr(\rho\sigma)$ using the cross-platform verification method requires $O(\max\{1/\varepsilon^2, \sqrt{d}/\varepsilon\})$ samples, compared to $\Omega(d^3)$ samples for full state tomography, where $\rho, \sigma \in \mathbb{C}^d$~\cite{anshu2022distributed}; see also Sec.~\ref{ssec:dist_inner_product}. \\

\implementations
While blind delegation and verification of quantum computation are still challenging to realize, several experimental building blocks have been demonstrated in the past years.
First blind quantum computations were performed using photons, in a server-client configuration where the client maintains the security of their data~\cite{barz2012demonstration,fisher2014quantum,greganti2016demonstration}.
The same photonic system was later used for the first demonstrated verification of a quantum computation, combining an interactive proof system with blind quantum computation~\cite{barz2013experimental}.
A proof-of-principle demonstration of blind quantum computing for classical clients achieved the task of factoring a small integer~\cite{huang2017experimental}.

Verification protocols are also being actively investigated on other platforms.
Interactive proofs of quantumness based on mid-circuit measurements have been studied with trapped ions~\cite{zhu2023interactive}, as well as a small-scale implementation of Mahadev's verification protocol for a single-qubit computations~\cite{stricker2024experimental}.
A trapped-ion quantum processor with up to $4 \times 4$ qubits was used to demonstrate verifiable quantum random sampling using measurement-based computation~\cite{ringbauer2025verifiable}.
A hybrid photon-matter implementation, based on a trapped-ion server and a photonic detection system at the client's side, analyzed and established bounds on information leakage through both quantum and classical channels~\cite{drmota2024verifiable}.
Recently, another small-scale blind quantum computing protocol was demonstrated using silicon-vacancy centers in diamond~\cite{wei2025universal}.

A proof-of-concept demonstration of robust self-testing has been achieved using polarization-entangled photon pairs~\cite{zhang2018experimentally}.
Parallel self-testing, in which two parties share two copies of a bipartite state or a tripartite configuration where a central node shares two independent states with peripheral nodes, was studied in a photonic system geared towards the certification of larger networks~\cite{agresti2021experimental}.
In another recent experiment, Bell-pair generation and measurements were self-tested in a system composed of two entangled superconducting circuits at a spatial separation of 30 m~\cite{storz2025complete}.

Cross-platform verification based on randomized measurements and its resource scaling has been experimentally investigated in superconducting and ion-trap platforms~\cite{zhu2022crossplatform}.
A similar protocol was used to compare two different quantum processes realized by distinct superconducting quantum processors~\cite{zheng2024crossplatform}.
In a recent experiment with two three-qubit modules connected by intermodular gates, a proof-of-principle demonstration of cross-platform verification based on Bell-basis measurements with superconducting circuits was achieved~\cite{dalton2025resourceefficient}.

\subsection{Quantum error mitigation \& correction}\label{ssec:qem-qec}

Achieving fault tolerance in quantum computing introduces significant overhead, making it probable that real-world applications will call for millions of physical qubits~\cite{gidney2021how,gidney2025how}.
So, quantum error correction may also benefit from modular architectures, \textit{e.g.}, through concatenation, where each module implements the same error-correcting code, and quantum links enable a concatenated code at the logical level of the original code~\cite{knill1996concatenated}.
Coherent access to multiple state copies, shared over a network, would also be useful in the context of error mitigation.
In this section, we briefly review progress on quantum error mitigation and correction in modular architectures.

\subsubsection{Quantum error mitigation}

Quantum error mitigation is targeted at current noisy quantum devices that lack the ability to perform quantum error correction.
Many methods have been devised that allow to improve the signal of a quantum computation, \emph{e.g.}, by trading in a higher number of circuit repetitions for a shorter circuit depth~\cite{cai2023quantum}.
In light of distributed quantum information processing, we consider a branch of error mitigation methods where multiple copies of a quantum state are used to suppress hardware noise.

One such method is \emph{virtual distillation}~\cite{huggins2021virtual}, or \emph{error suppression by derangement}~\cite{koczor2021exponential}, which uses $K$ copies of a quantum state $\rho$ to mitigate its incoherent part. Instead of estimating $\langle O \rangle = \tr(O\rho)$, this allows to compute a purified estimator
\begin{align}\label{eq:o-mitigated}
    \langle O \rangle_\text{mitigated} = \frac{\tr(O\rho^K)}{\tr(\rho^K)}.
\end{align}
The corresponding quantum circuit is shown in Fig.~\ref{fig:perm_test}, where the controlled-derangement operator $D$ is a permutation in which no element gets mapped onto itself.
This method is related to what~\cite{cotler2019quantum} denote as \emph{virtual cooling}:
with access to $K$ copies of a thermal state at temperature $T$, observables of a thermal state at a lower temperature $T/K$ can be estimated, based on the insight that a thermal state $\rho(T)$ obeys
\begin{equation}
    \rho(T/K) = \frac{\rho(T)^K}{\tr(\rho(T)^K)}.
\end{equation}
Another related protocol is \emph{echo verification}~\cite{cai2021resourceefficient,obrien2021error,huo2022dualstate}, where access is granted not to two separate copies in space, but to two copies separated in time (see also Sec.~\ref{ssec:dist_inner_product}).
These methods have similar resource requirements~\cite{cai2023quantum}.
In particular, the sampling costs of virtual distillation and error suppression by derangement generally scale exponentially in the number of copies $K$, because the denominator in Eq.~\eqref{eq:o-mitigated} is exponentially small in $K$ if $\rho$ is mixed.
The need for coherent copies may be circumvented using classical shadows which, however, comes at an overhead in required resources that scales exponentially in system size $N$~\cite{seif2023shadow}.

Echo verification and virtual distillation were experimentally applied to a range of quantum simulation problems using up to 20 superconducting qubits, finding error suppression by one to two orders of magnitude~\cite{obrien2023purificationbased}.

\subsubsection{Quantum error correction} \label{ssec:qec}

Quantum error correction (QEC) encodes logical qubits into collections of physical qubits, enabling the detection and correction of errors during computation~\cite{dur2007entanglement,devitt2013quantum,terhal2015quantum}.
As long as the physical error rate is below a threshold, enlarging the code reduces the logical error rate, which makes QEC the cornerstone of scalable, fault-tolerant quantum computing~\cite{aharonov2008faulttolerant}.
In the past years, progress has been rapid both theoretically and experimentally~\cite{campbell2024series}, with several self-contained introductions available~\cite{terhal2015quantum,roffe2019quantum,girvin2023introduction}.
In hardware architectures where qubits interact only with nearest neighbors on a planar lattice, planar codes and prominently the surface code~\cite{bravyi1998quantum,dennis2002topological,kitaev2003faulttolerant,fowler2012surface} and color code~\cite{steane1996error,bombin2006topological} stand out because they combine a comparatively high error threshold with planar connectivity.
Experimental milestones include implementations of surface-code primitives in superconducting circuits~\cite{barends2014superconducting,andersen2020repeated,krinner2022realizing,zhao2022realization,googlequantumai2023suppressing,googlequantumaiandcollaborators2025quantum} and neutral atoms~\cite{bluvstein2022quantum,bluvstein2024logical}, as well as instances of the color code with trapped ions~\cite{nigg2014quantum,ryan-anderson2021realization,ryan-anderson2022implementing,postler2022demonstration,paetznick2024demonstration,postler2024demonstration} and, more recently, superconducting qubits~\cite{lacroix2025scaling}.
At the same time, the demand for scaling towards millions of qubits has made modularity increasingly important.
Breaking large processors into smaller nodes interconnected through quantum links not only alleviates engineering constraints but also introduces qualitatively new challenges for maintaining fault tolerance.

One natural approach to modular fault tolerance is to run error-correcting codes locally within each node and to connect these nodes through quantum interconnects.
In such local QEC architectures, the effective threshold is determined not only by the performance of the local codes but also by the reliability of the interconnects.
One may expect this sensitivity to errors within the quantum link to significantly lower the effective fault-tolerance threshold.
Yet it has been theoretically analyzed and shown that fidelity requirements for quantum interconnects are less stringent than for local operations within the nodes~\cite{jiang2007distributed,nickerson2013topological,nickerson2014freely,li2016hierarchical,covey2023quantum,ramette2024faulttolerant,shalby2025optimized}.
One may distinguish between different strategies to connect different nodes with each other.

On a basic level, the quantum state-transfer techniques outlined in Sec.~\ref{sec:implementations} offer ways to connect separate modules with each other.
Beyond the results summarized there, there is an increasing interest in understanding, quantifying and controlling the types of errors introduced by various state transfer methods.
For example, some recent theoretical works have investigated the influence of the physical movement of qubits, or shuttling, on the error threshold of the surface code~\cite{siegel2024early,yenilen2025performance} and, as discussed previously, several experimental results have demonstrated instances of state transfer in various hardware platforms.
Despite this progress, the capabilities of today's interconnects are far from the desired communication rates and required fidelities.

One way of improving on this is via entanglement distillation, see Sec.~\ref{ssec:ent-dist}.
When entanglement is shared between modules in the form of Bell pairs, gate teleportation allows for operations across the modules.
To establish high-fidelity entanglement over noisy channels, entanglement distillation can be used to reduce many low-fidelity Bell pairs to fewer high-fidelity Bell pairs~\cite{bennett1996purification,deutsch1996quantum}.
While the corresponding protocols can produce entangled states with arbitrarily high fidelity in multiple rounds, the resource requirements are demanding and grow quickly.
In particular, the memory requirements increase exponentially with the number of rounds.
Memory requirements can be traded for increased temporal resources using (nested) entanglement pumping~\cite{briegel1998quantum,dur1999quantum,dur2003entanglement}.
The idea of turning noisy entangled states into high-fidelity links has been extended since then~\cite{campbell2007distributed,fujii2009entanglement}, leading to proposals for distributed architectures with relatively low quantum channel fidelities, even as low as $0.7$~\cite{fujii2012distributed}.

A widely adopted strategy for implementing two-qubit entangling gates between surface-code qubits is lattice surgery, where logical operations are implemented by merging code patches through joint stabilizer measurements~\cite{horsman2012surface}.
This method is native to two-dimensional topological stabilizer codes, which also include color codes~\cite{landahl2014quantum,thomsen2024lowoverhead}.
Because lattice surgery relies only on local interactions, it is well-suited to hardware with nearest-neighbor connectivity, and numerical analyses indicate that its overhead remains modest provided inter-module entanglement can be generated with sufficient speed and fidelity~\cite{ramette2024faulttolerant,jacinto2025network}.
One bottleneck of the method is its comparably slow speed, as it requires measurements on a subset of physical qubits, which can introduce new errors, such that multiple rounds of lattice surgery have to be performed to build confidence in the performed operation.
But owing to its feasibility and suitability for two-dimensional architectures, it is a powerful technique whose primitives have been realized in several laboratories.
Lattice-surgery experiments have been performed using trapped ions~\cite{erhard2021entangling,ryan-anderson2024highfidelity} and superconducting qubits~\cite{hetenyi2024creating,lacroix2025scaling,besedin2025realizing}.
Resource estimates have shown that quantum algorithms may be executed on distributed platforms via lattice surgery with little or no overhead compared to monolithic devices, provided that sufficiently fast and high-fidelity entanglement generation is available between modules~\cite{litinski2019game,gidney2021how,jacinto2025network}.

Alternatively, two-qubit logical gates may be realized transversally, by coupling every physical qubit in one code to its counterpart in another code.
Transversal operations thus act on physical qubits of one code block independently.
In platforms with reconfigurable connectivities, such as trapped ions~\cite{postler2022demonstration,ryan-anderson2024highfidelity} and Rydberg atoms~\cite{bluvstein2024logical}, transversal logical gates have been demonstrated.
On the other hand, transversal operations pose challenges for two-dimensional architectures with local connectivity, because of the nonlocal interactions they require~\cite{vasmer2019threedimensional,sahay2025error}.

Each of the previously highlighted and other unmentioned strategies, such as defect braiding~\cite{fowler2012surface}, comes with trade-offs in terms of communication and memory costs.
Using the surface code, a recent study has theoretically analyzed optimal strategies across different platforms and hardware parameters, discussing different networking regimes as a function of available memory and maximum rate of distributed physical Bell pairs~\cite{marqversen2025faulttolerant}:
transversal gates tend to outperform other methods when Bell rates are high, lattice surgery can be the only viable approach for small memories and low Bell rates, and for large memories and modest physical Bell rates, logical distillation may be preferable.

Another class of approaches to distributed QEC assumes individual code blocks being spread across multiple modules.
Code layouts for such distributed architectures have been studied theoretically in recent years~\cite{xu2022distributed,singh2024modular,babaie2025distributed,sutcliffe2025distributed,chandra2025distributed}.
With these designs, the number of non-local gates may be reduced when physical qubits from different code blocks are hosted on the same module, such that transversal gates may be executed locally~\cite{clayton2025distributed}.
While non-transversal gates remain a challenge in this setting, methods like magic-state injection or code switching may be considered.
While this class of approaches is conceptually appealing, it is not at a stage of experimental realization.

Looking further ahead, memory overheads may be lowered by packing many logical logical qubits into a single block of physical qubits.
In this context, it is noteworthy to mention recent progress with quantum low-density parity-check (LDPC) codes, which have asymptotically good parameters~\cite{breuckmann2021balanced,lin2022good,leverrier2022quantum,panteleev2022asymptotically,dinur2023good}.
LDPC codes are stabilizer codes in which the number of qubits participating in each check operator as well as the number of stabilizer checks that each qubits is part of are bounded by a constant~\cite{breuckmann2021quantum}.
The term \emph{good} refers to the desirable property of a code that is has constant encoding rate and constant relative distance; good quantum codes that are not LDPC codes have been known to exist since the early days of quantum error correction~\cite{calderbank1996good,ashikhmin2001asymptotically}.
Recent developments with quantum LDPC codes may pave the way towards quantum fault tolerance with a high coding efficiency~\cite{tremblay2022constantoverhead}, though a better understanding is needed of how to perform fast and parallelizable gates between the logical qubits they host~\cite{xu2025fast}.
A special family of quantum LDPC codes, known as bivariate bicycle (BB) codes and named after bivariate polynomials that they are based on, has recently been proposed and shown to exhibit high coding efficiency~\cite{bravyi2024highthreshold}.
In an experiment with superconducting qubits, the realization of BB code primitives was recently demonstrated~\cite{wang2025demonstration}.
Other notable recent theoretical progress includes, but is not limited to the study of quantum LDPC-based distillation~\cite{bonillaataides2025constantoverhead}, quantum LDPC codes for modular architectures~\cite{strikis2023quantum}, and the design of a distributed quantum memory based on flying qubits~\cite{tham2025distributed}.
Beyond such theoretical contributions, it remains interesting to see which experimental realizations will follow in the near future.

\subsection{Quantum machine learning \label{ssec:qml}}

Quantum machine learning (QML)~\cite{dong2008quantum, wittek2014quantum, aimeur2013quantum, biamonte2017quantum, schuld2015introduction, cerezo2022challenges, aimeur2006machine, dunjko2016quantumenhanced, dunjko2018machine, paparo2014quantum, saggio2021experimental} is an emerging field that combines quantum algorithms with machine learning techniques. 
It studies how quantum systems can represent, process and learn from available data.
By using the unique capabilities of quantum computers, the learning tasks can be improved by reducing runtime, memory requirements or sample complexities.
There are usually different kinds of QML paradigms such as quantum kernel methods~\cite{havlicek2019supervised, schuld2019quantum, huang2021power, schuld2021supervised, chatterjeegeneralized, schuld2020circuitcentric, rebentrost2014quantum} and variational models on parameterized quantum circuits~\cite{cerezo2021variational, bittel2021training, wang2021noiseinduced, beckey2022variational, xu2021variational, peruzzo2014variational, mcclean2016theory}.
It has also been theoretically demonstrated that for certain classes of problems, QML algorithms can offer computational advantages over their classical counterparts~\cite{liu2021rigorous, barthe2025quantum, molteni2025quantum, aimeur2013quantum, dunjko2016quantumenhanced, huang2021power, cerezo2022challenges, huang2021informationtheoretic, schuld2022quantum, saggio2021experimental}.

Given the immense scale of some problems (\emph{e.g.}, large datasets or complex models), it is often desirable to distribute tasks across multiple parties using either quantum~\cite{tang2023communicationefficient, neumann2022distributed,pira2023invitation,khait2023variational, gilboa2024exponential, gilboa2025consumable} or classical communication links~\cite{pira2023invitation, marshall2023high}. 
This yields speed-up and scalability to the QML algorithms due to the intrinsic parallelism of the distributed computing architecture.
In line with the general framework of distributed quantum computing, quantum links typically rely on entanglement partially shared among the participating parties such as Bell or GHZ states.
Alternatively, classical links based on LOCC, including mid-circuit measurements and conditional gate applications, can also be employed to enable collaborative QML among multiple parties. 

Distributed QML also offers advantages in scenarios where data security and client privacy are of great importance~\cite{sheng2017distributed, li2024blind, chen2021federated, larasati2022quantum, kwak2023quantum, ren2025quantum}.
A notable example is federated learning, where multiple clients jointly train a model using their local datasets without exposing them.
This collaborative model fits naturally into distributed QML and may benefit from enhanced security and scalability offered by quantum protocols~\cite{chen2021federated, larasati2022quantum, kwak2023quantum, ren2025quantum}. In classical federated learning, a central server coordinates training based on information exchanged with clients~\cite{li2020review, zhang2021survey, kairouz2021advances, li2020federated, mammen2021federated}.
However, untrusted communication channels may pose a risk, as adversaries may reverse-engineer the datasets by analyzing uploaded gradients. 
These issues can be efficiently addressed in the quantum picture to protect privacy. 
For example, techniques such as blind quantum computing can protect client data, which has been briefly reviewed in~\ref{sssec:bqc}.
One can also apply apply differential privacy~\cite{angrisani2023unifying, dwork2014algorithmic, hilton2012differential, dwork2008differential, dwork2006calibrating, angrisani2025quantum}, where clients add relatively proper noise to gradients before uploading them~\cite{li2021quantum, li2024privacypreserving, huang2022quantuma}.
Moreover, there are also approaches using quantum homomorphic encryption~\cite{liang2015quantum, zeuner2021experimental, li2024experimental}, which reduce communication complexity compared to blind quantum computing while preserving privacy~\cite{li2025quantum}. 
Additionally, other techniques such as quantum key distribution~\cite{zhang2023federated, sheng2017distributed, kaewpuang2023adaptive, xu2023privacypreserving} and post-quantum cryptography (PQC)~\cite{xu2023post} are also incorporated into quantum federated learning protocols to ensure secure and robust collaborative training.

\subsection{State preparation \label{ssec:state-preparation}}

In the context of optimizing adiabatic state preparation, the use of a distributed setup is proposed in~\cite{schiffer2022adiabatic}.
There, a simple phase-estimation inspired protocol is used on two devices to obtain a measure for the closeness of a quantum state to a target Hamiltonian ground state.
While a single-device approach requires sampling to estimate the measure, the two-device method leverages hypothesis testing to achieve exponentially fast convergence.

Another method of preparing an eigenstate of a given Hamiltonian $H$ is filtering~\cite{poulin2009preparing,ge2019faster,lu2021algorithms,irmejs2024efficient}.
Filtering can be implemented with a controlled unitary dynamics and a single auxiliary qubit (cf.~\cite{abrams1999quantum, xu2014demonlike, chen2020quantum}).
Similarly to iterative quantum phase estimation, this enables a spectral projection onto an eigenstate of the Hamiltonian.
Concretely, considering a Hadamard test, the energy variance of the initial state is reduced with every iteration.
The setup can be extended to a distributed setup, where two devices, Alice and Bob, perform identical operations in each iteration~\cite{schiffer2023quantum, liu2025preparing}. 
A coherent link (akin to a SWAP test) creates entanglement between Alice and Bob via the local auxiliary qubits. 
The resulting states will asymptotically converge to two identical eigenstates, distributed across both parties.
Tracing out one of the systems leaves the remaining state with an even lower energy variance, enabling faster convergence to an eigenstate.
Note that the postselection for the distributed filter is required, either weak or strong. 
If the measurement result at a given iteration falls outside the specified postselection criteria, the algorithm must be reset and restarted from the beginning. 
Strong postselection, while having stricter criteria that accelerate eigenstate convergence, incurs greater overhead due to the higher number of discarded outcomes that fail to meet the criteria.
This setup can also be extended to more than two devices, where the entangling operation on the respective auxiliary qubits is then a derangement operator, and the auxiliary degrees of freedoms are in a superposition of qudit states.

\subsection{Quantum metrology \label{ssec:q-metrology}}

The field of quantum metrology~\cite{giovannetti2006quantum} studies the precise measurement of physical parameters, thereby exploiting quantum effects such as entanglement between sensors.
It can thus be understood as a form of distributed quantum information processing, which is why we include a short description.
We note that as quantum metrology has developed into an independent and very fruitful area of research, any attempt of providing a complete description of the field would be beyond the scope of this review.
Rather, we point the reader to several reviews dedicated to quantum metrology and quantum sensing~\cite{giovannetti2004quantumenhanced, paris2009quantum, degen2017quantum, sidhu2020geometric, toth2014quantum}.
Additionally, we briefly introduce key concepts and highlight individual papers that are potentially relevant in the context of the previous sections.

To estimate a parameter $\theta$ from a quantum state $\rho(\theta)$, the precision of the estimation scales as $\Delta \theta_\text{SQL}\sim 1/\sqrt{n}$, for $n$ independent, \emph{i.e.}~unentangled, probes.
This is also known as the standard quantum limit (SQL) or shot noise limit.
A key objective of quantum metrology is to overcome the SQL by introducing entanglement between the probes. 
Using quantum states such as GHZ states, NOON states or spin-squeezed states allows to improve over the SQL towards $\Delta \theta\sim 1/n$, which is the fundamental limit for precision called the Heisenberg limit, originating from the quantum Cramér-Rao bound and the quantum Fisher information~\cite{humphreys2013quantum, lloyd2008enhanced, meyer2021fisher}.
The Laser Interferometer Gravitational-Wave Observatory (LIGO) famously used squeezed light to enhance the sensitivity of gravitational wave detection~\cite{aasi2013enhanced}.
In the context of distributed quantum information processing, one considers entanglement between spatially separated sensors. 
Applications include a quantum network of clocks~\cite{komar2014quantum}. 

Besides boosting the asymptotic scaling of the precision towards the Heisenberg limit, another advantage of spatially distributed quantum sensors is the estimation of nonlocal observables.
Such a nonlocal observable is a function of multiple parameters, which are respectively encoded in spatially separated probes, and cannot be accessed by a local probe for this reason. 
For an analysis of measurement strategies we refer to~\cite{eldredge2018optimal, proctor2018multiparameter}.
Privacy aspects of distributed sensing networks were investigated in~\cite{bugalho2025private}.
Finally, realistic implementations must take into account the effect of hardware imperfections that degrade measurement efficiency. 
Several approaches exist to incorporate quantum error-correction schemes to mitigate noise for sensing applications~\cite{zhuang2020distributed, kessler2014quantum, dur2014improved}.

\section{Conclusions \& Outlook \label{sec:outlook}}

Distributed quantum computing is increasingly recognized as a key architectural strategy for addressing scalability challenges of quantum hardware and control.
Rather than relying on large, monolithic devices, distributed quantum information processing envisions modular quantum processors connected via quantum and classical channels, enabling the execution of large-scale computations across spatially separated nodes.
This modularity offers several opportunities in the design of devices and algorithms.

As experimental platforms continue to mature and early demonstrations of distributed quantum computation emerge, it is timely to consider the theoretical limits, practical bottlenecks, and architectural design choices that will shape the field over the coming decade.
The transition to fault-tolerant quantum computing presents both challenges and opportunities.
Current proposals for fault tolerance, such as surface and quantum low-density parity-check (qLDPC) codes, yield demanding hardware requirements.
Distributing logical qubits across multiple nodes offers a potential route to modular fault-tolerant architectures that reduce local hardware demands and exploit teleportation-based gates and entanglement-assisted coding schemes.
However, identifying the regimes in which such approaches become advantageous remains an open challenge.
Quantitative comparisons, for example, in terms of total logical $T$-gate count, inter-node entanglement rate, or memory lifetimes, are needed to assess the practicality of distributed logical operations.

Many of the core building blocks of modular architectures are established in the bipartite setting.
Teleportation, entanglement swapping, Bell-basis measurements, and SWAP tests have been extensively studied and are beginning to be implemented across short-distance links in laboratories.
However, generalizing these primitives to multipartite settings is not trivial.
Theoretical proposals such as the permutation test, derangement-based projections, and recursive purification offer pathways toward scalable multi-copy operations and entanglement certification, but remain experimentally challenging due to their need for tight synchronization and precise multi-qubit control across distinct hardware.

Recent work has begun to delineate the complexity landscape of distributed quantum protocols under constrained resources such as, for example, limited quantum communication, restricted measurements, or access to only a small number of state copies.
A recurring theme is that access to multiple copies of a quantum state, or to a state and its conjugate, can enable learning and estimation tasks with exponentially fewer samples than single-copy or classical methods.
This advantage can persist even when inter-node communication is highly constrained.
A particularly active area of research is the characterization of quantum advantage in distributed settings for locally checkable problems.
While classical distributed algorithms are known to be efficient for many locally checkable programs, it remains open whether coherent quantum communication or multi-copy access can provably speed up coloring, labeling or other graph-based tasks in realistic network topologies.

As experiments push toward modular architectures, the need for lightweight and scalable verification protocols becomes increasingly acute.
Randomized measurement techniques, classical shadows, and Bell-sampling-based fidelity estimation have already shown promise for certifying entanglement and benchmarking operations in bipartite networks.
Extending these tools to heterogeneous and multipartite settings is a key challenge.
Interactive protocols for delegated computations will likely become central in practical deployments.
A particularly pressing need is for verification schemes that can function when node capabilities are asymmetric, such as networks where some devices have limited quantum memory or are restricted to classical communication.
These hybrid scenarios are expected to arise naturally in early quantum networks.

On the engineering side, several bottlenecks must be overcome.
High-fidelity entanglement distribution remains a major limiting factor, particularly in the multipartite regime.
Quantum memories with long coherence times and rapid, high-bandwidth input-output are needed to synchronize operations across probabilistic links.
Cross-platform integration, for instance, connecting superconducting qubits to photonic interconnects, or combining atomic ensembles with solid-state devices, requires advances in calibration, timing, and interfacing.
Moreover, distributed architectures demand entirely new layers of software:
for routing quantum information, scheduling entanglement generation and consumption, managing distributed error correction, and performing real-time circuit adaptation in response to probabilistic events across the network.

Despite these challenges, distributed quantum information processing is expected to emerge as an increasingly practical complement to monolithic quantum architectures.
Its modularity not only aligns with foreseeable hardware constraints but also unlocks qualitatively new capabilities like nonlocal operations across space, distributed state learning, and cooperative quantum sensing and metrology.
We expect the coming years to yield rigorous demonstrations of quantum advantage in distributed models, the development of a robust theoretical framework for understanding the power and limitations of distributed quantum systems and protocols, and the deployment of experimental testbeds.

\begin{acknowledgements}

We thank Jean-Claude Besse, Dongling Deng, \mbox{Vedran Dunjko}, Andreas Elben, Juan José García-Ripoll, Elies Gil-Fuster, Alexandru Gheorghiu, Richard Kueng,  Weikang Li, Norbert Linke, Stefano Paesani, Manuel Rispler, Lieven Vandersypen, Andreas Wallraff and Hengyun (Harry) Zhou for providing valuable feedback on the manuscript.

J.K.~acknowledges financial support by the Swiss State Secretariat for Education, Research and Innovation under contract number UeM019-11, and by ETH Zurich.
B.F.S.~acknowledges funding from the German Federal Ministry of Research, Technology and Space (BMFTR) via the project FermiQP (13N15889). 
Work at MPQ is part of the Munich Quantum Valley, which is supported by the Bavarian state government with funds from the Hightech Agenda Bayern Plus. 
This work is supported by the Dutch National Growth Fund (NGF), as part of the Quantum Delta NL programme.
This work has received support from the European Union’s Horizon Europe research and innovation programme through the ERC StG FINE-TEA-SQUAD (Grant No.~101040729), funded by the European Union.
Views and opinions expressed are however, those of the authors only and do not necessarily reflect those of the European Union or the European Commission. Neither the European Union nor the granting authority can be held responsible for them.

\end{acknowledgements}

\bibliographystyle{apsrev4-2-titles}
\bibliography{distributed_qc_bib}

\end{document}